\documentclass[12pt]{article}%
\usepackage[nosort]{cite}
\usepackage{graphicx}
\usepackage{multicol}
\usepackage{amsfonts}
\usepackage{amssymb}
\usepackage{amsmath}
\usepackage{heck}
\usepackage{afterpage}
\usepackage{setspace}
\usepackage{verbatim}
\usepackage{color}
\usepackage{longtable}
\usepackage{float}
\usepackage{subcaption}
\usepackage{epsfig}
\usepackage{epstopdf}
\usepackage{adjustbox}
\usepackage{tikz}
\usepackage[margin=1in]{geometry}
\usepackage{titletoc}
\usepackage{hyperref}%
\usepackage{mathrsfs}

\providecommand{\U}[1]{\protect\rule{.1in}{.1in}}
\newsavebox{\mysavebox}

\hypersetup{colorlinks,citecolor=black,filecolor=black,linkcolor=black,urlcolor=black,pdftex}
\usetikzlibrary{decorations.markings}

\numberwithin{equation}{section}

\newcommand{\be}{\begin{equation}}
\newcommand{\ee}{\end{equation}}

\usepackage{tikz}
\usetikzlibrary{arrows}
\usetikzlibrary{arrows.meta}
\usetikzlibrary{arrows,decorations.pathmorphing}
\usetikzlibrary{shapes.geometric,calc,arrows, positioning,shapes.misc,decorations.markings}
\tikzset{
  big arrow/.style={
    decoration={markings,mark=at position 1 with {\arrow[scale=2,#1]{>}}},
    postaction={decorate},
    shorten >=0.4pt},
  big arrow/.default=black}

\pgfdeclarelayer{edgelayer}
\pgfdeclarelayer{nodelayer}
\pgfsetlayers{edgelayer,nodelayer,main}
\tikzstyle{none}=[inner sep=0pt]

\tikzstyle{NodeCross}=[draw, shape=circle, cross out, inner sep=0pt, minimum size=6pt,line width=0.25mm]
\tikzstyle{Circle}=[draw, shape=circle, black, fill=black, inner sep=0pt, minimum size=6pt]
\tikzstyle{circle}=[draw, shape=circle, black, fill=black, inner sep=0pt, minimum size=16pt]
\tikzstyle{Star}=[draw, shape=star, fill=black, star points=8, inner sep=0pt, minimum size=8pt]
\tikzstyle{CircleRed}=[draw, shape=circle, black, fill=red, inner sep=0pt, minimum size=6pt]
\tikzstyle{StarP}=[draw={rgb,255: red,128; green,0; blue,128}, shape=star, fill={rgb,256: red,128; green,0; blue,128}, star points=8, inner sep=0pt, minimum size=12pt]
\tikzstyle{ShadedCircRed}=[draw=red, shape=circle, fill={rgb, 255: red,255; green,114; blue, 118}, inner sep=0pt, minimum size=80pt, line width=0.5mm, fill opacity=0.2]
\tikzstyle{ShadedCircRed2}=[draw=red, shape=circle, fill={rgb, 255: red,255; green,114; blue, 118}, inner sep=0pt, minimum size=10pt]
\tikzstyle{ShadedCircRed3}=[draw=black, shape=rectangle, fill={rgb, 255: red,255; green,114; blue, 118}, inner sep=0pt, minimum size=113pt, line width=0.25mm]
\tikzstyle{ShadedCirc}=[draw=red, shape=circle, fill=white, inner sep=0pt, minimum size=45pt,  fill opacity=1.0,  line width=0.5mm]
\tikzstyle{CircleBlue}=[draw, shape=circle, fill=blue, inner sep=0pt, minimum size=6pt]
\tikzstyle{BigCirclePurple}=[draw, shape=circle, fill={rgb,255: red,191; green,0; blue,191}, inner sep=0pt, minimum size=12pt]
\tikzstyle{CirclePurple}=[draw, shape=circle, fill={rgb,255: red,191; green,0; blue,191}, inner sep=0pt, minimum size=5pt]
\tikzstyle{EmptyCircle}=[draw, shape=circle, inner sep=0pt, minimum size=4pt]
\tikzstyle{GreenCircle}=[draw, shape=circle,  fill={rgb,255: red,80; green,200; blue,120}, inner sep=0pt, minimum size=8pt]
\tikzstyle{BrownCircle}=[draw, shape=circle,  fill={rgb,255: red,210; green,105; blue,30}, inner sep=0pt, minimum size=8pt]
\tikzstyle{CirclePurpleSmall}=[draw, shape=circle, fill={rgb,255: red,191; green,0; blue,191}, inner sep=0pt, minimum size=4pt]
\tikzstyle{BigCircleGreen}=[draw, shape=circle, fill={rgb,255: red,0; green,191; blue,0}, inner sep=0pt, minimum size=12pt]
\tikzstyle{BigCircleBlue}=[draw, shape=circle, fill={rgb,255: red,0; green,0; blue,191}, inner sep=0pt, minimum size=12pt]
\tikzstyle{BigCircleRed}=[draw, shape=circle, fill={rgb,255: red,191; green,0; blue,0}, inner sep=0pt, minimum size=12pt]
\tikzstyle{BrownCircleSmall}=[draw, shape=circle,  fill={rgb,255: red,210; green,105; blue,30}, inner sep=0pt, minimum size=6pt]

\tikzstyle{SmallCircleBrown}=[draw, shape=circle,  fill={rgb,255: red,210; green,105; blue,30}, inner sep=0pt, minimum size=5pt]

\tikzstyle{SmallCircleRed}=[draw, shape=circle, fill={rgb,255: red,191; green,0; blue,0}, inner sep=0pt, minimum size=6pt]

\tikzstyle{DashedLine}=[-, densely dashed, line width=0.25mm]
\tikzstyle{DottedLine}=[-, dotted, line width=0.25mm]
\tikzstyle{ThickLine}=[-, line width=0.25mm]
\tikzstyle{ArrowLineRight}=[-, -{Stealth[scale=1.25]}, line width=0.25mm, scale=5]
\tikzstyle{ArrowLineRed}=[-, draw={rgb,255: red,191; green,0; blue,0}, -{Stealth[scale=1.75]}, line width=0.1mm, scale=5]
\tikzstyle{RedLine}=[-, draw={rgb,255: red,191; green,0; blue,0}, fill=none, line width=0.5mm]
\tikzstyle{DashedLineThin}=[-, densely dashed, line width=0.125mm, fill=none, draw=black]
\tikzstyle{DottedRed}=[-, dotted, draw={rgb,255: red,191; green,0; blue,0}, fill=none, line width=0.25mm]
\tikzstyle{DashedRed}=[-, densely dashed, draw={rgb,255: red,191; green,0; blue,0}, fill=none, line width=0.25mm]
\tikzstyle{BlueLine}=[-, draw={rgb,255: red,0; green,0; blue,191}, fill=none, line width=0.5mm]
\tikzstyle{ArrowLineBlue}=[-, draw={rgb,255: red,0; green,0; blue,191}, -{Stealth[scale=1.75]}, line width=0.1mm, scale=5]
\tikzstyle{GreenDoubleArrow}=[<->, draw={rgb,155: red,0; green,255; blue,0},  line width= 0.5mm, scale=5]
\tikzstyle{RedDoubleArrow}=[<->, draw={rgb,255: red,255; green,0; blue,0},  line width= 0.5mm, scale=5]
\tikzstyle{BlueDottedLight}=[-, dotted, draw={rgb,255: red,0; green,0; blue,191}, fill=none, line width=0.3mm]
\tikzstyle{BrownLine}=[-, draw={rgb,255: red,210; green,105; blue,30}, fill=none, line width=0.5mm]
\tikzstyle{DottedRed}=[-, dotted, draw={rgb,255: red,191; green,0; blue,0}, fill=none, dotted, line width=0.5mm]
\tikzstyle{DottedPurple}=[-, dotted, draw={rgb,255: red,191; green,0; blue,191}, fill=none, dotted, line width=0.5mm]
\tikzstyle{BlueDottedLight}=[-, dotted, draw={rgb,255: red,0; green,0; blue,191}, fill=none, line width=0.5mm]
\tikzstyle{ArrowLinePurple}=[-, draw={rgb,255: red,191; green,0; blue,191}, -{Stealth[scale=1.75]}, line width=0.5mm, scale=5]
\tikzstyle{DashedLineGreen}=[-, densely dashed, draw={rgb,255: red,74; green,103; blue,65}, line width=0.25mm]
\tikzstyle{LineGreen}=[-, draw={rgb,255: red, 74; green,200; blue,65}, line width=0.5mm]
\tikzstyle{ArrowLineGreen}=[-, draw={rgb,255: red,0; green,191; blue,0}, -{Stealth[scale=1.75]}, line width=0.5mm, scale=5]
\tikzstyle{GreenLine}=[-, draw={rgb,255: red,0; green,191; blue,0}, fill=none, line width=0.5mm]
\tikzstyle{PurpleLine}=[-, draw={rgb,255: red,191; green,0; blue,191}, fill=none, line width=0.5mm]
\tikzstyle{PPurpleLine}=[-, draw={rgb,255: red,191; green,0; blue,191}, fill=none, line width=2.5mm]
\tikzstyle{DPurpleLine}=[-, dotted, draw={rgb,255: red,191; green,0; blue,191}, fill=none, line width=0.5mm]
\tikzstyle{SBrownLine}=[-, draw={rgb,255: red,191; green,0; blue,191}, fill=none, opacity=0.35, line width=2.5mm]
\tikzstyle{DottedBlue}=[-, dotted, draw=blue, fill=none, dotted, line width=0.5mm]

\tikzstyle{DashedPurpleLine}=[-, densely dashed, draw={rgb,255: red,191; green,0; blue,191}, fill=none, line width=0.5mm]

\tikzstyle{SmallCircleBlue}=[draw, shape=circle, fill=blue, inner sep=0pt, minimum size=5pt]
\tikzstyle{SmallCirclePurple}=[draw, shape=circle, fill={rgb,255: red,191; green,0; blue,191}, inner sep=0pt, minimum size=5pt]

\tikzset{snake it/.style={decorate, decoration=snake}}

\usetikzlibrary{decorations.markings}

\usetikzlibrary{hobby}

\tikzset{
dashstar/.style={
 dash pattern=on 5pt off 5pt,
 postaction={
  decorate,
  decoration={
   markings,
   mark=between positions 9pt and 1 step 10pt with {
     \node[color=red] {*};
   }
  }
 }
},
dashstarstar/.style={ 
 dash pattern=on 5pt off 10pt,
 postaction={
   decorate,
   decoration={
     markings,
     mark=between positions 10pt and 1
          step 15pt
           with {
            \node at (-2pt,0pt) {\pgfuseplotmark{star}};
            \node at (2pt,0pt) {\pgfuseplotmark{star}};
           }
   }
 }
}
}
\usepackage{pgfplots}
\pgfplotsset{compat=1.16}

\newcommand{\lb}{\left(}
\newcommand{\rb}{\right)}
\newcommand{\lbb}{\left[}
\newcommand{\rbb}{\right]}

\newcommand{\ba}{\begin{aligned}}
\newcommand{\ea}{\end{aligned}}
\newcommand{\Z}{{\mathbb Z}}
\newcommand{\R}{{\mathbb R}}

\renewcommand{\P}{{\mathbb P}}

\begin{document}

\date{April 2024}

\title{Generalized Symmetries of Non-Supersymmetric Orbifolds}

\institution{PENN}{\centerline{$^{1}$Department of Physics and Astronomy, University of Pennsylvania, Philadelphia, PA 19104, USA}}
\institution{PENNmath}{\centerline{$^{2}$Department of Mathematics, University of Pennsylvania, Philadelphia, PA 19104, USA}}

\authors{
Noah Braeger\worksat{\PENN}\footnote{e-mail: \texttt{braeger@sas.upenn.edu}},
Vivek Chakrabhavi\worksat{\PENN}\footnote{e-mail: \texttt{vivekcm@sas.upenn.edu}},\\[4mm]
Jonathan J. Heckman\worksat{\PENN,\PENNmath}\footnote{e-mail: \texttt{jheckman@sas.upenn.edu}}, and
Max H\"ubner\worksat{\PENN}\footnote{e-mail: \texttt{hmax@sas.upenn.edu}}
}

\abstract{We determine generalized symmetries for 4D theories engineered via type II strings on non-supersymmetric orbifold backgrounds $\mathbb{R}^{3,1} \times \mathbb{R}^6 / \Gamma$. Probe branes detect generalized symmetries via the adjacency matrix for fermionic degrees of freedom in an associated quiver gauge theory. In situations where the tachyons are sequestered away from the boundary $S^5 / \Gamma$, this exactly matches the result extracted from singular homology. In situations with an unsequestered tachyon which stretches out to the boundary, the presence of tachyonic pulses partitions up the space into several distinct sectors, and the net contribution again matches with the answer expected via quiver methods. For IIA backgrounds, the presence of a localized closed string tachyon leads to transitions in the spectrum of states, generalized symmetries, higher-group symmetries, as well as the level matrix of the associated symmetry topological field theory (SymTFT). For IIB backgrounds with a stack of spacetime filling probe D3-branes, the onset of a radiatively generated potential leads to similar considerations involving scale dependent transitions in the symmetries of the theory, including structures such as duality defects / interfaces.
}

\maketitle

\enlargethispage{\baselineskip}

\setcounter{tocdepth}{2}

\tableofcontents

\newpage

\section{Introduction}

The extra dimensions of string theory provide a general template for constructing and studying a wide variety of
interacting quantum systems. For example, this has led to the discovery of entirely new classes of quantum field theories,\footnote{See
e.g., the reviews \cite{Heckman:2018jxk, Argyres:2022mnu} and references therein.} as well as the development of new tools to study many systems at strong coupling.

A feature implicit in many such analyses is the use of supersymmetry. Indeed, especially at strong coupling, supersymmetry is very helpful in studying properties of such systems because various quantities of interest are still protected, e.g., by holomorphic structures.

But Nature is not supersymmetric (at least at currently probed energies).

Indeed, comparatively less progress has been made in the study of non-supersymmetric string backgrounds due to a number of interrelated issues. First of all, the absence of supersymmetry means that there will necessarily be less control over any putative strong coupling dynamics. Additionally, non-supersymmetric backgrounds are often accompanied by tachyonic excitations, which in turn leads to non-trivial time dependance.\footnote{In favorable circumstances it is possible to say more, but it is fair to say that many aspects of strings on non-supersymmetric backgrounds remain poorly understood.}

Given this state of affairs, it is natural to seek out other robust tools. Particularly promising from this standpoint is the discovery of generalized global symmetries \cite{Gaiotto:2014kfa} and their various categorical generalizations.\footnote{See, e.g., the reviews \cite{Cordova:2022ruw, Schafer-Nameki:2023jdn, Bhardwaj:2023kri, Luo:2023ive, Brennan:2023mmt, Shao:2023gho} and references therein.} In this framework, topological symmetry operators link / intersect with appropriate charged defects / operators. This topological formulation is quite powerful since it is insensitive to local deformations. As such, it provides a particularly robust way to access strong coupling phenomena in a wide variety of quantum systems, regardless of whether supersymmetry is present.

Now, in the context of string realizations of quantum field theories (QFTs), the main arena of application has thus far centered on systems
which are supersymmetric. Indeed, the best studied cases involve an extra-dimensional geometry of the form $X$ a conical geometry with the degrees of freedom of the QFT localized at the tip of the cone. Heavy defects of the QFT descend from branes wrapping relative cycles which stretch from the tip of the cone to the conformal boundary $\partial X$. Topological symmetry operators are obtained from magnetic dual branes which wrap cycles linking / intersecting with these heavy defects \cite{Apruzzi:2022rei, GarciaEtxebarria:2022vzq, Heckman:2022muc, Heckman:2022xgu, Dierigl:2023jdp, Cvetic:2023plv, Bah:2023ymy, Heckman:2024oot}.\footnote{It is worth noting here that while the heavy defects are typically assumed to preserve some supersymmetry, the topological operators are not BPS, and in fact, can even involve unstable / non-supersymmetric branes \cite{Dierigl:2023jdp}.}

The special case of type IIA on a supersymmetric orbifold $\mathbb{C}^3 / \Gamma$ for $\Gamma$ a finite subgroup of $SU(3)$ leads to a 4D $\mathcal{N} = 2$ QFT decoupled from gravity. For geometries which have collapsing curves and divisors, this engineers a
4D $\mathcal{N} = 2$ superconformal field theory (SCFT) of Argyres-Douglas type \cite{Argyres:1995jj}.\footnote{These can be obtained from dimensional reduction of the 5D SCFTs engineered via M-theory on the same singularity (see e.g, \cite{Seiberg:1996bd, Morrison:1996xf, Douglas:1996xp}).} Higher-form symmetries for these 4D theories and their lifts to 5D SCFTs follow from the calculation of the defect group \cite{DelZotto:2015isa} of the associated resolved geometry \cite{Albertini:2020mdx, Morrison:2020ool, Tian:2021cif} and can also be extracted directly from the spectrum of electric / magnetic particles obtained from wrapped branes \cite{DelZotto:2022fnw, DelZotto:2022ras}.\footnote{More precisely, in \cite{DelZotto:2022fnw} 5D SCFTs were engineered via supersymmetric orbifolds of the form $\mathbb{C}^3 / \Gamma$. The higher-form symmetries for these theories were then obtained by considering the spectrum of BPS particles in the Kaluza-Klein reduced theory, which is in turn captured by the BPS quiver of the associated 4D theory.} In more detail, the adjacency matrix for the quiver quantum mechanics of a probe D0-brane encodes the Dirac pairing for BPS states, which in turn fixes the 1-form symmetries of the system \cite{DelZotto:2022ras}. This same data is also captured in terms of the boundary homology $H_{\ast}(S^5 / \Gamma_{SU(3)})$. In cases where the group action on the boundary geometry has a fixed point locus, the resulting 4D theory has a non-trivial flavor symmetry, and there can then also be a non-trivial entwinement in the 0-form and 1-form symmetries, leading to a 2-group \cite{DelZotto:2022fnw, Cvetic:2022imb, DelZotto:2022joo}.

Similar considerations hold for type IIB backgrounds with a stack of $N$ D3-branes probing the orbifold singularity, i.e., by formally T-dualizing the type IIA probe D0-brane case. The \textit{same} geometry leads to $0$-form and $2$-form symmetries, and suitable tuning of the background axio-dilaton also leads to non-trivial topological duality defects,\footnote{See e.g., references \cite{Choi:2021kmx, Kaidi:2021xfk, Choi:2022zal, Kaidi:2022uux, Kaidi:2022cpf, Bashmakov:2022jtl, Heckman:2022xgu, Bashmakov:2022uek, Damia:2023ses}.} via constant axio-dilaton 7-branes at infinity \cite{Heckman:2022xgu}.

In this paper we show that this picture extends to non-supersymmetric type II string backgrounds of the form $\mathbb{R}^{3,1} \times \mathbb{R}^6 / \Gamma_{SU(4)}$ for $\Gamma_{SU(4)}$ a finite subgroup of $SU(4)$. At the level of the closed string spectrum the primary difference from the supersymmetric case is that we now expect to have a localized tachyon in a twisted sector of the closed string Hilbert space. This in turn means that the closed string background will dynamically resolve due to tachyon condensation.\footnote{The case of $\Gamma$ abelian was studied in \cite{Morrison:2004fr}, and related non-supersymmetric orbifold geometries have been considered in \cite{Adams:2001sv, Martinec:2001cf, Harvey:2001wm, Dabholkar:2001wn, Vafa:2001ra, Narayan:2009uy}.} We treat both the case of type IIA on this ``pure geometry'' as well as the case of type IIB with $N$ spacetime filling D3-branes. While our considerations hold for general finite $\Gamma_{SU(4)} \subset SU(4)$, we primarily present examples based on
abelian groups.

In the type IIA ``pure geometry'' case, the dynamic resolution of localized singularities means that the spectrum of heavy defects, as well as dynamical states which can screen these defects will have non-trivial time dependence. We explicitly track this dependence by again considering the quiver quantum mechanics of D-brane probes of the geometry. Much as in the supersymmetric case, this allows us to extract candidate electric and magnetic 1-form symmetries, as well as possible higher-categorical structures. The quiver based approach tracks the full basis of D-branes on this geometry, and thus leads to a natural collection of candidate defects and wrapped branes. After a tachyon fully condenses, we reach a new background, and the quiver quantum mechanics transitions to a new spectrum of generalized symmetries. This data can be captured in terms of the associated symmetry topological field theory (SymTFT),\footnote{See,
e.g., references
\cite{Heckman:2024oot, Reshetikhin:1991tc, Turaev:1992hq, Barrett:1993ab, Witten:1998wy,
Fuchs:2002cm, Kirillov:2010nh, Kapustin:2010if, Kitaev:2011dxc, Fuchs:2012dt,
Freed:2012bs, Heckman:2017uxe, Freed:2018cec, Gaiotto:2020iye, Apruzzi:2021nmk, Freed:2022qnc, Kaidi:2022cpf, Baume:2023kkf,Brennan:2024fgj, Antinucci:2024zjp, Bonetti:2024cjk, Apruzzi:2024htg, DelZotto:2024tae, GarciaEtxebarria:2024, Argurio:2024oym}.} but in which we make a step function approximation for the level matrix of the system which ``jumps'' at fixed times. There is an associated Euclidean worldvolume theory at each fixed timeslice where a jump occurs.

We present a number of examples to illustrate these general points. Quite remarkably, we find an exact match between the answer expected from quiver techniques and that expected from the topology of the boundary $S^5 / \Gamma$ in all cases where no tachyon is initially present on the boundary space. In situations where there is a tachyon present on the boundary, the net effect is to partition the $S^5 / \Gamma$ geometry up into distinct slices, and the net contribution over all the slices again matches to the answer expected from quiver based methods.

In the type IIB case with $N$ spacetime filling D3-branes at an orbifold singularity (see \cite{Douglas:1996sw, Kachru:1998ys, Lawrence:1998ja, Hanany:1998sd}), much of the geometric structure is similar, but the physical interpretation is somewhat different; there is now a non-trivial scale dependence in the 4D QFT, as captured by the flow of parameters for double trace operators and / or a radiatively generated Coleman-Weinberg potential \cite{Adams:2001jb, Dymarsky:2005uh, Dymarsky:2005nc, Horowitz:2007pr, Pomoni:2009joh}. This in turn triggers an instability away from the origin of field space. Precisely because the D3-brane probe is a quiver gauge theory, we can again read off the basis of fractional branes, including candidate heavy defects which can be screened by dynamical states. In this case, the associated SymTFT involves an evolution as a function of renormalization group (RG) scale, with transitions / jumps in the theory captured by interfaces. This is in line with the structure of a SymTree with a single branch \cite{Baume:2023kkf}. While it is more difficult to track the flow of parameters without supersymmetry, at least in the large $N$ / planar limit of the theory we again have a nearly vanishing beta function for the gauge couplings. As such, we expect that there are still duality interfaces / defects captured by (approximately) constant axio-dilaton 7-branes. That being said, the fusion rules for these duality defects still exhibit scale dependence; we track this by comparing the spectrum of duality interfaces / defects in the UV and IR.

The rest of this paper is organized as follows. In section \ref{sec:IIA} we give a broad sketch of our proposal in the IIA case. We present explicit IIA backgrounds illustrating these general points in sections \ref{sec:IIAexamples} and \ref{sec:NONAB}. In section \ref{sec:IIB} we turn to the related case of type IIB backgrounds with spacetime filling D3-branes. We present a summary and possible future directions in section \ref{sec:CONC}. Some additional examples, discussion, and background material is deferred to the Appendices.

\section{Type IIA on $\mathbb{R}^{3,1} \times \mathbb{R}^{6} / \Gamma$} \label{sec:IIA}

In this section we determine generalized symmetries for type IIA strings on the background $\mathbb{R}^{3,1} \times X$ with:
\begin{equation}
X = \mathbb{R}^{6} / \Gamma^{\mathbf{s}}
\end{equation}
where $\Gamma$ is a finite subgroup of $SU(4) \subset \mathrm{Spin}(6)$. Here, the superscript $\mathbf{s}$ indicates the group action on the $\mathbf{4}$ spinor representation of $SU(4) \cong \mathrm{Spin}(6)$; the group action on the vector representation $\mathbf{6}$ is induced from this. Supersymmetry is preserved when $\Gamma$ embeds in the $SU(3)$ factor of $(SU(3) \times U(1)) / \mathbb{Z}_3 \subset SU(4)$. Otherwise, we do not have a covariantly constant killing spinor, and supersymmetry is broken. We assume that the group action is chosen so that all bulk tachyons (i.e., tachyons in the untwisted sector) are projected out, in accord with having a type II background. The other possibility of a type 0 background is also interesting but will not be the focus of the present work.

The rest of this section is organized as follows. We begin by briefly reviewing some salient features in the special case where we retain supersymmetry. We then explain how these structures extend to the non-supersymmetric case. We present explicit examples illustrating these general points in sections \ref{sec:IIAexamples} and \ref{sec:NONAB}.

\subsection{Supersymmetric Case}

Consider first supersymmetric orbifold backgrounds. The 4D system retains eight real supercharges, i.e., $\mathcal{N} = 2$ supersymmetry.
On general grounds, the 10D background consists of the closed string modes, as well as localized ``QFT modes'' coming from branes wrapped on collapsed cycles of the geometry. 4D gravity is decoupled since the extra dimensions are non-compact. The vacuum moduli space matches to that of the orbifold geometry. Due to supersymmetry, we can start in the resolved geometry and then proceed to the orbifold. Observe that in the resolved geometry the reduction of the RR three-form potential on the various two-cycles yields a collection of $U(1)$ ``electric'' gauge fields of the Coulomb branch of this theory. We also have a magnetic dual basis of gauge fields given by reduction of the RR $5$-form potential on the various four-cycles. We find D2-branes  wrapped on collapsing two-cycles and D4-branes wrapped on collapsing four-cycles. In the limit where mutually non-local electric and magnetic degrees of freedom are both present we reach a strongly coupled 4D $\mathcal{N} = 2$ SCFT of Argyres-Douglas type \cite{Argyres:1995jj}. There is a natural lift of this configuration to M-theory. Indeed, starting from M-theory on $\mathbb{R}^{4,1} \times X$, we now get a 5D SCFT (see e.g, \cite{Seiberg:1996bd, Morrison:1996xf, Douglas:1996xp}). Reduction on a circle takes us to a 4D Kaluza-Klein theory, and in the limit where the circle shrinks to zero size we reach the 4D $\mathcal{N} = 2$ SCFT.\footnote{For further discussion of properties of the BPS spectrum, see references \cite{Closset:2018bjz, Closset:2019juk}.}

Our primary focus will be on the candidate 1-form electric and magnetic symmetries of the 4D theory, and their M-theory origin as 1-form electric and 2-form magnetic symmetries in the parent 5D theory. The general idea for determining these $p$-form symmetries is to first compute the associated defect group \cite{DelZotto:2015isa, GarciaEtxebarria:2019caf, Albertini:2020mdx, Morrison:2020ool} for the system. Given a $p$-brane which carries a conserved charge (and so it cannot decay to ``nothing'') one can consider wrapping it on a relative cycle of $H_{k+1}(X, \partial X)$. This gives rise to a defect in the QFT$_D$ spacetime with support on a subspace of dimension $p - k$. This defect can be partially screened by dynamical states, i.e., branes wrapped on compact cycles of $X$. Quotienting by this yields the collection of defects which cannot be screened. The end result is the defect group:\footnote{For some discussion on how to extend this to more general flux backgrounds via twisted K-theory, see e.g., references \cite{Heckman:2022xgu, Zhang:2024oas}.}
\begin{equation}
\mathbb{D} = \underset{n}{\oplus} \mathbb{D}^{(n)} \,\,\, \text{with} \,\,\, \mathbb{D}^{(n)} = \mathrm{Tor} \left(\,\underset{\mathrm{p-branes}}{\bigoplus}~ \underset{p - k = n}{\bigoplus} \frac{H_{k+1}(X, \partial X)}{H_{k+1}(X)}
\right),
\end{equation}
where the degree $n$ indicates the candidate $n$-form symmetry. Choosing a polarization (i.e., a collection of mutually commuting fluxes) leads to a specification of the spectrum of heavy defects, i.e., it determines a choice of absolute theory.\footnote{There can be obstructions to choosing some polarizations due to possible anomalies. While this is a complication in 5D systems, in 4D it is less of an issue. In any case, unless otherwise stated we shall implicitly assume an electric polarization.} In what follows we take all 
singular homology groups to have coefficients in $\mathbb{Z}$ unless otherwise stated.

In practice it can be somewhat involved to extract the defect group directly from an explicit resolution of the geometry $X$. 
In reference \cite{DelZotto:2022fnw} (see also \cite{Tian:2021cif}) two complementary methods were developed to extract this data without recourse to such resolution techniques. In these geometries, the defect group follows from:
\begin{equation}
\frac{H_{k+1}(X , \partial X)}{H_{k+1}(X)} \cong H_k(\partial X) = H_{k}(S^5 / \Gamma),
\end{equation}
where the group action on the boundary $S^5$ is induced from that on the bulk. Armstrong's theorem \cite{Armstrong_1968} tells us that the fundamental group $\pi_{1}(S^5 / \Gamma) \cong \Gamma /H$, where $H$ is the (normal) subgroup of $\Gamma$ which has a fixed point locus on the $S^5$. So, the abelianization of this group yields:
\begin{equation}
H_{1}(S^5 / \Gamma) = \mathrm{Ab}[\pi_1(S^5 / \Gamma)] = \mathrm{Ab}[\Gamma / H ].
\end{equation}
This is the (Pontryagin dual) of the electric 1-form symmetries, $\mathcal{A}^{(1)}_{\mathrm{mag}}$. Given a torsional cycle $\gamma \in H_{1}(S^5 / \Gamma)$, wrapping a D2-brane over the cone which stretches back to the tip of the singularity $\mathrm{Cone}(\gamma)$ yields a line defect in the 4D theory. This line defect is charged under $H_{1}(S^5 / \Gamma)^{\vee}$, the Pontryagin dual of $H_{1}(S^5 / \Gamma)$.\footnote{Recall that the Pontryagin dual of a finite abelian group $G$ is given by $G^{\vee} \equiv \mathrm{Hom}(G,U(1))$ which is isomorphic (though not canonically so) to the original group.}
The symmetry operator which acts on this line is given by a D4-brane wrapped on a linking cycle in $H_{3}(S^5 / \Gamma)$ \cite{Heckman:2022muc}. Similar considerations hold for the magnetic dual symmetries via the computation of $H_{3}(S^5 / \Gamma)$, although there can be some subtleties in situations where the singularity $\mathbb{C}^3 / \Gamma$ is not fixed point free, since this case leads to ``flavor symmetries'' in the QFT (see \cite{Cvetic:2022imb, DelZotto:2022joo} for further details). Summarizing, in the supersymmetric case, then, we can expect to encounter:
\begin{itemize}
  \item Codimension 6 Singularities: Tip of the Cone
  \item Codimension 4 Singularities: ``5-Branes'' of the form $\mathbb{C}^2 / \Gamma^{\prime}$,
\end{itemize}
and in both cases, the boundary topology of $S^5 / \Gamma$ produces an answer for the resulting generalized symmetries.

A complementary approach to extracting the higher-form symmetries is to directly construct the basis of electric / magnetic charged states, and their associated Dirac pairing. As explained in \cite{DelZotto:2022ras},
the Dirac pairing for the 4D theory appears in the SymTFT for the electric / magnetic 1-form symmetries:
\begin{equation}\label{eq:Kmatrix}
S_{\mathrm{5D}} = \frac{{K}_{ij}}{4 \pi} \int_{5D} C^{i} \wedge dC^{j},
\end{equation}
where the $C^{i}$ are 2-form potentials which should be viewed as background fields in the 4D system. Here, we also allow for the possibility that some of the $U(1)$'s of the theory are associated with flavor symmetries, i.e., their magnetic duals are absent.
As found in \cite{DelZotto:2022ras} (see also \cite{Tian:2021cif}), the data of the electric and magnetic 1-form
symmetries can be read off directly from the torsion of the cokernel of this pairing:
\begin{equation}
\mathbb{D}^{(1)} = \mathrm{Tor}(\mathrm{Coker}(K)) = \mathcal{A}^{(1)}_{\mathrm{elec}} \oplus \mathcal{A}^{(1)}_{\mathrm{mag}}.
\end{equation}
One reaches an absolute theory by choosing a polarization of $\mathbb{D}^{(1)}$.\footnote{It is worth noting that in the analogous computation for 5D SCFTs, the contributions split up in terms of wrapped M2-branes and M5-branes, generating respectively candidate 1-form and 2-form symmetries. There can be obstructions to choosing the magnetic polarization in the 5D system, but in the 4D theory obtained from circle compactification these complications are not present. For further discussion on subtleties with polarizations in 5D theories, see e.g., \cite{Cvetic:2023pgm, DelZotto:2024tae, Cheesesteak}.}

How then do we determine the matrix $K_{ij}$ in practice? For 4D $\mathcal{N} = 2$ theories (include their 4D KK cousins) this data follows directly from the associated BPS quiver of electric / magnetic bound states. One way to access this data is to consider the worldvolume theory of a probe D0-brane near the singularity in question. This leads to a supersymmetric quiver quantum mechanics which retains four real supercharges. We get a basis of ``fractional branes'' associated with irreducible representations of $\Gamma$, and connecting ``open strings'' associated with bifundamental matter. The key point for us is that the adjacency matrix for the quiver $A_{ij}$ is closely related to the matrix $K_{ij}$ of line (\ref{eq:Kmatrix}):
\begin{equation}
K_{ij} = A_{ij} - A_{ji}.
\end{equation}
So in other words, determining the adjacency matrix of the quiver is enough to compute the associated higher-form symmetries. Let us comment here that while we have used the D0-brane probe theory to access the Dirac pairing, we can of course entertain more general probe particle states. In this more general setting we still get the same adjacency matrices but the ranks of the gauge groups will be different.

As one would expect, the geometric method based on computing $H_{\ast}(S^5 / \Gamma)$ and the quiver method based on computing $\mathrm{Tor}(\mathrm{Coker}(K))$ exactly match, and this was explicitly verified in a number of examples in reference \cite{DelZotto:2022fnw}. The reason that one should a priori have expected a match is that the matrix $K_{ij}$ is also a linear map on the basis of generators in the associated relative K-theory group $K^{0}(X, \partial X)$ with integer coefficients. The resulting quotient of heavy defects versus screened objects thus follows:\footnote{Evaluating via the Chern character map and dualizing this sequence leads to an analogous expression in homology, but with all entries dualized and all arrows reversed. More precisely, the Chern character map only maps to cohomology with rational coefficients and annihilates torsion of the K-theory classes. However, for the spaces $X$ considered in this paper both $K^0(X,\partial X)$ and $K^0(X)$ do not contain torsional elements, and the Chern character map therefore maps these strictly to integral representatives in (rational) cohomology. The corresponding integral cohomology groups of the pair $(X,\partial X)$ and $X$ are also free of torsion, and therefore their lifts to integral cohomology are unique. With two entries in the sequence thus mapped to integral cohomology, the mapping to cohomology of the third, $K^0(\partial X)$, which generally is torsional, can be inferred from exactness. This also implies that these are isomorphic to the torsional integral cohomology groups (even though the Chern character map does not supply an isomorphism).}
\begin{equation}
0 \rightarrow K^{0}(X, \partial X) \xrightarrow[]{\,K_{ij}\,}
K^{0}(X) \rightarrow K^{0}(\partial X) \rightarrow 0.
\end{equation}

Our discussion so far has focussed on the 1-form symmetries of the 4D theory, but in many cases there can be other symmetries which can also entwine with these structures. For example, precisely when the group action of $\Gamma_{SU(3)}$ on $\mathbb{C}^3$ has a fixed locus on the boundary $S^5$, we find additional non-isolated singularities which are locally of the form $\mathbb{C}^2 / \Gamma_{SU(2)}$ with $\Gamma_{SU(2)}$ a finite subgroup of $SU(2)$. This is interpreted as a 6D Super Yang-Mills theory sector. In general, there could be multiple simple flavor group factors, with global form correlated via geometric effects. The global form of the flavor group is directly tied to the higher-form symmetries of the 4D system, as captured via the long exact sequence:
\begin{equation}
\label{eq:2gpGroups}
0 \rightarrow \mathcal{A} \rightarrow \widetilde{\mathcal{A}} \rightarrow \widetilde{G} \rightarrow  G \rightarrow 1,
\end{equation}
with $\mathcal{A}$ and $G$ the true 1-form symmetry and 0-form symmetry, respectively, and $\widetilde{\mathcal{A}}$ and $\widetilde{G}$ the ``naive'' 1-form and 0-form symmetry in which we neglect possible correlations between these structures, as captured by the presence of a 2-group.\footnote{For applications of 2-groups in QFTs, see e.g., references \cite{Kapustin:2013uxa,Cordova:2018cvg,Cordova:2020tij} as well as \cite{Sati:2008eg,Baez:2005sn,Fiorenza:2012tb,Fiorenza:2010mh,Sati:2009ic}. For a helpful account of the interplay between 2-groups and line-changing operators in 4D QFTs, see references \cite{Bhardwaj:2021wif, Lee:2021crt}.}

In the context of 5D SCFTs (and thus implicitly their reduction to 4D SCFTs), 2-group symmetries were investigated in \cite{Apruzzi:2021vcu, Cvetic:2022imb, DelZotto:2022joo}. As conjectured in \cite{DelZotto:2022fnw} and explicitly proved in \cite{Cvetic:2022imb},
the existence of a 2-group structure is directly tied to having a non-split short exact sequence (in line with the analysis of \cite{Apruzzi:2021vcu}):
\begin{equation}
\label{eq:2gp}
0 \rightarrow \mathrm{Ab}[\Gamma / H]^{\vee} \rightarrow \mathrm{Ab}[\Gamma]^{\vee} \rightarrow \mathcal{C}^{\vee} \rightarrow 0,
\end{equation}
where $\mathcal{C}$ is the kernel of the map on the centers of the Lie groups:
\begin{equation}
\mathcal{C} = \mathrm{ker} (Z(\widetilde{G}) \rightarrow Z(G)).
\end{equation}

\subsection{Non-Supersymmetric Case}

Let us now turn to the non-supersymmetric case, i.e., we now consider type IIA string theory on the background $\mathbb{R}^{3,1} \times \mathbb{R}^6 / \Gamma^{\mathbf{s}}$. On general grounds, we do not expect this system to engineer a conformal field theory simply because the absence of supersymmetry in such backgrounds is typically (i.e., in all known examples) correlated with a tachyon in a twisted sector of the closed string Hilbert space. So, whereas we have an exact moduli space of vacua in the supersymmetric setting, in the non-supersymmetric setting we can expect some of these scalars to have a non-trivial potential which triggers a rolling / automatic (possibly only partial) resolution of the geometry. Precisely this issue was studied in references \cite{Adams:2001sv, Harvey:2001wm, Vafa:2001ra, Morrison:2004fr} where it was found that in many cases the end result of tachyons condensing is the transition to a locally supersymmetric background (which nevertheless might still have singularities).

Given this situation, we ought not expect to have an isolated QFT sector. Nevertheless, we can still study the spectrum of charged objects such as wrapped branes by calculating the associated global symmetries of this system. Indeed, 4D gravity is still switched off, and in many cases the candidate symmetry operators involve a collection of branes ``at infinity'' far from the location of the localized condensed tachyon. For these reasons we still expect to be able to reliably calculate global symmetries in this non-supersymmetric setting. To organize our analysis, we shall first fix a characteristic timescale $t_{\ast}$. We begin by establishing some basic features of the early time $t \ll t_{\ast}$ symmetries, returning to the late time behavior later.

We have two complementary approaches we can use to study the spectrum of defects and possible screening effects at each stage of evolution, namely quiver based methods and the topology of the boundary space $S^5 / \Gamma$. Compared with the supersymmetric case, the structure of non-supersymmetric backgrounds will generically involve three distinct singularity types:
\begin{itemize}
  \item Codimension 6 Singularities: Tip of the Cone
  \item Codimension 4 Singularities: ``5-Branes'' of the form $\mathbb{R}^4 / \Gamma^{\prime}$
  \item Codimension 2 Singularities: ``7-Branes'' of the form $\mathbb{R}^{2} / \Gamma^{\prime \prime}$.
\end{itemize}
Since the codimension 4 and 2 singularities extend to the boundary $S^5 / \Gamma$, there can a priori be localized tachyons on these subspaces.

A general issue we therefore face is that unless the tachyonic degrees of freedom are initially sequestered near the tip of the cone $\mathbb{R}^6 / \Gamma$, the geometric interpretation of ``branes wrapping cycles at infinity'' will also need to be treated with more care. On the other hand, when no tachyons are initially present in the boundary $S^5 / \Gamma$ we expect an exact match between the quiver and geometry based analyses.

With this in mind, we first develop the treatment of higher-form symmetries based on quiver based methods, and then return to the geometric analysis. We then discuss further time dependent considerations.

\subsubsection{Defect Group via Quivers}

To extract the defect group, we return to the Dirac pairing of electric / magnetic states in the theory. The spectrum of possible charges is in turn obtained from the spectrum of possible quiver quantum mechanics theories as realized by wrapped fractional branes of the IIA extra-dimensional geometry. The special case of a D0-brane probe particle already detects the entire basis of fractional branes, as well as the spectrum of open strings which stretch between these objects. As such, it suffices to study the D0-brane quantum mechanical theory;\footnote{We briefly comment on the relevant time scales. Tachyon condensation is a stringy effect, and its characteristic time scale $t_T\sim \ell_s$ is set by the string length scale. In contrast, being a D-brane, the length scale of the D0-brane is $\ell_{\text{D0}}\sim g_s\ell_s$. In the weak coupling limit $g_s \ll 1$. Therefore, we have that $\ell_{D0}\sim g_s\ell_s \ll \ell_s \sim \ell_T$. In this manner, the D0-brane probe is sensitive to time-dependent features of the unstable background, and its quantum mechanical theory can probe the symmetries of the resulting spacetime theory as a function of time.} we obtain other particle-like states by modifying the choice of gauge groups.

Now, the quiver quantum mechanics for a probe D0-brane follows from the general procedure given in \cite{Douglas:1996sw, Kachru:1998ys, Lawrence:1998ja, Hanany:1998sd}. We have a collection of gauge groups in correspondence with irreducible representations of $\Gamma$, and connecting lines between the nodes indicating bifundamental matter. Because there is no supersymmetry, we have two adjacency matrices, one for fermionic degrees of freedom, i.e., $A_{ij}^{F}$, and one for bosonic degrees of freedom $A_{ij}^{B}$. The interaction terms for these degrees of freedom follow from orbifold projection of interaction terms present in the D0-brane probe of $\mathbb{R}^{3,1} \times \mathbb{R}^6 / \Gamma$.

We claim that the adjacency matrix for the fermionic degrees of freedom $A_{ij}^{F}$ encodes the Dirac pairing:
\begin{equation}\label{eq:DIRACferm}
K_{ij} = A_{ij}^{F} - A_{ji}^{F}.
\end{equation}
Observe that this is in accord with the special case where we have a supersymmetric background.
With this in place we can then extract the defect group for 1-form symmetries via the considerations presented in reference \cite{DelZotto:2022ras}:
\begin{equation}
\mathbb{D}^{(1)} = \mathrm{Tor}(\mathrm{Coker}(K)) = \mathcal{A}^{(1)}_{\mathrm{elec}} \oplus \mathcal{A}^{(1)}_{\mathrm{mag}}.
\end{equation}

\paragraph{Dirac Pairing}

We now turn to a derivation of equation (\ref{eq:DIRACferm}). To this end, we first (briefly) review how to extract the quiver quantum mechanics theory for probe branes of the type IIA singularity. A helpful starting point is to actually begin in type IIB string theory with spacetime filling branes probing the singularity $\mathbb{R}^{6} / \Gamma$. Working on the 4D spacetime $\mathbb{R}_{t} \times T^3$ and dimensionally reducing / T-dualizing, we reach the quiver quantum mechanics for probe particles in the IIA background.

We extract the quiver following the general procedure given in \cite{Douglas:1996sw, Kachru:1998ys, Lawrence:1998ja, Hanany:1998sd} (see also \cite{McKay, KronheimerNakajima}). Each irreducible representation $\gamma_i \in \mathrm{Rep} (\Gamma)$ specifies a fractional brane, which in geometric terms we identify with a $\Gamma$-equivariant sheaf on $\mathbb{R}^6$. For each irreducible representation we get a corresponding quiver node, i.e., a gauge group $U(n_i)$ as associated with a representation $\mathcal{R} = \mathbb{C}^{n_i} \gamma_i$, where $\Gamma$ acts trivially on the $\mathbb{C}^{n_i}$ factor. The special case of $n$ mobile D3-branes corresponds to taking $n_i = n \mathrm{dim} \gamma_i$.

The connectivity of the quiver involves bifundamentals between the different gauge groups. Fermions between gauge group $U(n_i)$ and $U(n_j)$ will be labeled as $\psi^{i,j}$ and bosons will be labelled as $\phi^{i,j}$. By abuse of notation we shall often also have a multiplicity, which we explicitly indicate, as appropriate. The fermions and bosons descend from modes present in the unorbifolded parent theory. In particular, fermions transform in the $\mathbf{4}$ of $\mathrm{Spin}(6)$ and bosons transform in the $\mathbf{6}$ of $\mathrm{Spin}(6)$. Consequently there is an induced group action of $\Gamma \subset SU(4)$ on these representations. Indeed, for a representation $\mathcal{R}$ of $SU(4)$, we get an induced representation via the embedding of $\Gamma$ on $SU(4)$. The adjacency matrix for the quiver follows from the tensor product:
\begin{equation}
\mathcal{R} \otimes \gamma_{i} = \underset{j}{\bigoplus} A^{\mathcal{R}}_{ij} \gamma_{j}.
\end{equation}
A helpful formula for extracting the adjacency matrices follows from the character formula (see e.g., the discussion in Appendix C of reference \cite{DelZotto:2022fnw}):
\begin{equation}\label{adjmat}
A_{ij}^\mathcal{R} = \frac{1}{|\Gamma|}\sum_\alpha r_\alpha \chi(\mathcal{R})^\alpha \chi(\gamma_i)^\alpha \overline{\chi(\gamma_i)}^\alpha
\end{equation}
where $r_\alpha$ counts the dimension of the $\alpha$ conjugacy class, $\chi$ denotes the character, and the bar means complex conjugate.
The fermionic adjacency matrix is obtained by setting $\mathcal{R} = \mathbf{4}$, and the bosonic adjacency matric is obtained by setting $\mathcal{R} = \mathbf{6}$:
\begin{equation}
A_{ij}^{F} = A_{ij}^{\mathbf{4}} \,\,\, \text{and} \,\,\, A_{ij}^{B} = A_{ij}^{\mathbf{6}}.
\end{equation}

Consider next the dimensional reduction on a $T^3$. Each of the scalars directly descends to a scalar, and each of the Weyl fermions descends to a complex doublet. The 4D gauge boson splits up as a 1D vector potential and three adjoint-valued scalars which rotate as a vector of the spacetime $SO(3)$:
\begin{equation}
V^{4D}_{j} = v_j \oplus \overrightarrow{x}_j,
\end{equation}
i.e., the $\overrightarrow{x}_{j}$ specify adjoint-valued positions of the constituent probe particles of the IIA background. 
Finally, the time dependent resolution parameters enter as dynamical ``driving parameters'' in the quiver quantum mechanics 
(in the same sense as reference \cite{Adams:2001sv}).

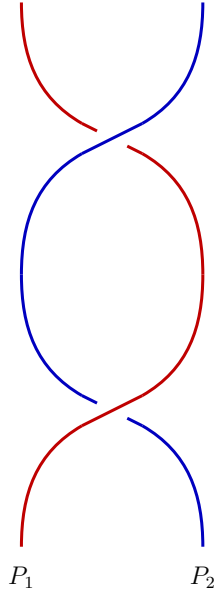
\begin{figure}
    \centering
    \scalebox{0.8}{
    \begin{tikzpicture}
    \begin{pgfonlayer}{nodelayer}
		\node [style=none] (0) at (-1.5, -2.25) {};
		\node [style=none] (1) at (-0.5, -0.25) {};
		\node [style=none] (2) at (0.5, 0.25) {};
		\node [style=none] (3) at (1.5, 2.25) {};
		\node [style=none] (4) at (-0.5, 0.25) {};
		\node [style=none] (5) at (0.5, -0.25) {};
		\node [style=none] (6) at (-1.5, 2.25) {};
		\node [style=none] (7) at (1.5, -2.25) {};
		\node [style=none] (8) at (-0.25, 0.125) {};
		\node [style=none] (9) at (0.25, -0.125) {};
		\node [style=none] (10) at (-1.5, 2.25) {};
		\node [style=none] (11) at (-0.5, 4.25) {};
		\node [style=none] (12) at (0.5, 4.75) {};
		\node [style=none] (13) at (1.5, 6.75) {};
		\node [style=none] (14) at (-0.5, 4.75) {};
		\node [style=none] (15) at (0.5, 4.25) {};
		\node [style=none] (16) at (-1.5, 6.75) {};
		\node [style=none] (17) at (1.5, 2.25) {};
		\node [style=none] (18) at (-0.25, 4.625) {};
		\node [style=none] (19) at (0.25, 4.375) {};
		\node [style=none] (20) at (-1.5, -2.75) {$P_1$};
		\node [style=none] (21) at (1.5, -2.75) {$P_2$};
	\end{pgfonlayer}
	\begin{pgfonlayer}{edgelayer}
		\draw [style=RedLine, in=-150, out=90] (0.center) to (1.center);
		\draw [style=RedLine] (1.center) to (2.center);
		\draw [style=RedLine, in=-90, out=30] (2.center) to (3.center);
		\draw [style=BlueLine, in=-30, out=90] (7.center) to (5.center);
		\draw [style=BlueLine, in=-90, out=150] (4.center) to (6.center);
		\draw [style=BlueLine] (5.center) to (9.center);
		\draw [style=BlueLine] (4.center) to (8.center);
		\draw [style=BlueLine, in=-150, out=90] (10.center) to (11.center);
		\draw [style=BlueLine] (11.center) to (12.center);
		\draw [style=BlueLine, in=-90, out=30] (12.center) to (13.center);
		\draw [style=RedLine, in=-30, out=90] (17.center) to (15.center);
		\draw [style=RedLine] (15.center) to (19.center);
		\draw [style=RedLine] (18.center) to (14.center);
		\draw [style=RedLine, in=-90, out=150] (14.center) to (16.center);
	\end{pgfonlayer}
\end{tikzpicture}
    }
    \caption{Depiction of monodromy for a pair of worldlines for particles $P_1$ and $P_2$.}
    \label{fig:monodromy}
\end{figure}

We now derive equation (\ref{eq:DIRACferm}). The main idea will be to consider a pair of particles, each with its own quiver quantum-mechanics. Each such particle specifies a worldline in the 4D spacetime, so we can consider the effect of monodromy (see figure \ref{fig:monodromy}). Consider one of these particles. The position of each constituent fractional brane in this particle is specified by the background value of $\overrightarrow{x}_{j}$ in the overall $U(1)$ factor of $U(n^j)$. Since we are assuming these fractional branes have all coalesced at a single location, we refer to this whole configuration as $\overrightarrow{x}$, in the obvious notation. By the same token, we can also introduce another particle with collective coordinate $\overrightarrow{y}$. In general, we can expect there to be non-trivial interactions between these particles. Some of these effects can be captured by starting with a higher rank quiver quantum mechanics in which the collective coordinate appears as $\overrightarrow{X} = \mathrm{diag}(\overrightarrow{x}, \overrightarrow{y})$. Observe as we move $\overrightarrow{x}$ away from $\overrightarrow{y}$, some of the open string degrees of freedom will pick up a mass. In the limit where the separation $\overrightarrow{r} = \overrightarrow{x} - \overrightarrow{y}$ is very large, we can therefore integrate out these degrees of freedom. Our plan will be to study the change in this two-particle wave function $\Psi[\overrightarrow{r}]$ as we rotate the position $\overrightarrow{r} \mapsto M \cdot \overrightarrow{r}$ with $\mathbb{M} \in SO(3)$. We claim that under a full $2 \pi$ rotation along a fixed axis, the wave function $\Psi[\overrightarrow{r}]$ can pick up an overall Berry phase \cite{Berry:1984jv}:
\begin{equation}
\Psi[\mathbb{M}_{2 \pi} \cdot \overrightarrow{r}] = e^{i \theta} \Psi[\overrightarrow{r}].
\end{equation}
This phase encodes the Dirac pairing for electric / magnetic states, and single-valuedness of the wavefunction enforces Dirac quantization.

The $\overrightarrow{r}$ dependence of the quiver quantum mechanics appears through the dimensional reduction of the covariant derivative of the D3-brane probe quiver gauge theory. Indeed, for a fermionic degree of freedom $\psi$ and a bosonic degree of freedom $\phi$, the 4D Lagrangian contains the mass terms:
\begin{equation}\label{eq:massterms}
\mathcal{L} \supset \psi^{\dag} \overline{\sigma}_{a} r^{a} \psi + \phi^{\dag} r_{a} r^{a} \phi + ...,
\end{equation}
where $a = 1,2,3$ indexes the spatial directions and here we package the fermionic degrees of freedom in terms of the dimensional reduction of 4D left-handed Weyl spinors. In fact, the Berry phase in the supersymmetric case was implicitly determined e.g., in \cite{Denef:2002ru}. The main issue we need to address is how things might change in the absence of supersymmetry.

The main point is already visible from the interaction term of line (\ref{eq:massterms}): since this interaction term is quadratic in the bosonic and fermionic fields, the response to a rotation in $\overrightarrow{r}$ will follow from the one-loop determinants of the massive modes. Note, however, that the bosonic mass term has no dependence on $\mathbb{M} \in SO(3)$ rotations at all (since it is a dot product). As such, the only possible contribution to the monodromy can come from the fermionic degrees of freedom. This is enough to establish the main claim, since we can now simply reapply the same reasoning used in the supersymmetric context.

Nevertheless, it is also instructive to track through the Berry phase contribution more directly. By inspection of line
(\ref{eq:massterms}), we observe that the effective Hamiltonian is of the form:
\begin{equation}
\widehat{H}_{\mathrm{eff}} = \mu \overrightarrow{r} \cdot \overrightarrow{\sigma} + ...,
\end{equation}
where the ``...'' are terms which do not contribute to the Berry phase. Focusing on just this first term, we 
have the Hamiltonian for a two-level system with $\overrightarrow{r}$ playing the role of a magnetic field. 
See \cite{TongNotes} for a pedagogical treatment of the resulting Berry phase.

The upshot is that for a pair of particles with respective gauge groups $\{U(n^i) \}_i$ and $\{U(m^j) \}_j$ the resulting Dirac pairing is simply:
\begin{equation}
K_{ij} n^i m^j = (A_{ij}^{F} - A_{ji}^{F}) n^i m^j,
\end{equation}
in the obvious notation. Consequently, we have established equation \eqref{eq:DIRACferm}, as claimed.

\subsubsection{Defect Group via Geometry: Sequestered Tachyons}

We now provide a complementary method for determining the defect group based on the topology of the boundary space $S^5 / \Gamma$.
From the general ``branes at infinity'' for topological symmetry operators, we expect that we can extract generalized symmetries provided
the boundary is far away from the dynamics of tachyon condensation, i.e., the case where all tachyons are sequestered. We turn to the case of unsequestered tachyons after this.

Since we are assuming that all tachyons are sequestered, we are restricted to codimension 6 and codimension 4 singularities, where the ``flavor-brane'' codimension 4 singularities are of the special form $\mathbb{C}^2 / \Gamma^{\prime}$ with some local supersymmetry preserved (otherwise there would be a tachyon present in this configuration as well). Indeed, in these cases $\Gamma^{\prime}$ must be a finite subgroup of $SU(2)$ of ADE type, and this in turn specifies the ADE type of a localized 6D Super Yang-Mills theory which wraps a non-compact (relative) cycle in $\mathbb{R}^{6} / \Gamma$.

Let us now turn to the spectrum of defects. We focus on heavy defects realized by wrapped Dp-branes for $p$ even so that they carry a conserved charge. Observe that since we have a time dependent resolution parameter (via tachyon condensation), a wrapped brane stretching from the boundary $S^5 / \Gamma$ to the tip of the cone will still persist, but the objects which can potentially screen this defect might change as a function of time. Nevertheless, sufficiently far from such transition points, we can still calculate the analog of a defect group. Doing so, we can still extract candidate electric and magnetic 1-form symmetries. For example, we get electric line defects from D2-branes wrapped on $\mathrm{Cone}(\gamma)$ for $\gamma \in H_{1}(S^5 / \Gamma)$ and so the electric 1-form symmetry follows from Armstrong's theorem:\footnote{In principle there could be other non-geometric contributions, but we neglect this in what follows.}
\begin{equation}
\mathcal{A}^{(1)}_{\mathrm{elec}} = H_{1}[S^5 / \Gamma]^{\vee} \cong \mathrm{Ab}[\Gamma / H]^{\vee}.
\end{equation}
Observe also that there is still a 2-group structure whenever we have a non-split short exact sequence:
\begin{equation}
0 \rightarrow \mathrm{Ab}[\Gamma / H]^{\vee} \rightarrow \mathrm{Ab}[\Gamma]^{\vee} \rightarrow \mathcal{C}^{\vee} \rightarrow 0.
\end{equation}
This interpretation holds because we have assumed that the codimension 4 singularities are locally supersymmetric, i.e., they engineer
6D Super Yang-Mills sectors (``flavor branes'').

\subsubsection{Defect Group via Geometry: Unsequestered Tachyons}

We now turn to the more general case where we have unsequestered tachyons. From the perspective of the boundary topology, we can again proceed to compute $H_{\ast}(S^5 / \Gamma)$, much as we would in the sequestered (as well as supersymmetric) case. That being said, the presence of twisted sector tachyons in the boundary geometry means that this topology will itself undergo dynamical transitions so we must exercise more caution in reading off the data of the defect group in this case. In principle these tachyons can originate from both the codimension 4 and codimension 2 singularities since both stretch ``out to infinity''.\footnote{See e.g., references \cite{Adams:2001sv, Harvey:2001wm, Vafa:2001ra, Martinec:2002wg} for some analyses of these cases.} Let us briefly discuss each possibility in turn.

In the case of non-supersymmetric codimension 4 singularities, the local geometry will now be of the form $\mathbb{R}^{4} / \Gamma^{\prime}$ with $\Gamma^{\prime}$ a finite subgroup of $\mathrm{Spin}(4)$ which does not embed in an $SU(2)$ subfactor of $SU(2)_L \times SU(2)_R \cong \mathrm{Spin}(4)$. In these cases we cannot give a ``flavor-brane'' interpretation of this singularity, but at least group theoretically we can still speak of a 2-group-like structure whenever the short exact sequence:
\begin{equation}
0 \rightarrow \mathrm{Ab}[\Gamma / H]^{\vee} \rightarrow \mathrm{Ab}[\Gamma]^{\vee} \rightarrow \mathcal{C}^{\vee} \rightarrow 0,
\end{equation}
does not split.

In the case of codimension 2 singularities the local geometry is of the form $\mathbb{R}^2 / \Gamma^{\prime \prime}$ and so the geometry does not support a covariantly constant spinor. Indeed, in perturbative string theory these backgrounds all have a tachyon.\footnote{Contrast this with F-theory backgrounds where we can switch on an axio-dilaton profile to retain supersymmetry. In the weakly coupled IIA setting, no such loophole is available.} Observe that a codimension 2 singularity of $\mathbb{R}^6 / \Gamma$ also specifies a codimension subspace of the boundary $S^5 / \Gamma$. As such, once the tachyon pulse begins to expand the resulting bubble will fill out a codimension $1$ subspace, partitioning the $S^5 / \Gamma$ into distinct sectors. For each connected component we can calculate a corresponding defect group and ask whether this matches to the answer computed via the quiver based method.

Clearly, this analysis depends on the choice of group $\Gamma$ as well as the choice of group action; for multiple codimension $2$ loci the precise partitioning of the space will also involve determining which tachyon grows most quickly. To bypass these subtleties, we now specialize to the case $\Gamma = \mathbb{Z}_N$, but in which we allow for the possibility of a codimension $2$ singularity.

The result from considering a number of abelian examples is the empirically obtained formula based on the quiver based method:
\be \label{eq:CoolFormula}
\mathbb{D}^{(1)} = \mathrm{Tor} ( \mathrm{Coker} K ) \cong (\Gamma/H)^2\oplus \lb  \bigoplus_{i} \, (\Gamma/H_{\mathscr{S}_{i}})^{|H_i|-1}\rb\,.
\ee
Here $H$ is the subgroup of $\Gamma\cong\Z_N$ generated by all elements with fixed points on $S^5$ and we have $H_1(S^5/\Gamma)\cong \Gamma/H$. We label by $\mathscr{S}_i$ the closure of a codimension 2 singular locus in $S^5/\Gamma$, it is an irreducible closed component of the full singular locus and in particular $\mathscr{S}_i$ contains both a generic codimension 2 locus together with possible enhancement loci along codimension 2 subloci in $\mathscr{S}_i$. The subgroup $H_i\subset \Gamma$ is generated by all elements with codimension 2 fixed loci in $S^5$. The subgroup $H_{\mathscr{S}_i}\subset \Gamma$ is generated by all elements which lead to any of the singularities contained in $\mathscr{S}_i$. In particular $H_i\subset H_{\mathscr{S}_i}$ and $H_{\mathscr{S}_i}$ is obtained from $H_i$ by adding elements only associated with the enhancement locus of $\mathscr{S}_i$. The order $|H_i|$ is odd and consequently the sum always splits into isomorphic electric and magnetic contributions, and we have $|H_i|=1$ in the absence of codimension 2 singularities. In the supersymmetric case codimension 2 singularities do not arise and the formula reduces to the supersymmetric result \cite{DelZotto:2022fnw}. A final comment here is that we expect that for $\Gamma$ non-abelian we expect a similar formula to hold where we instead take the abelianization of all available groups.

Let us now provide some further motivation for equation (\ref{eq:CoolFormula}).
Reading from left to right,
the first contribution of line (\ref{eq:CoolFormula}) derives from geometry $H_1(S^5/\Gamma)\cong \Gamma/H$ when $\Gamma\cong \Z_N$ and the corresponding line defects are constructed via D2-branes wrapped over cones of cycles in $H_1(S^5/\Gamma)$.
The other contributions appear to arise from tachyon pulses ``partitioning up'' the geometry into individual pieces.
Indeed, following the discussion of codimension 2 singularities $\mathscr{S}_i$ given in \cite{Adams:2001sv},
the orbifold $\mathbb{R}^2/\Z_{2\ell+1}$ decays via a series of dilaton pulses associated with the sequence of deficit angles
\be
\mathbb{R}^2/\Z_{2\ell+1}~\rightarrow~\mathbb{R}^2/\Z_{2\ell-1}~\rightarrow~\dots~\rightarrow~\mathbb{R}^2/\Z_{3} ~\rightarrow~\mathbb{R}^2\,.
\ee
Given the starting point $2\ell+1$, there are $\ell$ such transitions. The singularity of the initial geometry $\mathbb{R}^2/\Z_{2\ell+1}$ is driven to a geometry containing $\ell$ concentric circles across which the deficit angle jumps. Each cylinder segment between two adjacent circles is modelled on a geometry in the above sequence.

Let us discuss equation \eqref{eq:CoolFormula} when $S^5/\Gamma$ contains an isolated codimension 2 singularity, folded by say $H_1$.
Then the faithfulness of the action implies that $|H_1|$ and $|\Gamma/H_1|$ are coprime, and consequently the following sequence splits
\be
1~\rightarrow~H_1~\rightarrow~\Gamma~\rightarrow~\Gamma/H_1~\rightarrow ~1\,.
\ee
This sequence governs how the singularity model / normal geometry $\mathbb{R}^2/H_1$ is fibered over the singular locus $\mathscr{S}_1$. The local model for the codimension 2 singularity $\mathscr{S}_1$ in $S^5/\Gamma$ is now:
\be
\mathbb{R}^2/H_1\times \mathscr{S}_1 \,, \qquad \mathscr{S}_1=S^3/(\Gamma/H_1)\,.
\ee
The contribution to the electric 1-form symmetry not captured by singular homology can be suggestively rewritten as:
\be
(\Gamma/H_1)^{(|H_1|-1)/2}=H_1(\mathscr{S}_1)^{\ell_1}
\ee
where we have reparameterized $|H_1|=2{\ell_1}+1$. We interpret this as noting that when $\mathbb{R}^2/H_i$ decays via dilaton pulses then each of the circles across which the deficit angle jumps contributes one torsional 1-cycle, a copy of the generator of $H_1(\mathscr{S}_1)$, which via a D2-brane wrapping results in a electric line defect. This 1-cycle is ``stuck'' in the dilaton pulse. We refer the interested reader to Appendix \ref{app:COOLNESS} for additional discussion and examples.

\subsection{Time Dependent Considerations}

One of the important distinctions with the supersymmetric case is that there will inevitably be some time dependence in our analysis. We turn to some general features of how this impacts our analysis. At early times, i.e., $t \ll t_{\ast}$, we have the original singularity. At late times, i.e., $t \gg t_{\ast}$, tachyon condensation has occurred and the singularity will have been (partially) resolved. In principle there can be multiple stages to this resolution process so we indicate these characteristic timescales as:
\begin{equation}
t_{\mathrm{start}} \equiv t_{0} < t_{1} < t_{2} < ... < t_{I} \equiv t_{\mathrm{end}} .
\end{equation}
The local neighborhood around the singularity will therefore have a similar sequence:
\begin{equation}
\mathbb{R}^6 / \Gamma_{0}^{\mathbf{s}_0} , \mathbb{R}^6 / \Gamma_{1}^{\mathbf{s}_1} ,..., \mathbb{R}^6 / \Gamma_{I}^{\mathbf{s}_I}.
\end{equation}
In the type II case this endpoint preserves supersymmetry \cite{Morrison:2004fr}.

In between each transition we can study the spectrum of defects and symmetry operators, and thus extract a corresponding defect group. We denote this sequence as:
\begin{equation}
\mathbb{D}_{0}, \mathbb{D}_{1},...,\mathbb{D}_{I}.
\end{equation}

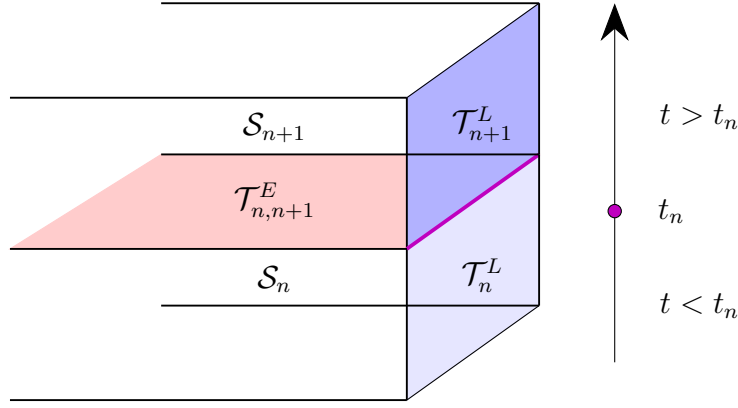
\begin{figure}
    \centering
    \scalebox{1.0}{
    \begin{tikzpicture}
	\begin{pgfonlayer}{nodelayer}
		\node [style=none] (0) at (-1.75, 3.75) {};
		\node [style=none] (1) at (0, 5) {};
		\node [style=none] (2) at (-1.75, -0.25) {};
		\node [style=none] (3) at (0, 1) {};
		\node [style=none] (4) at (-1.75, 1.75) {};
		\node [style=none] (5) at (0, 3) {};
		\node [style=none] (6) at (-7, 1.75) {};
		\node [style=none] (7) at (-5, 3) {};
		\node [style=none] (8) at (-7, 3.75) {};
		\node [style=none] (9) at (-5, 5) {};
		\node [style=none] (10) at (-7, -0.25) {};
		\node [style=none] (11) at (-5, 1) {};
		\node [style=none] (13) at (2.125, 1) {$t<t_{n}$};
		\node [style=none] (14) at (1, 2.75) {};
		\node [style=none] (15) at (1.75, 2.25) {$t_{n}$};
		\node [style=none] (16) at (1, 4.75) {};
		\node [style=none] (17) at (2.125, 3.5) {$t>t_{n}$};
        \node [style=none] (19) at (-3.5, 1.35) {$\mathcal{S}_n$};
		\node [style=none] (20) at (-3.5, 3.35) {$\mathcal{S}_{n+1}$};
		\node [style=none] (21) at (-3.5, 2.375) {$\mathcal{T}^{E}_{{n,n+1}}$};
		\node [style=none] (22) at (-0.75, 1.4) {$\mathcal{T}^{L}_{n}$};
		\node [style=none] (23) at (-0.75, 3.4) {$\mathcal{T}^{L}_{n+1}$};
  \node [style=CirclePurple] (24) at  (1, 2.25) {};
	\end{pgfonlayer}
	\begin{pgfonlayer}{edgelayer}
        \filldraw[fill=red!20, draw=red!20]  (-7, 1.75) -- (-1.75, 1.75) -- (0, 3) -- (-5, 3) -- cycle;
        \filldraw[fill=blue!10, draw=blue!10]  (-1.75, 1.75) -- (0, 3) -- (0, 1) -- (-1.75, -0.25) -- cycle;
        \filldraw[fill=blue!30, draw=blue!30]  (-1.75, 1.75) -- (0, 3) -- (0, 5) -- (-1.75, 3.75) -- cycle;
		\draw (0.center) to (1.center);
		\draw (2.center) to (3.center);
		\draw [style=ThickLine] (0.center) to (4.center);
		\draw [style=ThickLine] (4.center) to (2.center);
		\draw [style=PurpleLine] (4.center) to (5.center);
		\draw [style=ThickLine] (5.center) to (1.center);
		\draw [style=ThickLine] (5.center) to (3.center);
		\draw [style=ThickLine] (10.center) to (2.center);
		\draw [style=ThickLine] (11.center) to (3.center);
		\draw [style=ThickLine] (6.center) to (4.center);
		\draw [style=ThickLine] (7.center) to (5.center);
		\draw [style=ThickLine] (8.center) to (0.center);
		\draw [style=ThickLine] (9.center) to (1.center);
        \draw [-{Stealth[length=5mm]}] (1,0.25) -- (1,5);
	\end{pgfonlayer}
\end{tikzpicture}
    }
    \caption{Depiction of the SymTFT layout for time $t$ smaller and larger than $t_{n}$. We approximate the configuration via a step function jump at $t_n$, at which time the Euclidean interface theory, $\mathcal{T}^E_{n,n+1}$, and the late time Lorentzian theory $\mathcal{T}^L_{i+1}$ branch off from the early time Lorentzian theory $\mathcal{T}^L_{n}$. $\mathcal{S}_{n}$ and $\mathcal{S}_{n+1}$ are the associated 5D SymTFTs.}
    \label{fig:symtft}
\end{figure}

In each such regime, we can also introduce an auxiliary 5D Symmetry TFT with level matrix $K^{(n)}_{ij}$ for $t_{n-1} < t < t_{n}$. As we cross from $K^{(n)}_{ij}$ to $K_{ij}^{(n+1)}$ we get a 4D Euclidean interface $\mathcal{T}^E_{n,n+1}$ which need not be topological. Indeed, this interface theory is given by the Euclidean path integral of the 4D system with boundary conditions dictated by the jump (see figure \ref{fig:symtft}).

It is instructive to compare this sort of interface with the ``SymTrees'' of reference \cite{Baume:2023kkf}. In the context of SymTree theories, one also has non-topological interfaces, but these are localized in the extra dimensions of the bulk symmetry theory. Here, the junction theory is instead a Euclidean theory localized at a particular timeslice.

The Euclidean theory on this timeslice simply consists of all the degrees of freedom of the IIA configuration which are in the process of
decoupling due to decompactification induced from tachyon condensation. Indeed, with an explicit geometry in hand we can directly track how the basis of fractional branes changes across a transition, and thus also determine which candidate $U(1)^{(0)}_{\mathrm{elec}}$ and $U(1)^{(0)}_{\mathrm{mag}}$ gauge symmetries are no longer present. The junction theory simply enforces a boundary condition which matches the two SymTFTs, much as in \cite{Baume:2023kkf}. See figure \ref{fig:symtft} for a depiction of this matching.

The IIA configurations are purely geometric and as such much of the SymTree analysis carries over. Consider for example the first step $\mathbb{R}^6 / \Gamma_{0}^{\mathbf{s}_0}\rightarrow \mathbb{R}^6 / \Gamma_{1}^{\mathbf{s}_1}$ which is understood as partially resolving $\mathbb{R}^6 / \Gamma_{0}^{\mathbf{s}_0}$ to a space which contains, among others, a singularity modelled on $\mathbb{R}^6 / \Gamma_{1}^{\mathbf{s}_1}$, and subsequently taking the local limit centered on $\mathbb{R}^6 / \Gamma_{1}^{\mathbf{s}_1}$. From here, excise from the partially resolved geometry a small ball centered on the singularity modelled on $\mathbb{R}^6 / \Gamma_{1}^{\mathbf{s}_1}$. This results in a manifold with boundary $X_{0,1}$, the boundaries are $S^5 / \Gamma_{1}^{\mathbf{s}_0}$ and $S^5 / \Gamma_{1}^{\mathbf{s}_1}$, and as such $X_{0,1}$ realizes a cobordism between ``infinity" at early and late times. The junction theory $\mathcal{T}^{E}_{0,1}$ is the Euclidean relative theory obtained from IIA on $\mathbb{R}^{3,1} \times X_{0,1}$.
We defer much of the computational details of such construction to upcoming work \cite{Cheesesteak}.

We conclude this section by discussing one key feature of $\mathcal{T}^E_{0,1}$, which demonstrates that, in general, there is no subgroup relation between the defect groups $\mathbb{D}_{0}, \mathbb{D}_{1}$. We will comment  later in section \ref{sec:IIB} on the defects and symmetry operators found in the initial local model and their fate at later times. As mentioned above, the bordism characterizes the degrees of freedom of $\mathcal{T}^E_{0,1}$ as those resulting from compactifying IIA theory on $X_{0,1}$. For this compactification, we would require the cohomology ring $H^*(X_{0,1})$. Of particular interest here, for example in the study of 1-form symmetries, would be the group $H^2(X_{0,1})$, which contains classes to expand the RR-form potential $C_3^{\text{RR}}$ (which couples to the D2-branes used to construct 1-form symmetry defects) to produce abelian spacetime gauge fields. This cohomology group can be computed from the Mayer-Vietoris long exact sequence for the covering  $ X_{0,1}\cup \mathbb{R}^6 / \Gamma_{1}^{\mathbf{s}_1}$. Here, we have presented the partially resolved geometry as a union of the new local model and ``the rest", which is the bordism. Setting up the sequence, note that the partially resolved geometry (as we are dealing with toric resolutions) contains free 2- and 4-cycles. In contrast, the new local model $\mathbb{R}^6 / \Gamma_{1}^{\mathbf{s}_1}$ does not. By exactness, these free class must be associated with free classes in $X_{0,1}$. From this, we learn that the bordisms we are considering carry away at least one $U(1)$ field. We see that a line defect in $\mathbb{D}_{0}$ transitions to a line defect in $\mathbb{D}_{1}$, which is dressed by a Wilson line of this $U(1)$ field. We are required to specify this dressing to relate lines in $\mathbb{D}_{0}$ to those in $\mathbb{D}_{1}$, which we will discuss in the context of the IIB frame in section \ref{sec:IIB}.

\section{IIA Examples: $\mathbb{Z}_N$ Orbifolds} \label{sec:IIAexamples}

In Section \ref{sec:IIA} we presented a prescription for determining the generalized symmetries for type IIA strings on backgrounds of the form
\begin{equation}
    \R^{3,1}\times \R^6/\Gamma^\mathbf{s}
\end{equation}
where $\Gamma$ is a finite subgroup of $SU(4) \subset \mathrm{Spin}(6)$. In this section we will show by way of example how our method works in practice.

The examples we consider are mainly drawn from reference \cite{Morrison:2004fr} (see also \cite{Lee:2003ar}) where the tachyon condensation process is mapped to explicit partial resolutions of the singular geometry. The examples studied there involve $\Gamma = \mathbb{Z}_N$, where a holomorphic presentation of the geometry is chosen so as to make use of methods from toric geometry. Let us emphasize, however, that the considerations presented in section \ref{sec:IIA} hold for general $\Gamma$; the only complication in studying examples in the non-abelian case is in performing all explicit resolutions and tracking tachyon condensation in such cases, a task we defer to future work. Indeed, as we have already emphasized, to extract the defect group both the quiver based method and the method based on the boundary geometry of $S^5 / \Gamma$ do not require any knowledge of partial resolutions; it is only when we turn to explicit time dependent phenomena that we require this more detailed information.

To track the explicit evolution of twisted sector closed string tachyon condensation we need to specify the group action on all of the worldsheet fields. To this end, introduce a basis of four vectors $e_{i}$ for $i=1,...,4$ for the $\mathbf{4}$. Then, the basis for the $\mathbf{6}$ (treated as a complex representation) is $e_{i} \wedge e_{j}$ for $i \neq j$. For $\zeta = \exp(2 \pi i / N)$ a generator of $\mathbb{Z}_N$, the group action of weight $\mathbf{s} = (s_1,s_2,s_3,s_4)$ on the two representations is induced from:
\begin{align}\label{eq:weights}
\mathbf{4} & : e_{i} \mapsto \zeta^{s_i} e_i \\
\mathbf{6} & : e_{i} \wedge e_{j} \mapsto \zeta^{s_i + s_j} e_{i} \wedge e_{j}.
\end{align}
Fixing a complex structure for $\mathbb{C}^3 = \mathbb{R}^6$ we can also specify an action on the $\mathbf{3}$ of $SU(3)$, i.e., the vector representation. Introducing basis vectors $h_a = e_{a} \wedge e_4$ for $a = 1,2,3$ we also have an induced group action:
\begin{equation}\label{eq:holo}
\mathbf{3} : h_{a} \mapsto \zeta^{s_a + s_4} h_{a}.
\end{equation}
These considerations suffice to fully fix the worldsheet CFT, i.e., we simply gauge by $\Gamma$ (with actions as specified above). This also suffices to specify the worldvolume theory of probe D-branes in this background.

An important subtlety with this procedure is that we still need to implement the GSO projection to produce a worldsheet theory which has a modular invariant 1-loop partition function. In the case of a supersymmetric background this is implicitly determined once we specify the action on the holomorphic basis of line (\ref{eq:holo}). Since we no longer have supersymmetry, we need to verify that our GSO projection has eliminated bulk (i.e., untwisted sector) tachyons, namely, that we are in type II string theory rather than type 0 string theory.

One way to establish this is to start from the action of line (\ref{eq:holo})
on the holomorphic coordinates and then build a suitable spin lift.
Following \cite{Adams:2001sv}, let $J_{i}$ denote the spin $1/2$ generators of rotations in the three directions.
Then, the action on spacetime fermions are generated by:
\begin{equation}
r_{\text{ferm}} = \exp\left(\frac{2 \pi i}{N} \underset{a = 1,2,3}{\sum} (s_a + s_4) J_a \right).
\end{equation}
The condition that we have landed in the type II rather than type 0 string means we do not gauge by $(-1)^{F}_{\text{spacetime}}$, i.e., we require $(r_{\text{ferm}})^{N} = 1$, namely:
\begin{equation}\label{eq:sumo1}
\exp \left(\pi i (s_1 + s_2 + s_3 + 3s_4) \right) = 1.
\end{equation}
Choosing integer representatives for the $s_{i}$, this amounts to the condition:
\begin{equation}\label{eq:sumo2}
s_1 + s_2 + s_3 + 3s_4 = 0 \, \mod 2.
\end{equation}
Observe, however, that the choice of integer representation superficially appears to suffer from an ambiguity; for $N$ odd, the shift $s_{i} \rightarrow s_{i} + N$ would seem to produce an inconsistent solution. All that has happened, however, is that we have reorganized the Hilbert space and the GSO projection now takes us to the type 0 theory where we have no spacetime fermions in the untwisted sector, and we also have a bulk tachyon.

With this in mind, we shall opt to always pick integral weights $s_{i}$ so that the conditions of (\ref{eq:sumo1}) and (\ref{eq:sumo2}) explicitly hold, and to make this manifest we allow both positive and negative values. We stress that at the level of extracting the quiver gauge theory and the geometry $S^5 / \Gamma$ (where we work mod $N$ anyway) these distinctions play no role; it is really in tracking the tachyon condensation of the type II theory that we need this further data.

To simplify the toric geometry analysis (and to closely follow the presentation given in \cite{Morrison:2004fr}) it will prove useful, whenever possible, to present the target space geometry as $\mathbb{C}^3 / \mathbb{Z}_{N}$ with holomorphic 
weights $(1,p,q)_{\mathrm{hol}}$ namely the group action of line
(\ref{eq:holo}) is used to define an equivalent action on holomorphic coordinates $(Z_1,Z_2,Z_3)$ of $\mathbb{C}^3$:
\begin{equation}\label{eq:eltoro}
(Z_1,Z_2,Z_3) \mapsto (\omega Z_1, \omega^{p} Z_2, \omega^{q} Z_3),
\end{equation}
where $\omega = \zeta^{m}$ is fixed by the convention that the
action on one of the holomorphic coordinates (possibly after an $SU(3)$ rotation) has weight one (namely, on $Z_1$).\footnote{Sometimes this is not possible, but this choice will be available in all the examples we consider.} Note that in making this change of basis the action on the spactime fermions (induced from the spin lift) is left implicit; it is again fixed by the condition that the GSO projection eliminates all bulk tachyons.

Much as in \cite{Lee:2003ar, Morrison:2004fr} we sort candidate tachyonic operators according to chiral / anti-chiral rings. In the RNS formalism we can introduce three separate sectors $Z_1$, $Z_2$ and $Z_3$ and due to the structure of the orbifold theory correspondingly construct chiral / anti-chiral rings for each coordinate separately, e.g., $c_1$ and $\overline{c_{1}}$ for the chiral / anti-chiral ring of $Z_1$. The operators of lowest scaling dimension dominate the flow, and much as in \cite{Lee:2003ar, Morrison:2004fr} we assume that tachyon condensation can be analyzed sequentially by first determining the endpoint of a given deformation before the other operator deformations dominate. In a given unstable orbifold, the most relevant tachyon(s) as determined by the R-charge of the operator will belong to one (or more) of the (anti-)chiral rings (see Appendix \ref{app:WORLDSHEET} for more discussion on worldsheet considerations). We pick a convention where the most dominant tachyon is in the chiral $(c_1,c_2,c_3)$ ring. This process can be somewhat elaborate, but as noted in reference \cite{Morrison:2004fr}, the endpoint after all tachyons have condensed is a supersymmetric background. When this background is a singular target space it admits marginal deformations which we can interpret geometrically as resolution parameters.\footnote{In the case of type 0 backgrounds the endpoint of tachyon condensation can sometimes result in a geometry with terminal singularities, i.e., those which do not admit a crepant resolution
\cite{ Morrison:2004fr}.}

The explicit examples we analyze are chosen to exhibit different
possible phenomena associated with tachyon condensation,
and the generalized symmetries of these backgrounds with different codimension singularities:
\begin{itemize}
    \item Codim. $6$: $\R^6/\Z_N^{(1,1,1,-3)}$, for $N>3$ odd,
    \item Codim. $6$: $\R^6/\Z_{17}^{(4,6,-7,-3)}$,
    \item Codim. $6$: $\R^6/\Z_9^{(2,3,-4,-1)}$,
    \item Codim. $6$ and $4$: $\R^6/\Z_9^{(3,5,-6,-2)}$,
    \item Codim. $6$ (multiple tachyons): $\R^6/\Z_{23}^{(4,7,-8,-3)}$,
    \item Codim. $6$ and $2$: $\R^6/\Z_9^{(-4,-2,1,5)}$
\end{itemize}
where in the above, the notation $\mathbb{Z}_{N}^{s_1,s_2,s_3,s_4}$ indicates the action of the group $\mathbb{Z}_N$ on the four components of the $\mathbf{4}$ spinor representation as in line (\ref{eq:weights}). The use of negative weights is in accord with our discussion of the GSO projection near lines (\ref{eq:sumo1}) and (\ref{eq:sumo2}).

For each case we compute the defect group both before and after tachyon condensation. We do this via the quiver based method as well as the method based on the geometry of the boundary space $S^5 / \Gamma$. As expected, we find an exact match when all tachyons are initially localized at the tip of the cone. In the case with a codimension $2$ singularity there is a tachyon present in the boundary $S^5 / \Gamma$ we find a simple generalization which works this case as well (equation (\ref{eq:CoolFormula})).

\subsection{Codim. $6$: $\R^6/\Z_N^{(1,1,1,-3)}$, for $N>3$ odd}\label{NonSUSYEx}

In this subsection, we consider orbifolds of the form $\R^6/\Z_N^{(1,1,1,-3)}$, for odd $N>3$.  We begin by determining the defect group of the 4D theory before the onset of any tachyon condensation.

Letting $g$ denote a generator of $\Z_N$, and $\zeta$ the $N^{th}$ root of unity, we have the following action in the \textbf{4} of $SU(4)$:
\begin{equation}
    r(g^n) = \diag(\zeta^n, \zeta^n, \zeta^n, \zeta^{-3n})
\end{equation}
This yields the following action in the \textbf{6} of $SO(6)$:
\begin{equation} \label{(1,1,1)Boson}
    R(g^n) =\diag(\zeta^{2n},\zeta^{2n},\zeta^{2n},\zeta^{-2n},\zeta^{-2n},\zeta^{-2m})
\end{equation}

\begin{figure}[]
\usetikzlibrary{arrows}
\centering
\begin{tikzpicture}
\node[draw=none, minimum size=7cm,regular polygon,regular polygon sides=5] (a) {};

\foreach \x in {1,2,...,5}
  \fill (a.corner \x) circle[radius=6pt, fill = none];

\begin{scope}[very thick,decoration={
    markings,
    mark=at position 0.85 with {\arrow[scale = 1.5, >=stealth]{>>>}}}
    ]
\draw[postaction = {decorate}] (a.corner 1) to (a.corner 2);
\draw[postaction = {decorate}] (a.corner 2) to (a.corner 3);
\draw[postaction = {decorate}] (a.corner 3) to (a.corner 4);
\draw[postaction = {decorate}] (a.corner 4) to (a.corner 5);
\draw[postaction = {decorate}] (a.corner 5) to (a.corner 1);
\end{scope}
\begin{scope}[very thick,decoration={
    markings,
    mark=at position 0.75 with {\arrow[scale = 1.5, >=stealth]{>}}}
    ]
\draw[postaction = {decorate}] (a.corner 1) to (a.corner 3);
\draw[postaction = {decorate}] (a.corner 2) to (a.corner 4);
\draw[postaction = {decorate}] (a.corner 3) to (a.corner 5);
\draw[postaction = {decorate}] (a.corner 4) to (a.corner 1);
\draw[postaction = {decorate}] (a.corner 5) to (a.corner 2);
\end{scope}
\end{tikzpicture}\qquad
 \;\;\;\;\;\;\;\;
\begin{tikzpicture}
\node[draw=none, minimum size=7cm,regular polygon,regular polygon sides=5] (a) {};

\foreach \x in {1,2,...,5}
  \fill (a.corner \x) circle[radius=6pt, fill = none];

\begin{scope}[very thick,decoration={
    markings,
    mark=at position 0.85 with {\arrow[scale = 1.5, >=stealth]{>>>}}}
    ]
\draw[postaction = {decorate}][dashed] (a.corner 1) to (a.corner 3);
\draw[postaction = {decorate}][dashed] (a.corner 2) to (a.corner 4);
\draw[postaction = {decorate}][dashed] (a.corner 3) to (a.corner 5);
\draw[postaction = {decorate}][dashed] (a.corner 4) to (a.corner 1);
\draw[postaction = {decorate}][dashed] (a.corner 5) to (a.corner 2);
\end{scope}
\end{tikzpicture}

\caption{\label{Z5quivers} Left/Right: Fermionic/Bosonic quiver for $\R^6/\Z_5^{(1,1,1,-3)}$.}
\end{figure}
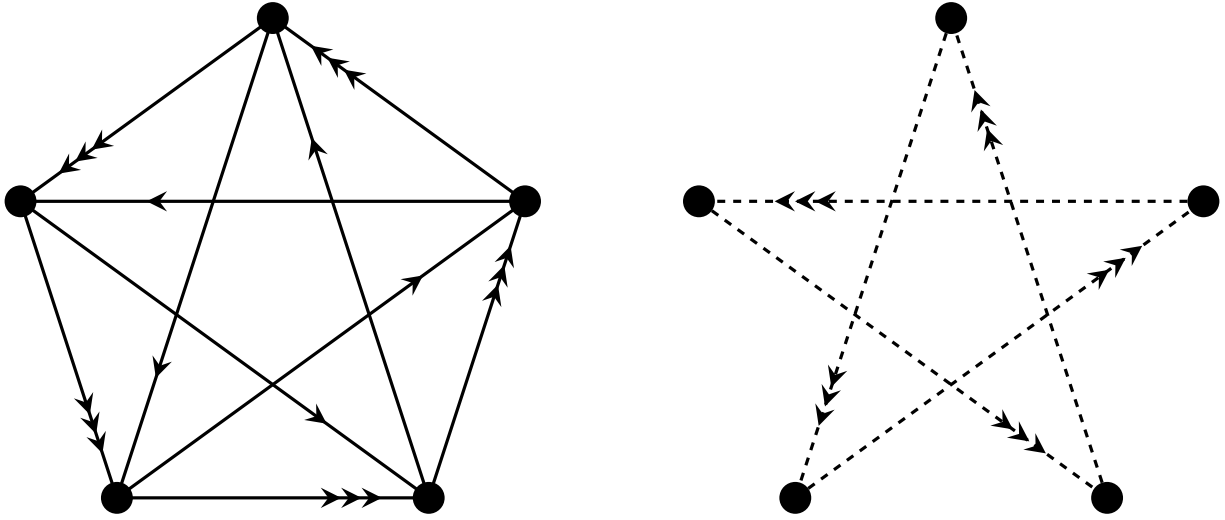

The D0-brane probe results in a quiver quantum mechanics with $N$ nodes. Along the boundary of the quiver, we have bifundamental fermions $\psi^{i,i+1}$ for $i = 1,...,N$ (indexing mod $N$) each with multiplicity $3$. There are also bifundamental fermions $\psi^{i, i-3}$ for $i = 1,...,N$ (indexing mod $N$) each with multiplicity $1$. For the scalar sector, there are are bifundamental scalars $\phi^{i,i+2}$, each with multiplicity $3$. We illustrate this in the case of $N = 5$ in figure \ref{Z5quivers}.

We are interested in the electric and magnetic $1$-form symmetries of the 4D theory. For this  we need only consider the fermionic quiver. Consider again the case of $N = 5$. The adjacency matrix for the fermionic quiver in figure \ref{Z5quivers} is given by
\begin{equation}
K = \begin{pmatrix}
    0 & 3 & 1 & -1 & -3 \\
    -3 & 0 & 3 & 1 & -1 \\
    -1 & -3 & 0 & 3 & 1 \\
    1 & -1 & -3 & 0 & 3 \\
    3 & 1 & -1 & -3 & 0
\end{pmatrix}
\end{equation}
The torsional generators of the defect group for the 4D theory are determined from $\text{Coker}(K)$.  From taking the Smith normal form of $K$, we find that
\begin{equation} \label{(1,1,1)DefG}
    \mathcal{A}^{(1)}_{\mathrm{elec}} \oplus \mathcal{A}^{(1)}_{\mathrm{mag}} \cong \text{Tor}(\text{Coker}(K)) \cong \Z_5 \oplus \Z_5
\end{equation}
In the more general case of $\R^6/\Z_N^{(1,1,1,-3)}$, for odd $N >3$, we follow an identical procedure and find that the defect group is $\Z_N \oplus \Z_N$.

This result is also predicted by the geometry. Indeed, consider the action of $\Gamma = \Z_N$ on $S^5$ induced by the bosonic action given in \eqref{(1,1,1)Boson}. It is clear that this action is fixed point free. Hence, using Armstrong's theorem, we expect a factor in the defect group given by
\begin{equation}
    H_1(S^5/\Gamma) \cong \text{Ab}[\pi_1(\Gamma/H)] \cong  \Z_N
\end{equation}
That is, $\mathcal{A}^{(1)}_{\mathrm{elec}} \cong \Z_N$, and we have agreement between the defect group computed by the quiver and $H_1(S^5/\Gamma)$.


We now move to study the time dependent nature of our analysis. In line with our discussion near line (\ref{eq:eltoro})
we switch to a holomorphic coordinate system where we can track the toric geometry explicitly,
i.e., $\mathbb{C}^3 / \mathbb{Z}_N^{(1,1,1-N})$. The most relevant GSO-preserved tachyon is $T_1$ in the  chiral ring $(c_1,c_2,c_3)$.  Here we let $T_j$ denote a tachyonic operator in the $j^{th}$ twisted sector. Note that there are other GSO preserved tachyons, but $T_1$ is the most relevant and, as we will see, resolves our orbifold to smooth space upon condensing. The condensation of $T_1$ is studied using the fact that $\mathbb{C}^3/\Z_N$ with weights $(1,1,1-N)$ is a toric variety. Following the discussion in Appendix \ref{app:WORLDSHEET}, this orbifold is a toric variety whose fan $\Sigma$ is given by the vertices:\footnote{To distinguish the weights of the group action on the holomorphic coordinates from the three-component vectors of the toric fan we adopt the notation $(\bullet, \bullet, \bullet)$ and $((\bullet , \bullet , \bullet))$, respectively.}
\begin{equation}
\alpha_1 = ((N,-1,N-1)), \;\; \alpha_2 = ((0,1,0)), \;\; \alpha_3 = ((0,0,1))
\end{equation}
As a lattice point in the toric diagram, $T_1$ corresponds to $T_1 = ((1,0,1))$. Condensation of $T_1$ corresponds to blowing up $\Sigma $ by $T_1$. This gives the residual subcones $C[T_1,\alpha_1,\alpha_2]$, $C[T_1,\alpha_1,\alpha_3]$, and $C[T_1,\alpha_2,\alpha_3]$, all of which describe patches of the resolved geometry. The orbifold conformal field theories described by each of these subcones correspond to smooth spaces. That is, there are no residual singularities associated to the above subcones. Hence, the endpoint of the most relevant tachyon sequence is smooth, as expected.

Using standard techniques in toric geometry (see e.g. \cite{Hori:2003ic}), we find that the geometry of the resultant space after $T_1$ condenses is $\mathcal{O}(-N)\rightarrow\P^2$.

\subsection{Codim. $6$: $\R^6/\Z_{17}^{(4,6,-7,-3)}$}\label{sec:smthBordism}

In this subsection, we consider the orbifold $\R^6/\Z_{17}^{(4,6,-7,-3)}$. We proceed in an analogous way as that for the previous example. We begin by determining the defect group of our orbifold and then move on to study how the geometry and defect group change as the tachyons of our theory condense.

Letting $g$ denote a generator of $\Z_{17}$ and $\zeta$ a $17^{th}$ root of unity, we choose the following action in the \textbf{4} of $SU(4)$:
\begin{equation}
    r(g^n) = \diag(\zeta^{4n}, \zeta^{6n}, \zeta^{10n}, \zeta^{14n})
\end{equation}
This yields the following action in the \textbf{6} of $SO(6)$:
\begin{equation} \label{(4,6,10)BosAc}
    R(g^n) =\diag(\zeta^{16n},\zeta^{14n},\zeta^{10n},\zeta^{-16n},\zeta^{-14n},\zeta^{-10n})
\end{equation}
We find that the D0-brane probe results in a quiver quantum mechanics with $17$ nodes. Furthermore, there are bifundamental fermions (each with multiplicity one) given by $\psi^{i,i+4}$, $\psi^{i,i+6}$, $\psi^{i,i+10}$, and $\psi^{i,i+14}$. For the scalar sector, there are bifundamental bosons (each with multiplicity one) given by $\phi^{i,i+16}$, $\phi^{i,i+14}$, and $\phi^{i,i+10}$. The resultant fermionic and bosonic quivers are given in figure \ref{(2,3,5)Quivs}.

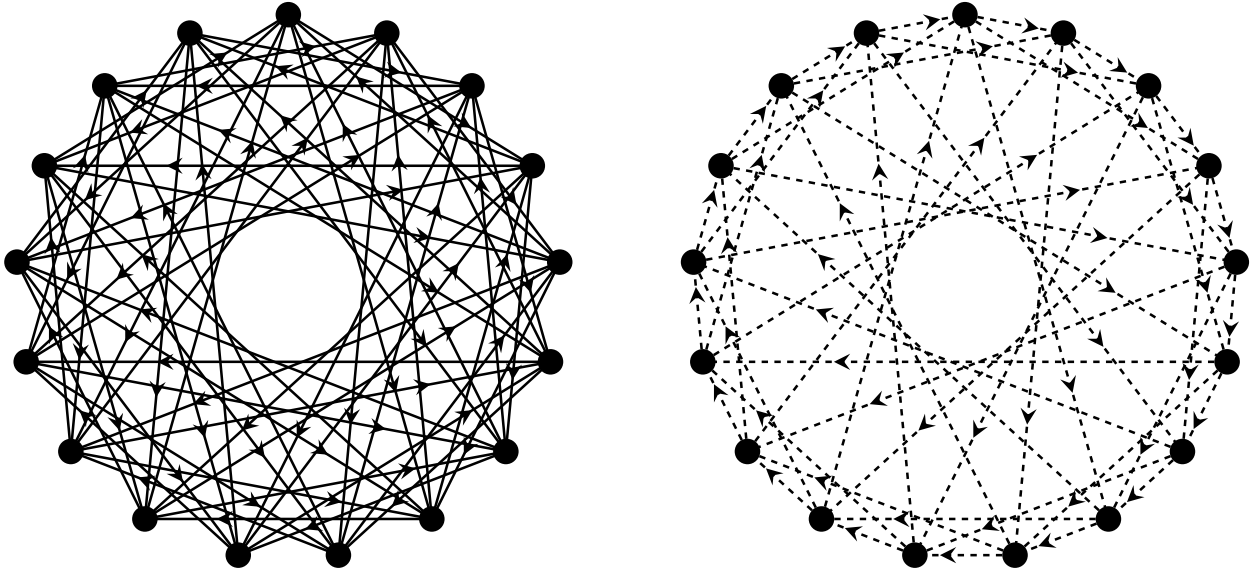
\begin{figure}[]
\usetikzlibrary{arrows}
\centering
\scalebox{0.8}{
\begin{tikzpicture}
\node[draw=none,minimum size=9cm,regular polygon,regular polygon sides=17] (a) {};

\foreach \x in {1,2,...,17}
  \fill (a.corner \x) circle[radius=6pt];

\begin{scope}[very thick,decoration={
    markings,
    mark=at position 0.75 with {\arrow[scale = 1.5, >=stealth]{>}}}
    ]
\draw[postaction = {decorate}] (a.corner 1) to (a.corner 5);
\draw[postaction = {decorate}] (a.corner 2) to (a.corner 6);
\draw[postaction = {decorate}] (a.corner 3) to (a.corner 7);
\draw[postaction = {decorate}] (a.corner 4) to (a.corner 8);
\draw[postaction = {decorate}] (a.corner 5) to (a.corner 9);
\draw[postaction = {decorate}] (a.corner 6) to (a.corner 10);
\draw[postaction = {decorate}] (a.corner 7) to (a.corner 11);
\draw[postaction = {decorate}] (a.corner 8) to (a.corner 12);
\draw[postaction = {decorate}] (a.corner 9) to (a.corner 13);
\draw[postaction = {decorate}] (a.corner 10) to (a.corner 14);
\draw[postaction = {decorate}] (a.corner 11) to (a.corner 15);
\draw[postaction = {decorate}] (a.corner 12) to (a.corner 16);
\draw[postaction = {decorate}] (a.corner 13) to (a.corner 17);
\draw[postaction = {decorate}] (a.corner 14) to (a.corner 1);
\draw[postaction = {decorate}] (a.corner 15) to (a.corner 2);
\draw[postaction = {decorate}] (a.corner 16) to (a.corner 3);
\draw[postaction = {decorate}] (a.corner 17) to (a.corner 4);
\draw[postaction = {decorate}] (a.corner 1) to (a.corner 7);
\draw[postaction = {decorate}] (a.corner 2) to (a.corner 8);
\draw[postaction = {decorate}] (a.corner 3) to (a.corner 9);
\draw[postaction = {decorate}] (a.corner 4) to (a.corner 10);
\draw[postaction = {decorate}] (a.corner 5) to (a.corner 11);
\draw[postaction = {decorate}] (a.corner 6) to (a.corner 12);
\draw[postaction = {decorate}] (a.corner 7) to (a.corner 13);
\draw[postaction = {decorate}] (a.corner 8) to (a.corner 14);
\draw[postaction = {decorate}] (a.corner 9) to (a.corner 15);
\draw[postaction = {decorate}] (a.corner 10) to (a.corner 16);
\draw[postaction = {decorate}] (a.corner 11) to (a.corner 17);
\draw[postaction = {decorate}] (a.corner 12) to (a.corner 1);
\draw[postaction = {decorate}] (a.corner 13) to (a.corner 2);
\draw[postaction = {decorate}] (a.corner 14) to (a.corner 3);
\draw[postaction = {decorate}] (a.corner 15) to (a.corner 4);
\draw[postaction = {decorate}] (a.corner 16) to (a.corner 5);
\draw[postaction = {decorate}] (a.corner 17) to (a.corner 6);
\draw[postaction = {decorate}] (a.corner 1) to (a.corner 11);
\draw[postaction = {decorate}] (a.corner 2) to (a.corner 12);
\draw[postaction = {decorate}] (a.corner 3) to (a.corner 13);
\draw[postaction = {decorate}] (a.corner 4) to (a.corner 14);
\draw[postaction = {decorate}] (a.corner 5) to (a.corner 15);
\draw[postaction = {decorate}] (a.corner 6) to (a.corner 16);
\draw[postaction = {decorate}] (a.corner 7) to (a.corner 17);
\draw[postaction = {decorate}] (a.corner 8) to (a.corner 1);
\draw[postaction = {decorate}] (a.corner 9) to (a.corner 2);
\draw[postaction = {decorate}] (a.corner 10) to (a.corner 3);
\draw[postaction = {decorate}] (a.corner 11) to (a.corner 4);
\draw[postaction = {decorate}] (a.corner 12) to (a.corner 5);
\draw[postaction = {decorate}] (a.corner 13) to (a.corner 6);
\draw[postaction = {decorate}] (a.corner 14) to (a.corner 7);
\draw[postaction = {decorate}] (a.corner 15) to (a.corner 8);
\draw[postaction = {decorate}] (a.corner 16) to (a.corner 9);
\draw[postaction = {decorate}] (a.corner 17) to (a.corner 10);
\draw[postaction = {decorate}] (a.corner 1) to (a.corner 15);
\draw[postaction = {decorate}] (a.corner 2) to (a.corner 16);
\draw[postaction = {decorate}] (a.corner 3) to (a.corner 17);
\draw[postaction = {decorate}] (a.corner 4) to (a.corner 1);
\draw[postaction = {decorate}] (a.corner 5) to (a.corner 2);
\draw[postaction = {decorate}] (a.corner 6) to (a.corner 3);
\draw[postaction = {decorate}] (a.corner 7) to (a.corner 4);
\draw[postaction = {decorate}] (a.corner 8) to (a.corner 5);
\draw[postaction = {decorate}] (a.corner 9) to (a.corner 6);
\draw[postaction = {decorate}] (a.corner 10) to (a.corner 7);
\draw[postaction = {decorate}] (a.corner 11) to (a.corner 8);
\draw[postaction = {decorate}] (a.corner 12) to (a.corner 9);
\draw[postaction = {decorate}] (a.corner 13) to (a.corner 10);
\draw[postaction = {decorate}] (a.corner 14) to (a.corner 11);
\draw[postaction = {decorate}] (a.corner 15) to (a.corner 12);
\draw[postaction = {decorate}] (a.corner 16) to (a.corner 13);
\draw[postaction = {decorate}] (a.corner 17) to (a.corner 14);
\end{scope}

\end{tikzpicture}\qquad\qquad
\usetikzlibrary{arrows}
\centering
\begin{tikzpicture}
\node[draw=none,minimum size=9cm,regular polygon,regular polygon sides=17] (a) {};

\foreach \x in {1,2,...,17}
  \fill (a.corner \x) circle[radius=6pt];

\begin{scope}[very thick,decoration={
    markings,
    mark=at position 0.75 with {\arrow[scale = 1.5, >=stealth]{>}}}
    ]
\tikzset{edge/.style = {->,> = latex}}
\draw[postaction = {decorate}][dashed] (a.corner 1) to (a.corner 17);
\draw[postaction = {decorate}][dashed] (a.corner 2) to (a.corner 1);
\draw[postaction = {decorate}][dashed] (a.corner 3) to (a.corner 2);
\draw[postaction = {decorate}][dashed] (a.corner 4) to (a.corner 3);
\draw[postaction = {decorate}][dashed] (a.corner 5) to (a.corner 4);
\draw[postaction = {decorate}][dashed] (a.corner 6) to (a.corner 5);
\draw[postaction = {decorate}][dashed] (a.corner 7) to (a.corner 6);
\draw[postaction = {decorate}][dashed] (a.corner 8) to (a.corner 7);
\draw[postaction = {decorate}][dashed] (a.corner 9) to (a.corner 8);
\draw[postaction = {decorate}][dashed] (a.corner 10) to (a.corner 9);
\draw[postaction = {decorate}][dashed] (a.corner 11) to (a.corner 10);
\draw[postaction = {decorate}][dashed] (a.corner 12) to (a.corner 11);
\draw[postaction = {decorate}][dashed] (a.corner 13) to (a.corner 12);
\draw[postaction = {decorate}][dashed] (a.corner 14) to (a.corner 13);
\draw[postaction = {decorate}][dashed] (a.corner 15) to (a.corner 14);
\draw[postaction = {decorate}][dashed] (a.corner 16) to (a.corner 15);
\draw[postaction = {decorate}][dashed] (a.corner 17) to (a.corner 16);
\draw[postaction = {decorate}][dashed] (a.corner 1) to (a.corner 11);
\draw[postaction = {decorate}][dashed] (a.corner 2) to (a.corner 12);
\draw[postaction = {decorate}][dashed] (a.corner 3) to (a.corner 13);
\draw[postaction = {decorate}][dashed] (a.corner 4) to (a.corner 14);
\draw[postaction = {decorate}][dashed] (a.corner 5) to (a.corner 15);
\draw[postaction = {decorate}][dashed] (a.corner 6) to (a.corner 16);
\draw[postaction = {decorate}][dashed] (a.corner 7) to (a.corner 17);
\draw[postaction = {decorate}][dashed] (a.corner 8) to (a.corner 1);
\draw[postaction = {decorate}][dashed] (a.corner 9) to (a.corner 2);
\draw[postaction = {decorate}][dashed] (a.corner 10) to (a.corner 3);
\draw[postaction = {decorate}][dashed] (a.corner 11) to (a.corner 4);
\draw[postaction = {decorate}][dashed] (a.corner 12) to (a.corner 5);
\draw[postaction = {decorate}][dashed] (a.corner 13) to (a.corner 6);
\draw[postaction = {decorate}][dashed] (a.corner 14) to (a.corner 7);
\draw[postaction = {decorate}][dashed] (a.corner 15) to (a.corner 8);
\draw[postaction = {decorate}][dashed] (a.corner 16) to (a.corner 9);
\draw[postaction = {decorate}][dashed] (a.corner 17) to (a.corner 10);
\draw[postaction = {decorate}][dashed] (a.corner 1) to (a.corner 15);
\draw[postaction = {decorate}][dashed] (a.corner 2) to (a.corner 16);
\draw[postaction = {decorate}][dashed] (a.corner 3) to (a.corner 17);
\draw[postaction = {decorate}][dashed] (a.corner 4) to (a.corner 1);
\draw[postaction = {decorate}][dashed] (a.corner 5) to (a.corner 2);
\draw[postaction = {decorate}][dashed] (a.corner 6) to (a.corner 3);
\draw[postaction = {decorate}][dashed] (a.corner 7) to (a.corner 4);
\draw[postaction = {decorate}][dashed] (a.corner 8) to (a.corner 5);
\draw[postaction = {decorate}][dashed] (a.corner 9) to (a.corner 6);
\draw[postaction = {decorate}][dashed] (a.corner 10) to (a.corner 7);
\draw[postaction = {decorate}][dashed] (a.corner 11) to (a.corner 8);
\draw[postaction = {decorate}][dashed] (a.corner 12) to (a.corner 9);
\draw[postaction = {decorate}][dashed] (a.corner 13) to (a.corner 10);
\draw[postaction = {decorate}][dashed] (a.corner 14) to (a.corner 11);
\draw[postaction = {decorate}][dashed] (a.corner 15) to (a.corner 12);
\draw[postaction = {decorate}][dashed] (a.corner 16) to (a.corner 13);
\draw[postaction = {decorate}][dashed] (a.corner 17) to (a.corner 14);
\end{scope}

\end{tikzpicture}} \caption{\label{(2,3,5)Quivs} Left/Right: Fermionic/Bosonic quiver for $\R^6/\Z_{17}^{(4,6,-7,-3)}$.}
\end{figure}

We are interested in the electric and magnetic $1$-form symmetries of the 4D theory. We need only consider the fermionic quiver and its corresponding adjacency matrix $K$. Indeed, the torsional generators of the defect group of the 4D theory are determined from $\text{Coker}(K)$:
\begin{equation} \label{(1,2,-5)DefG}
\mathbb{D}^{(1)} = \mathcal{A}^{(1)}_{\mathrm{elec}}\oplus \mathcal{A}^{(1)}_{\mathrm{mag}} \cong \text{Tor}(\text{Coker}(K)) = \Z_{17}\oplus \Z_{17}
\end{equation}

This result is predicted by the geometry. Consider the action of $\Gamma = \Z_{17}$ on $S^5$ induced by the bosonic action in \eqref{(4,6,10)BosAc}. It is clear that this action is fixed point free. Hence,  through an application of Armstrong's theorem, we find that
\begin{equation}
       H_1(S^5/\Gamma) \cong \text{Ab}[\pi_1(\Gamma/H)] \cong  \Z_{17}
\end{equation}
That is, $\mathcal{A}^{(1)}_{\mathrm{{elec}}}\cong \Z_{17}$.


We now move to study the time dependent nature of our analysis. In particular, we study how the geometry and defect group of our theory evolve with tachyon condensation. As before, we follow the procedure of \cite{Morrison:2004fr}. The orbifold we have been considering, $\R^6/\Z_{17}^{(4,6,-7,-3)}$, is, in the notation of \cite{Morrison:2004fr}, given by $\mathbb{C}^3/\Z_{17}$, where $\Z_{17}$ acts in accordance to the weights $(1,3,-10)_{\mathrm{hol}}$. The chiral ring $(c_1, c_2, c_3)$ of operators has two tachyons $T_1$ and $T_6$ with R-charges $R_1 = \frac{11}{17}$ and $R_6 = \frac{15}{17}$, respectively, that survive the Type II GSO projection. Here we made use of line \eqref{eq:oprcharge} in Appendix \ref{app:WORLDSHEET} to determine the R-charges. While there are GSO-preserved tachyons in the other rings, the most relevant tachyon in this theory is $T_1$ from the $(c_1 , c_2 , c_3)$ ring.

\begin{figure}[]
\usetikzlibrary{arrows}
\centering
\begin{tikzpicture}
\node[draw=none, minimum size=6cm,regular polygon,regular polygon sides=7] (a) {};

\foreach \x in {1,2,...,7}
  \fill (a.corner \x) circle[radius=6pt, fill = none];

\begin{scope}[very thick,decoration={
    markings,
    mark=at position 0.8 with {\arrow[scale = 1.5, >=stealth]{>>}}}
    ]
\draw[postaction = {decorate}] (a.corner 1) to (a.corner 5);
\draw[postaction = {decorate}] (a.corner 2) to (a.corner 6);
\draw[postaction = {decorate}] (a.corner 3) to (a.corner 7);
\draw[postaction = {decorate}] (a.corner 4) to (a.corner 1);
\draw[postaction = {decorate}] (a.corner 5) to (a.corner 2);
\draw[postaction = {decorate}] (a.corner 6) to (a.corner 3);
\draw[postaction = {decorate}] (a.corner 7) to (a.corner 4);
\end{scope}
%
\begin{scope}[very thick,decoration={
    markings,
    mark=at position 0.7 with {\arrow[scale = 1.5, >=stealth]{>}}}
    ]
\draw[postaction = {decorate}] (a.corner 1) to (a.corner 7);
\draw[postaction = {decorate}] (a.corner 2) to (a.corner 1);
\draw[postaction = {decorate}] (a.corner 3) to (a.corner 2);
\draw[postaction = {decorate}] (a.corner 4) to (a.corner 3);
\draw[postaction = {decorate}] (a.corner 5) to (a.corner 4);
\draw[postaction = {decorate}] (a.corner 6) to (a.corner 5);
\draw[postaction = {decorate}] (a.corner 7) to (a.corner 6);
\end{scope}

\end{tikzpicture}\qquad \qquad
\begin{tikzpicture}
\node[draw=none,minimum size=6cm,regular polygon,regular polygon sides=3] (a) {};

\foreach \x in {1,2,3}
  \fill (a.corner \x) circle[radius=6pt, fill = none];

\begin{scope}[very thick,decoration={
    markings,
    mark=at position 0.5 with {\arrow[scale = 1.5, >=stealth]{>>>}}}
    ]
\draw[postaction = {decorate}] (a.corner 1) to (a.corner 2);
\draw[postaction = {decorate}] (a.corner 2) to (a.corner 3);
\draw[postaction = {decorate}] (a.corner 3) to (a.corner 1);
\end{scope}
%

\end{tikzpicture}
\caption{\label{(1,3,-2)Quivs} Left: Quiver for $\R^6/\Z_{7}^{(-3,-1,0,4)}$. Right: Quiver for $\R^6/\Z_3^{(1,1,1,0)}$}
\end{figure}
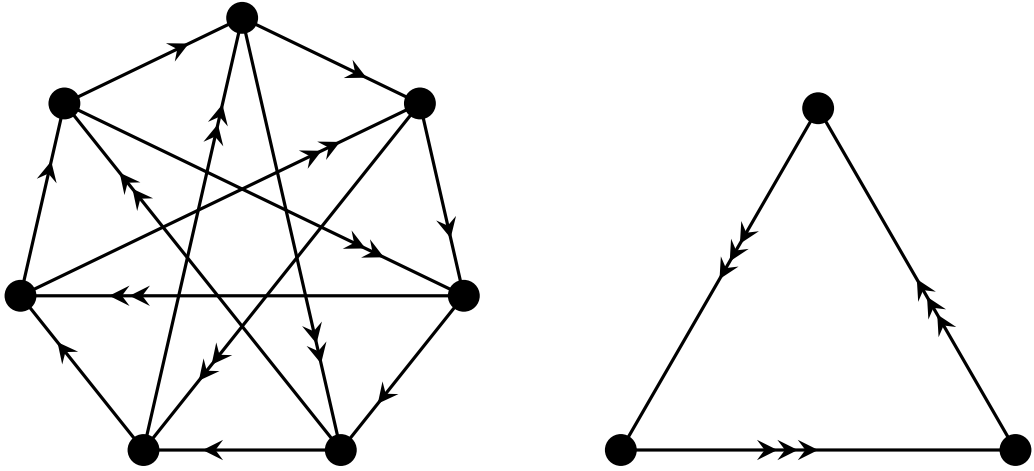

\begin{figure}
\centering
\scalebox{0.725}{
\begin{tikzpicture}
	\begin{pgfonlayer}{nodelayer}
		\node [style=Star] (0) at (-4, 0) {};
		\node [style=none] (1) at (-0.75, 0) {};
		\node [style=none] (2) at (0.75, 0) {};
		\node [style=none] (3) at (0, 0.5) {$T_1$};
		\node [style=none] (4) at (-4, -3.25) {\large $S^5/\Z_{17}$};
		\node [style=none] (5) at (-5, 0) { $\mathbb{C}^3/\Z_{17}$};
		\node [style=none] (6) at (5.5, -1) { $S^5/\Z_3$};
		\node [style=none] (7) at (6.5, 0) {};
		\node [style=none] (8) at (1.5, 0) {};
		\node [style=none] (9) at (4, 2.5) {};
		\node [style=none] (10) at (4, -2.5) {};
		\node [style=none] (11) at (4, -3.25) {\large $S^5/\Z_{17}$};
		\node [style=none] (16) at (-1.5, 0) {};
		\node [style=none] (17) at (-6.5, 0) {};
		\node [style=none] (18) at (-4, 2.5) {};
		\node [style=none] (19) at (-4, -2.5) {};
		\node [style=none] (20) at (4.75, -1) {};
		\node [style=none] (21) at (3.25, -1) {};
		\node [style=none] (22) at (4, -0.25) {};
		\node [style=none] (23) at (4, -1.75) {};
		\node [style=Star] (26) at (4, -1) {};
		\node [style=none] (27) at (5.5, 1) {$S^5/\Z_7$};
		\node [style=none] (28) at (4.75, 1) {};
		\node [style=none] (29) at (3.25, 1) {};
		\node [style=none] (30) at (4, 1.75) {};
		\node [style=none] (31) at (4, 0.25) {};
		\node [style=Star] (32) at (4, 1) {};
		\node [style=none] (34) at (8.75, 0) {};
		\node [style=none] (53) at (-3.25, -6.5) {};
		\node [style=none] (54) at (-3.25, -8) {};
		\node [style=none] (55) at (1.75, -5) {};
		\node [style=none] (56) at (1.75, -6.5) {};
		\node [style=none] (57) at (1.75, -8) {};
		\node [style=none] (58) at (1.75, -9.5) {};
		\node [style=none] (59) at (2.875, -5.75) {\large $S^5/\Z_7$};
		\node [style=none] (60) at (2.875, -8.75) {\large $S^5/\Z_3$};
		\node [style=none] (61) at (-4.5, -7.25) {\large $S^5/\Z_{17}$};
		\node [style=none] (62) at (-1.25, -7.3) {};
		\node [style=none] (63) at (-0.5, -7.3) {};
		\node [style=none] (64) at (-0.25, -7.25) {};
		\node [style=none] (65) at (-1.5, -7.25) {};
		\node [style=none] (66) at (-7.75, 0) {\large (i)};
		\node [style=none] (67) at (6, -6.5) {};
		\node [style=none] (68) at (6, -8) {};
		\node [style=none] (87) at (-0.75, -10) {\large (ii)};
		\node [style=none] (89) at (4, -10.5) {};
        \node [style=none] (90) at (2.25, 0) {};
		\node [style=none] (91) at (3, 0) {};
		\node [style=none] (92) at (3.25,  0.05) {};
		\node [style=none] (93) at (2,  0.05) {};
	\end{pgfonlayer}
	\begin{pgfonlayer}{edgelayer}
		\draw [style=ArrowLineRight] (1.center) to (2.center);
		\draw [style=none, bend left=45] (8.center) to (9.center);
		\draw [style=none, bend left=45] (9.center) to (7.center);
		\draw [style=none, bend left=45] (7.center) to (10.center);
		\draw [style=none, bend left=45] (10.center) to (8.center);
		\draw [bend left=45] (17.center) to (18.center);
		\draw [bend left=45] (18.center) to (16.center);
		\draw [bend left=45] (16.center) to (19.center);
		\draw [bend right=315] (19.center) to (17.center);
		\draw [bend right] (17.center) to (16.center);
		\draw [style=DottedLine, bend left] (17.center) to (16.center);
		\draw [bend left=45] (21.center) to (22.center);
		\draw [bend left=45] (22.center) to (20.center);
		\draw [bend left=45] (20.center) to (23.center);
		\draw [bend right=315] (23.center) to (21.center);
		\draw [bend right] (21.center) to (20.center);
		\draw [style=DottedLine, bend left] (21.center) to (20.center);
		\draw [bend left=45] (29.center) to (30.center);
		\draw [bend left=45] (30.center) to (28.center);
		\draw [bend left=45] (28.center) to (31.center);
		\draw [bend right=315] (31.center) to (29.center);
		\draw [bend right] (29.center) to (28.center);
		\draw [style=DottedLine, bend left] (29.center) to (28.center);
		\draw [style=ThickLine, bend left=270, looseness=0.75] (53.center) to (54.center);
		\draw [style=ThickLine, bend right=90, looseness=0.75] (55.center) to (56.center);
		\draw [style=ThickLine, bend left=270, looseness=0.75] (57.center) to (58.center);
		\draw [style=ThickLine, bend left=90, looseness=0.75] (55.center) to (56.center);
		\draw [style=ThickLine, bend left=90, looseness=0.75] (57.center) to (58.center);
		\draw [style=DottedLine, bend left=90, looseness=0.75] (53.center) to (54.center);
		\draw [style=ThickLine, in=-180, out=0] (53.center) to (55.center);
		\draw [style=ThickLine, bend left=270, looseness=2.50] (56.center) to (57.center);
		\draw [style=ThickLine, in=0, out=180] (58.center) to (54.center);
		\draw [style=ThickLine, bend left=15] (62.center) to (63.center);
		\draw [style=ThickLine, bend right=15] (65.center) to (64.center);
        \draw [style=ThickLine, bend left=15] (90.center) to (91.center);
		\draw [style=ThickLine, bend right=15] (93.center) to (92.center);
	\end{pgfonlayer}
\end{tikzpicture}
}
\caption{ We sketch the initial geometry and blowups thereof as induced by tachyon condensation. In (i) we show how the initial isolated singularity $\mathbb{C}^3/\mathbb{Z}_{17}$ resolves to a $\mathbb{C}^3/\mathbb{Z}_{3}$ and $\mathbb{C}^3/\mathbb{Z}_{7}$ singularity via condensation of $T_1$. The resolved geometry realizes bordisms between the boundaries of various local models as sketched in (ii).}
\label{fig:twoisolatedsingularities}
\end{figure}
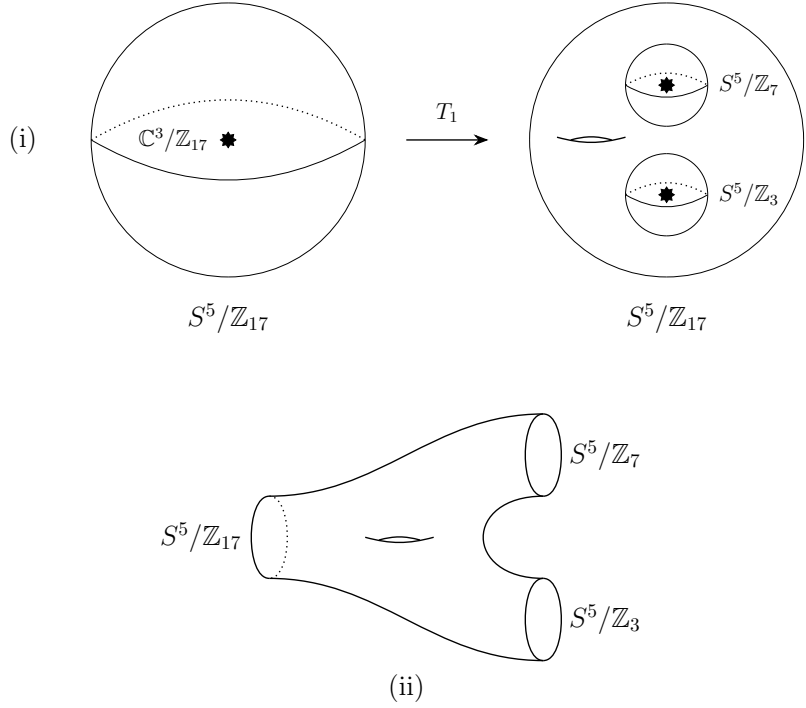

Following the discussion in Appendix \ref{app:WORLDSHEET},
the orbifold $\mathbb{C}^3/\Z_{17}$ with weights $(1,3,-10)$
is a toric variety whose fan is described by the vertices:
\begin{equation}
    \alpha_1 = ((17,-3,10)), \;\; \alpha_2 = ((0,1,0)), \;\; \alpha_3 = ((0,0,1))
\end{equation}
In terms of the toric diagram, the tachyons correspond to the lattice vectors $T_1 = ((1, 0, 1))$ and $T_6 = ((6, -1, 4))$. The tachyons blow up the singularity in order of relevance, i.e. first $T_1$ and then $T_6$. Condensation of $T_1$ corresponds to blowing up the toric fan by the lattice vector $T_1$. This gives the residual subcones $C[T_1,\alpha_1,\alpha_2]$, $ C[T_1,\alpha_2,\alpha_3]$, and $C[T_1, \alpha_1, \alpha_3]$, which corresponds to $\mathbb{C}^3/\Z_7$ with weights $(1,3,4)_{\mathrm{hol}}$, a smooth space, and $\mathbb{C}^3/\Z_3$ with weights $(1,1,1)_{\mathrm{hol}}$, respectively. Here we have made use of the fact that each of our subcones represents a new orbifold conformal field theory that are locally decoupled from the other theories corresponding to the other subcones. In our notation we have:
\begin{equation}\label{firstcondense}
    C[T_1,\alpha_1,\alpha_2] \sim \R^6/\Z_7^{(-3,-1,0,4)} \qquad C[T_1,\alpha_1,\alpha_3] \sim \R^6/\Z_3^{(1,1,1,0)}
\end{equation}
We remark here that the orbifolds in line \eqref{firstcondense} are both supersymmetric and their singularities are isolated (see figure \ref{fig:twoisolatedsingularities}).

Before we consider the remaining tachyon $T_6$, let us first briefly describe the quivers associated with
the subcones $C[T_1,\alpha_1,\alpha_2]$ and $C[T_1,\alpha_1,\alpha_3]$. We summarize in figure \ref{(1,3,-2)Quivs} the content of the quivers.
Note that both $C[T_1,\alpha_1,\alpha_2]$ and $C[T_1,\alpha_1,\alpha_3]$ are supersymmetric, so their fermionic and bosonic quivers are the same.

We can also study the defect groups for the subsystems $C[T_1,\alpha_1,\alpha_2] $ and $C[T_1,\alpha_1,\alpha_3]$ by computing the Smith normal form of the adjacency matrix for their respective quivers. In doing so, we find that $C[T_1,\alpha_1,\alpha_2] $ has defect group
\begin{equation}
    \mathcal{A}^{(1)}_{\mathrm{elec}} \oplus \mathcal{A}^{(1)}_{\mathrm{mag}} \cong \Z_7 \oplus \Z_7
\end{equation}
and $C[T_6,\alpha_1,\alpha_3]$ has defect group
\begin{equation}
    \mathcal{A}^{(1)}_{\mathrm{elec}} \oplus \mathcal{A}^{(1)}_{\mathrm{mag}} \cong \Z_3 \oplus \Z_3
\end{equation}
It is straightforward to check that both of these defect groups match what is predicted by the geometry.

We now return to the remaining tachyon $T_6$. Notice that $T_6$ is inside $C[T_1,\alpha_1,\alpha_3]$ since $T_6 = \frac{1}{3}(T_1 + \alpha_1 + \alpha_3)$. We recall from \cite{Morrison:2004fr} that R-charges of the subsequent tachyons remaining in the residual geometries get renormalized after a given tachyon has condensed; the specific renormalization of a particular subsequent tachyon depends on which of the three decoupled subcones it lies within. In our case, we find that  $T_6$  has a renormalized R-charge equal to one after $T_1 $ condenses. That is, the operator $T_6$ becomes marginal after $T_1$ condenses. Further subdividing our fan by $T_6$ results in the subcones: $C[T_6,\alpha_1,\alpha_3]$, $C[T_1,T_6,\alpha_1]$, $C[T_1,T_6,\alpha_3]$, $C[T_1,\alpha_1,\alpha_2]$, and $C[T_1,\alpha_2,\alpha_3]$. Notice that the latter two were previously considered. Furthermore, it is a straightforward exercise to show that the ``new'' cones, $C[T_6,\alpha_1,\alpha_3]$, $C[T_1, T_6, \alpha_1]$, and  $C[T_1,T_6,\alpha_3]$ are smooth. The only non-trivial thing left to consider is $C[T_1,\alpha_1,\alpha_2]$. However, this orbifold is supersymmetric and resolves to a smooth space via generic metric blowup modes. See \cite{Morrison:2004fr} for further details. We conclude that the endpoint of the most relevant tachyon sequence is smooth, as expected.

\subsection{Codim. $6$: $\R^6/\Z_9^{(2,3,-4,-1)}$}

We next consider the orbifold $\R^6/\Z_9^{(2,3,-4,-1)}$.  We begin by determining the defect group of the 4D theory at early times.
Letting $g$ denote a generator of $\Z_9$ and $\zeta $ a primitive $9^{th}$ root of unity,
we choose the following action on the {\bf 4} of $SU(4)$:
\begin{equation}
    r(g^n) = \diag(\zeta^{2n}, \zeta^{3n}, \zeta^{5n}, \zeta^{8n}).
\end{equation}
This yields the following action on the {\bf 6} of $SO(6)$:
\begin{equation} \label{(2,3,-4)Bos}
    R(g^n) =\diag(\zeta^{8n},\zeta^{7n},\zeta^{5n},\zeta^{-8n},\zeta^{-7n},\zeta^{-5n}).
\end{equation}
The resulting quivers are given in figure \ref{(2,3,-4)Quivs}.

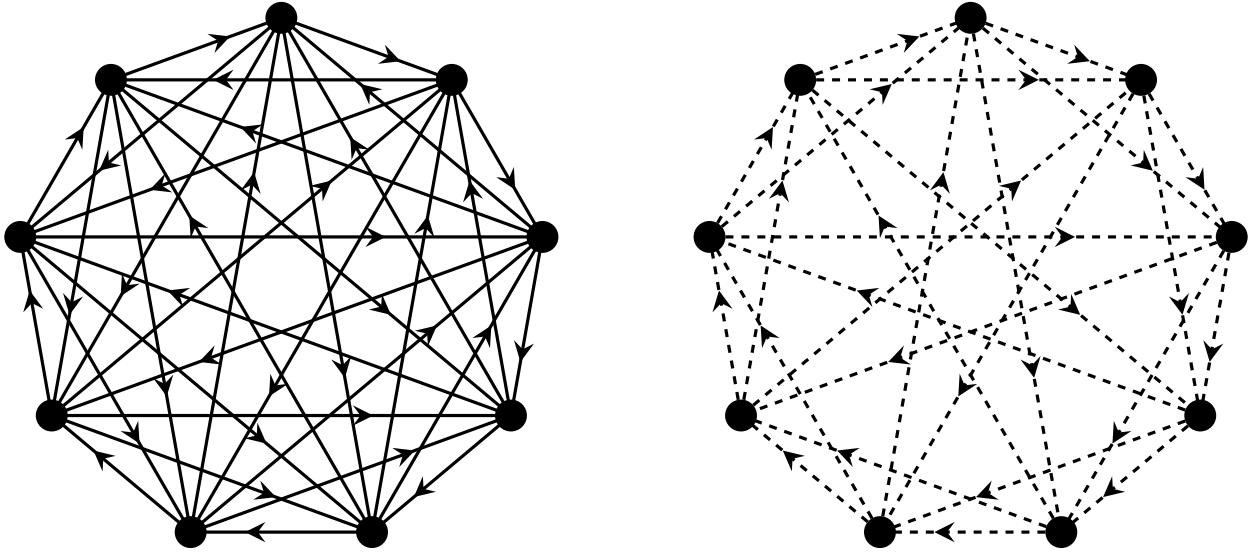
\begin{figure}[]
\usetikzlibrary{arrows}
\centering
\begin{tikzpicture}
\node[draw=none,minimum size=7cm,regular polygon,regular polygon sides=9] (a) {};

\foreach \x in {1,2,...,9}
  \fill (a.corner \x) circle[radius=6pt, fill = none];

\begin{scope}[very thick,decoration={
    markings,
    mark=at position 0.7 with {\arrow[scale = 1.5, >=stealth]{>}}}
    ]
\draw[postaction = {decorate}] (a.corner 1) to (a.corner 3);
\draw[postaction = {decorate}] (a.corner 2) to (a.corner 4);
\draw[postaction = {decorate}] (a.corner 3) to (a.corner 5);
\draw[postaction = {decorate}] (a.corner 4) to (a.corner 6);
\draw[postaction = {decorate}] (a.corner 5) to (a.corner 7);
\draw[postaction = {decorate}] (a.corner 6) to (a.corner 8);
\draw[postaction = {decorate}] (a.corner 7) to (a.corner 9);
\draw[postaction = {decorate}] (a.corner 8) to (a.corner 1);
\draw[postaction = {decorate}] (a.corner 9) to (a.corner 2);
\draw[postaction = {decorate}] (a.corner 1) to (a.corner 4);
\draw[postaction = {decorate}] (a.corner 2) to (a.corner 5);
\draw[postaction = {decorate}] (a.corner 3) to (a.corner 6);
\draw[postaction = {decorate}] (a.corner 4) to (a.corner 7);
\draw[postaction = {decorate}] (a.corner 5) to (a.corner 8);
\draw[postaction = {decorate}] (a.corner 6) to (a.corner 9);
\draw[postaction = {decorate}] (a.corner 7) to (a.corner 1);
\draw[postaction = {decorate}] (a.corner 8) to (a.corner 2);
\draw[postaction = {decorate}] (a.corner 9) to (a.corner 3);
\draw[postaction = {decorate}] (a.corner 1) to (a.corner 6);
\draw[postaction = {decorate}] (a.corner 2) to (a.corner 7);
\draw[postaction = {decorate}] (a.corner 3) to (a.corner 8);
\draw[postaction = {decorate}] (a.corner 4) to (a.corner 9);
\draw[postaction = {decorate}] (a.corner 5) to (a.corner 1);
\draw[postaction = {decorate}] (a.corner 6) to (a.corner 2);
\draw[postaction = {decorate}] (a.corner 7) to (a.corner 3);
\draw[postaction = {decorate}] (a.corner 8) to (a.corner 4);
\draw[postaction = {decorate}] (a.corner 9) to (a.corner 5);
\draw[postaction = {decorate}] (a.corner 1) to (a.corner 9);
\draw[postaction = {decorate}] (a.corner 2) to (a.corner 1);
\draw[postaction = {decorate}] (a.corner 3) to (a.corner 2);
\draw[postaction = {decorate}] (a.corner 4) to (a.corner 3);
\draw[postaction = {decorate}] (a.corner 5) to (a.corner 4);
\draw[postaction = {decorate}] (a.corner 6) to (a.corner 5);
\draw[postaction = {decorate}] (a.corner 7) to (a.corner 6);
\draw[postaction = {decorate}] (a.corner 8) to (a.corner 7);
\draw[postaction = {decorate}] (a.corner 9) to (a.corner 8);
\end{scope}

\end{tikzpicture} \qquad \qquad
\begin{tikzpicture}
\node[draw=none,minimum size=7cm,regular polygon,regular polygon sides=9] (a) {};

\foreach \x in {1,2,...,9}
  \fill (a.corner \x) circle[radius=6pt, fill = none];

\begin{scope}[very thick,decoration={
    markings,
    mark=at position 0.7 with {\arrow[scale = 1.5, >=stealth]{>}}}
    ]
\draw[postaction = {decorate}][dashed] (a.corner 1) to (a.corner 6);
\draw[postaction = {decorate}][dashed] (a.corner 2) to (a.corner 7);
\draw[postaction = {decorate}][dashed] (a.corner 3) to (a.corner 8);
\draw[postaction = {decorate}][dashed] (a.corner 4) to (a.corner 9);
\draw[postaction = {decorate}][dashed] (a.corner 5) to (a.corner 1);
\draw[postaction = {decorate}][dashed] (a.corner 6) to (a.corner 2);
\draw[postaction = {decorate}][dashed] (a.corner 7) to (a.corner 3);
\draw[postaction = {decorate}][dashed] (a.corner 8) to (a.corner 4);
\draw[postaction = {decorate}][dashed] (a.corner 9) to (a.corner 5);
%
\draw[postaction = {decorate}][dashed] (a.corner 1) to (a.corner 8);
\draw[postaction = {decorate}][dashed] (a.corner 2) to (a.corner 9);
\draw[postaction = {decorate}][dashed] (a.corner 3) to (a.corner 1);
\draw[postaction = {decorate}][dashed] (a.corner 4) to (a.corner 2);
\draw[postaction = {decorate}][dashed] (a.corner 5) to (a.corner 3);
\draw[postaction = {decorate}][dashed] (a.corner 6) to (a.corner 4);
\draw[postaction = {decorate}][dashed] (a.corner 7) to (a.corner 5);
\draw[postaction = {decorate}][dashed] (a.corner 8) to (a.corner 6);
\draw[postaction = {decorate}][dashed] (a.corner 9) to (a.corner 7);
%
\draw[postaction = {decorate}][dashed] (a.corner 1) to (a.corner 9);
\draw[postaction = {decorate}][dashed] (a.corner 2) to (a.corner 1);
\draw[postaction = {decorate}][dashed] (a.corner 3) to (a.corner 2);
\draw[postaction = {decorate}][dashed] (a.corner 4) to (a.corner 3);
\draw[postaction = {decorate}][dashed] (a.corner 5) to (a.corner 4);
\draw[postaction = {decorate}][dashed] (a.corner 6) to (a.corner 5);
\draw[postaction = {decorate}][dashed] (a.corner 7) to (a.corner 6);
\draw[postaction = {decorate}][dashed] (a.corner 8) to (a.corner 7);
\draw[postaction = {decorate}][dashed] (a.corner 9) to (a.corner 8);
\end{scope}

\end{tikzpicture} \caption{\label{(2,3,-4)Quivs} Left/Right: Fermionic/Bosonic quiver  $\R^6/\Z_9^{(2,3,-4,-1)}$.}

\end{figure}

We are interested in computing the defect group for the 4D theory. Hence, we need only consider the fermionic quiver and its corresponding adjacency matrix. Taking the Smith normal form, we find that the defect group is given by
\begin{equation} \label{(2,3,-4)DefG}
   \mathcal{A}^{(1)}_{\mathrm{elec}} \oplus \mathcal{A}^{(1)}_{\mathrm{mag}} \cong    \text{Tor}(\text{Coker}(K)) \cong \Z_9\oplus \Z_9
\end{equation}

The geometry accounts for this result. Indeed, consider the action of $\Gamma = \Z_9$ on $S^5$ induced from the bosonic action in \eqref{(2,3,-4)Bos}. We see that this action is fixed point free. Through an application of Armstrong's theorem, we then find that the bosonic data contributes a factor to the defect group in \eqref{(2,3,-4)DefG} given by
\begin{equation}
        H_1(S^5/\Gamma) \cong \text{Ab}[\pi_1(\Gamma/H)] \cong  \Z_9
\end{equation}

Consider next tachyon condensation. In the holomorphic presentation of the geometry this is given by
$\mathbb{C}^3/\Z_9$ where the action of $\Z_9$ is specified by the weights $(1,2,-5)_{\mathrm{hol}}$.
There is one relevant tachyon in the $(c_1, c_2, c_3)$ ring that survives the chiral GSO projection, $T_1$.
While there are GSO-preserved tachyons in the other rings, the most relevant tachyon
is $T_1$ in the $(c_1 , c_2 , c_3)$ ring.

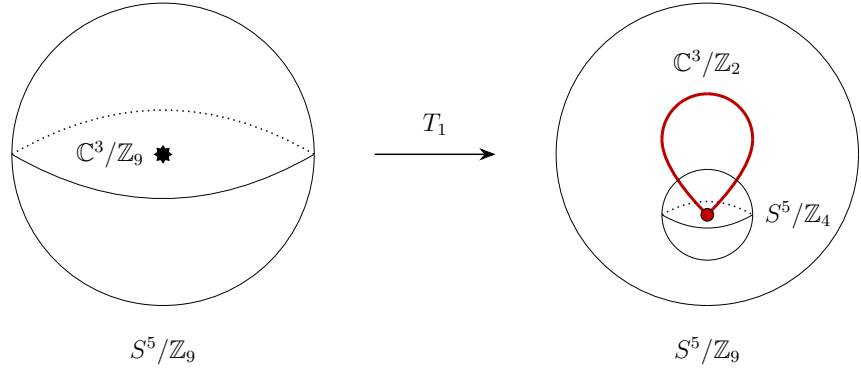
\begin{figure}
    \centering
    \scalebox{0.8}{	
    \begin{tikzpicture} \begin{pgfonlayer}{nodelayer}
		\node [style=Star] (3) at (-4, -1) {};
		\node [style=none] (4) at (-0.5, -1) {};
		\node [style=none] (5) at (1.5, -1) {};
		\node [style=none] (9) at (0.5, -0.5) {$T_1$};
		\node [style=none] (10) at (-4, -4.25) {$S^5/\Z_9$};
		\node [style=none] (12) at (-4.875, -1) {$\mathbb{C}^3/\Z_9$};
		\node [style=none] (26) at (6.5, -2) {$S^5/\Z_4$};
		\node [style=none] (27) at (7.5, -1) {};
		\node [style=none] (28) at (2.5, -1) {};
		\node [style=none] (29) at (5, 1.5) {};
		\node [style=none] (30) at (5, -3.5) {};
		\node [style=none] (31) at (5, -4.25) {$S^5/\Z_9$};
		\node [style=none] (32) at (5, 0) {};
		\node [style=none] (33) at (5, -2) {};
		\node [style=none] (34) at (4.25, -0.75) {};
		\node [style=none] (35) at (5.75, -0.75) {};
		\node [style=none] (44) at (-1.5, -1) {};
		\node [style=none] (45) at (-6.5, -1) {};
		\node [style=none] (46) at (-4, 1.5) {};
		\node [style=none] (47) at (-4, -3.5) {};
		\node [style=none] (48) at (5.75, -2) {};
		\node [style=none] (49) at (4.25, -2) {};
		\node [style=none] (50) at (5, -1.25) {};
		\node [style=none] (51) at (5, -2.75) {};
		\node [style=SmallCircleRed] (52) at (5, -2) {};
		\node [style=none] (53) at (5, 0.5) {$\mathbb{C}^3/\Z_2$};
	\end{pgfonlayer}
	\begin{pgfonlayer}{edgelayer}
		\draw [style=ArrowLineRight] (4.center) to (5.center);
		\draw [style=none, bend left=45] (28.center) to (29.center);
		\draw [style=none, bend left=45] (29.center) to (27.center);
		\draw [style=none, bend left=45] (27.center) to (30.center);
		\draw [style=none, bend left=45] (30.center) to (28.center);
		\draw [style=RedLine, in=-180, out=90] (34.center) to (32.center);
		\draw [style=RedLine, in=90, out=0] (32.center) to (35.center);
		\draw [style=RedLine, in=45, out=-90, looseness=0.75] (35.center) to (33.center);
		\draw [style=RedLine, in=-90, out=135, looseness=0.75] (33.center) to (34.center);
		\draw [bend left=45] (45.center) to (46.center);
		\draw [bend left=45] (46.center) to (44.center);
		\draw [bend left=45] (44.center) to (47.center);
		\draw [bend right=315] (47.center) to (45.center);
		\draw [bend right] (45.center) to (44.center);
		\draw [style=DottedLine, bend left] (45.center) to (44.center);
		\draw [bend left=45] (49.center) to (50.center);
		\draw [bend left=45] (50.center) to (48.center);
		\draw [bend left=45] (48.center) to (51.center);
		\draw [bend right=315] (51.center) to (49.center);
		\draw [bend right] (49.center) to (48.center);
		\draw [style=DottedLine, bend left] (49.center) to (48.center);
	\end{pgfonlayer}
\end{tikzpicture}}
    \caption{We sketch the geometry $\R^6/\Z_9^{(2,3,-4,-1)}$ after the blow up. Here $\mathbb{C}^3/\Z_2=\mathbb{C}\times \mathbb{C}^2/\Z_2$ and the $\mathbb{C}^2/\Z_2$ singularity is along all of $\mathbb{C}$ which compactifies into a teardrop (red) and enhances at infinity (red dot).}
    \label{fig:ExampleZ9Sketch}
\end{figure}

The orbifold $\mathbb{C}^3/\Z_9$ with weights $(1,2,-5)_{\mathrm{hol}}$ is a toric variety with fan generated by the vertices
\begin{equation}
    \alpha_1 = ((9,-2,5)), \;\; \alpha_2 = ((0,1,0)), \;\; \alpha_3 = ((0,0,1)).
\end{equation}
The tachyon corresponds to the lattice vector $T_1 = ((1, 0, 1))$.  Condensation of $T_1$ gives the residual subcones $C[T_1,\alpha_1,\alpha_2]$, $ C[T_1,\alpha_2,\alpha_3]$, and $C[T_1, \alpha_3,\alpha_1]$, which correspond to $\mathbb{C}^3/\Z_4$ with weights $(-1,1,2)_{\mathrm{hol}}$, $\mathbb{C}^3$, i.e., flat space, and $\mathbb{C}^3/\Z_2$ with weights $(1,0,-1)_{\mathrm{hol}}$, respectively. Notice that the orbifolds $\mathbb{C}^3/\Z_4$ and $\mathbb{C}^3/\Z_2$ exhibit a non-isolated singularity. In both cases, there is a non-compact curve supporting a $\mathbb{C}^2/\Z_2$ singularity. Gluing patches these loci compactify to a teardrops worth of $\mathbb{C}^2/\Z_2$ singularities (see figure \ref{fig:ExampleZ9Sketch}).

After a suitable change in complex structure, we observe that $\mathbb{C}^3/\Z_2$ with weights $(1,0,-1)$ and $\mathbb{C}^3/\Z_4$ with weights $(-1,1,2)$ are in fact supersymmetric backgrounds. We conclude then that the endpoint of the most relevant tachyon sequence in this Type II theory includes flat and supersymmetric spaces, for which the latter have singularities that are resolved by generic metric blowup modes.

Consider next the quivers associated with the two subcones, $C[T_1,\alpha_1, \alpha_2]$ and $C[T_1,\alpha_1,\alpha_3]$.
In our usual notation
\begin{equation}
    C[T_1, \alpha_1,\alpha_2] \sim \R^6/\Z_4^{(-2,0,1,1)}, \qquad  C[T_1, \alpha_1,\alpha_3] \sim \R^6/\Z_2^{(1,0,-1,0)}
\end{equation}
We summarize the quivers in figure \ref{(-1,1,2)Quivs}, and we remark that the fermionic and bosonic quivers are identical in these cases due to supersymmetry.

\begin{figure}[]
\usetikzlibrary{arrows}
\centering
\begin{tikzpicture}
\node[draw=none,minimum size=6cm,regular polygon,regular polygon sides=4] (a) {};

\foreach \x in {1,2,...,9}
  \fill (a.corner \x) circle[radius=6pt, fill = none];

\begin{scope}[very thick,decoration={
    markings,
    mark=at position 0.5 with {\arrow[scale = 1.5, >=stealth]{>}}}
    ]
\draw[postaction = {decorate}][bend right] (a.corner 1) to (a.corner 3);
\draw[postaction = {decorate}][bend right] (a.corner 2) to (a.corner 4);
\draw[postaction = {decorate}][bend right] (a.corner 3) to (a.corner 1);
\draw[postaction = {decorate}][bend right] (a.corner 4) to (a.corner 2);
\end{scope}
\begin{scope}[very thick,decoration={
    markings,
    mark=at position 0.5 with {\arrow[scale = 1.5, >=stealth]{>}}}
    ]
\draw[postaction = {decorate}] (a.corner 1) to (a.corner 4);
\draw[postaction = {decorate}] (a.corner 2) to (a.corner 1);
\draw[postaction = {decorate}] (a.corner 3) to (a.corner 2);
\draw[postaction = {decorate}] (a.corner 4) to (a.corner 3);
\end{scope}

\end{tikzpicture} \qquad
\begin{tikzpicture}

\node[draw = none] (0) at  (-2,1){};
\node[draw = none] (1) at (2,1){};

\foreach \x in {0,1}
  \fill (\x) circle[radius=6pt, fill = none];

\begin{scope}[very thick,decoration={
    markings,
    mark=at position 0.5 with {\arrow[scale = 1.5, >=stealth]{>>}}}
    ]
\draw[postaction = {decorate}][bend right] (0) to (1);
\draw[postaction = {decorate}][bend right] (1) to (0);
\end{scope}

\end{tikzpicture}
\caption{\label{(-1,1,2)Quivs} Left: Quiver for $\R^6/\Z_4^{(-2,0,1,1)}$. Right: Quiver for $\R^6/\Z_2^{(1,0,-1,0)}$.}
\end{figure}
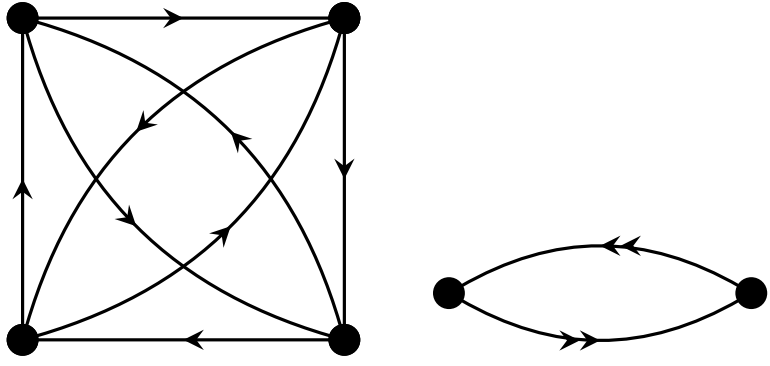

We now determine the defect group for the cones $C[T_1,\alpha_1, \alpha_2]$ and $C[T_1,\alpha_1,\alpha_3]$. Taking the Smith normal form, we find that the defect group for $C[T_1,\alpha_1,\alpha_2]$ is $\mathbb{Z}_2 \oplus \mathbb{Z}_2$, and that of $C[T_1,\alpha_1,\alpha_3]$ is trivial, in accord with geometric expectations.

In this case we also have a codimension 4 singularity which generates a ``flavor brane'' locus associated with an $A_1$ singularity which is locally of the form $\mathbb{C}^2 / \mathbb{Z}_2$. Much as in \cite{Cvetic:2022imb, DelZotto:2022joo}, we find that in an electric polarization
there is an $SO(3)$ flavor symmetry which combines with the $1$-form symmetry to generate a 2-group via the non-split long exact sequence:
\begin{equation}
0 \rightarrow \Z_2 \rightarrow \Z_4 \rightarrow SU(2) \rightarrow SO(3) \rightarrow 1.
\end{equation}

Summarizing, we see that at late times, the tachyon condensation generates 
an emergent $2$-group involving a $\mathbb{Z}_2$ 1-form symmetry and an $SO(3)$ flavor symmetry.

\subsection{Codim. $6$ and $4$: $\R^6/\Z_9^{(3,5,-6,-2)}$}

In this subsection, we consider the orbifold $\R^6/\Z_9^{(3,5,-6,-2)}$. We first determine the defect group.
Let the action on the \textbf{4} of $SU(4)$ be given by
\begin{equation}
    r(g^n) = \diag(\zeta^{3n},\zeta^{5n},\zeta^{3n},\zeta^{7n})
\end{equation}
 where $g$ is the generator of $\Z_9$ and $\zeta$ is a primitive $9^{th}$ root of unity. This determines the action in the \textbf{6} of $SO(6)$:
 \begin{equation}\label{(3,5,-6,-2)BosAc}
     R(g^n) = \diag(\zeta^{8n},\zeta^{6n},\zeta^{8n},\zeta^{-8n},\zeta^{-6n},\zeta^{-8n})
 \end{equation}
The fermionic and bosonic quivers are given in figure \ref{(3,5,-6,-2)Quivs}.

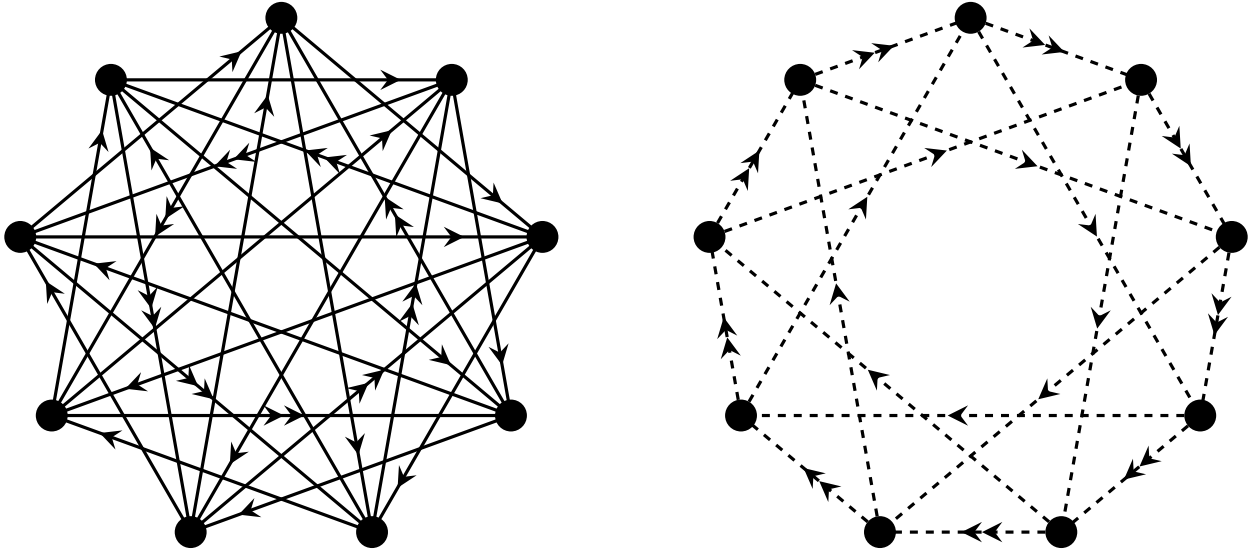
\begin{figure}[]
\usetikzlibrary{arrows}
\centering
\begin{tikzpicture}
\node[draw=none,minimum size=7cm,regular polygon,regular polygon sides=9] (a) {};

\foreach \x in {1,2,...,9}
  \fill (a.corner \x) circle[radius=6pt, fill = none];

\begin{scope}[very thick,decoration={
    markings,
    mark=at position 0.55 with {\arrow[scale = 1.5, >=stealth]{>>}}}
    ]
\draw[postaction = {decorate}] (a.corner 1) to (a.corner 4);
\draw[postaction = {decorate}] (a.corner 2) to (a.corner 5);
\draw[postaction = {decorate}] (a.corner 3) to (a.corner 6);
\draw[postaction = {decorate}] (a.corner 4) to (a.corner 7);
\draw[postaction = {decorate}] (a.corner 5) to (a.corner 8);
\draw[postaction = {decorate}] (a.corner 6) to (a.corner 9);
\draw[postaction = {decorate}] (a.corner 7) to (a.corner 1);
\draw[postaction = {decorate}] (a.corner 8) to (a.corner 2);
\draw[postaction = {decorate}] (a.corner 9) to (a.corner 3);
\end{scope}
\begin{scope}[very thick,decoration={
    markings,
    mark=at position 0.85 with {\arrow[scale = 1.5, >=stealth]{>}}}
    ]
\draw[postaction = {decorate}] (a.corner 1) to (a.corner 6);
\draw[postaction = {decorate}] (a.corner 2) to (a.corner 7);
\draw[postaction = {decorate}] (a.corner 3) to (a.corner 8);
\draw[postaction = {decorate}] (a.corner 4) to (a.corner 9);
\draw[postaction = {decorate}] (a.corner 5) to (a.corner 1);
\draw[postaction = {decorate}] (a.corner 6) to (a.corner 2);
\draw[postaction = {decorate}] (a.corner 7) to (a.corner 3);
\draw[postaction = {decorate}] (a.corner 8) to (a.corner 4);
\draw[postaction = {decorate}] (a.corner 9) to (a.corner 5);
%
\draw[postaction = {decorate}] (a.corner 1) to (a.corner 8);
\draw[postaction = {decorate}] (a.corner 2) to (a.corner 9);
\draw[postaction = {decorate}] (a.corner 3) to (a.corner 1);
\draw[postaction = {decorate}] (a.corner 4) to (a.corner 2);
\draw[postaction = {decorate}] (a.corner 5) to (a.corner 3);
\draw[postaction = {decorate}] (a.corner 6) to (a.corner 4);
\draw[postaction = {decorate}] (a.corner 7) to (a.corner 5);
\draw[postaction = {decorate}] (a.corner 8) to (a.corner 6);
\draw[postaction = {decorate}] (a.corner 9) to (a.corner 7);
\end{scope}

\end{tikzpicture} \qquad \qquad
\begin{tikzpicture}
\node[draw=none,minimum size=7cm,regular polygon,regular polygon sides=9] (a) {};

\foreach \x in {1,2,...,9}
  \fill (a.corner \x) circle[radius=6pt, fill = none];

\begin{scope}[very thick,decoration={
    markings,
    mark=at position 0.55 with {\arrow[scale = 1.5, >=stealth]{>}}}
    ]
\draw[postaction = {decorate}][dashed] (a.corner 1) to (a.corner 7);
\draw[postaction = {decorate}][dashed] (a.corner 2) to (a.corner 8);
\draw[postaction = {decorate}][dashed] (a.corner 3) to (a.corner 9);
\draw[postaction = {decorate}][dashed] (a.corner 4) to (a.corner 1);
\draw[postaction = {decorate}][dashed] (a.corner 5) to (a.corner 2);
\draw[postaction = {decorate}][dashed] (a.corner 6) to (a.corner 3);
\draw[postaction = {decorate}][dashed] (a.corner 7) to (a.corner 4);
\draw[postaction = {decorate}][dashed] (a.corner 8) to (a.corner 5);
\draw[postaction = {decorate}][dashed] (a.corner 9) to (a.corner 6);
\end{scope}
\begin{scope}[very thick,decoration={
    markings,
    mark=at position 0.55 with {\arrow[scale = 1.5, >=stealth]{>>}}}
    ]
\draw[postaction = {decorate}][dashed] (a.corner 1) to (a.corner 9);
\draw[postaction = {decorate}][dashed] (a.corner 2) to (a.corner 1);
\draw[postaction = {decorate}][dashed] (a.corner 3) to (a.corner 2);
\draw[postaction = {decorate}][dashed] (a.corner 4) to (a.corner 3);
\draw[postaction = {decorate}][dashed] (a.corner 5) to (a.corner 4);
\draw[postaction = {decorate}][dashed] (a.corner 6) to (a.corner 5);
\draw[postaction = {decorate}][dashed] (a.corner 7) to (a.corner 6);
\draw[postaction = {decorate}][dashed] (a.corner 8) to (a.corner 7);
\draw[postaction = {decorate}][dashed] (a.corner 9) to (a.corner 8);
\end{scope}

\end{tikzpicture} \caption{\label{(3,5,-6,-2)Quivs} Left/Right: Fermionic/Bosonic quiver for $\R^6/\Z_9^{(3,5,-6,-2)}$.}
\end{figure}
Computing the Smith normal form of the fermionic adjacency matrix for the fermionic quiver in figure \ref{(3,5,-6,-2)Quivs} yields the defect group of the 4D theory:
\begin{equation}
    \mathcal{A}^{(1)}_{\mathrm{elec}} \oplus \mathcal{A}^{(1)}_{\mathrm{mag}} \cong    \text{Tor}(\text{Coker}(K)) \cong\Z_3\oplus \Z_3
\end{equation}
The geometry accounts for this result. Indeed, the action of $\Gamma = \Z_9$ on $S^5$ induced from the bosonic action in \eqref{(3,5,-6,-2)BosAc} has fixed points. By Armstrong's theorem, we find
\begin{equation}
    H_1(S^5/\Gamma) \cong \textnormal{Ab}[\pi_1(\Gamma/H)]\cong \Z_3
\end{equation}

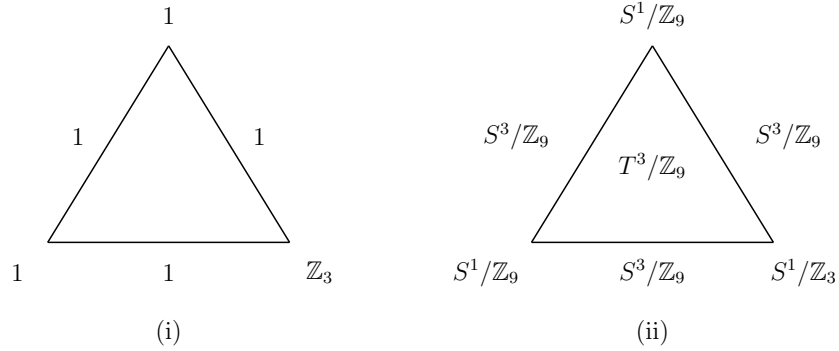
\begin{figure}
    \centering
    \scalebox{0.8}{\begin{tikzpicture}
	\begin{pgfonlayer}{nodelayer}
		\node [style=none] (4) at (-6, -1) {};
		\node [style=none] (5) at (-4, 2.25) {};
		\node [style=none] (6) at (-2, -1) {};
		\node [style=none] (7) at (-4, -2.5) {(i)};
		\node [style=none] (11) at (2, -1) {};
		\node [style=none] (12) at (4, 2.25) {};
		\node [style=none] (13) at (6, -1) {};
		\node [style=none] (14) at (4, -2.5) {(ii)};
		\node [style=none] (18) at (-4, 2.75) {$1$};
		\node [style=none] (19) at (-6.5, -1.5) {$1$};
		\node [style=none] (20) at (-1.5, -1.5) {$\Z_3$};
		\node [style=none] (21) at (-2.5, 0.75) {};
		\node [style=none] (22) at (-5.5, 0.75) {$1$};
		\node [style=none] (23) at (-4, -1.5) {$1$};
        \node [style=none] (24) at (-2.5, 0.75) {$1$};
		\node [style=none] (25) at (4, 0.25) {$T^3/\Z_{9}$};
		\node [style=none] (26) at (4, 2.75) {$S^1/\Z_{9}$};
		\node [style=none] (27) at (1.25, -1.5) {$S^1/\Z_{9}$};
		\node [style=none] (28) at (6.5, -1.5) {$S^1/\Z_{3}$};
		\node [style=none] (29) at (6.25, 0.75) {$S^3/\Z_{9}$};
		\node [style=none] (30) at (1.75, 0.75) {$S^3/\Z_{9}$};
		\node [style=none] (31) at (4, -1.5) {$S^3/\Z_{9}$};
	\end{pgfonlayer}
	\begin{pgfonlayer}{edgelayer}
		\draw [style=ThickLine] (4.center) to (6.center);
		\draw [style=ThickLine] (6.center) to (5.center);
		\draw [style=ThickLine] (5.center) to (4.center);
		\draw [style=ThickLine] (11.center) to (13.center);
		\draw [style=ThickLine] (13.center) to (12.center);
		\draw [style=ThickLine] (12.center) to (11.center);
	\end{pgfonlayer}
\end{tikzpicture}}
    \caption{In (i) we give the subgroups $H_k,H_{ij}$ and in (ii) singular loci for the quotient $S^5/\Gamma$ with $\Gamma=\Z_{9}$ and weights $(3,5,-6,-2)$.  }
    \label{fig:Labeleg4}
\end{figure}

A summary of the geometric data which contribute to the defect group follows from
the fibration $S^5/\Z_N \rightarrow \Delta$ as in figure \ref{fig:Labelling}.
See figure \ref{fig:Labeleg4} for a summary of the salient features of the toric geometry.

We now move on to study tachyon condensation in our orbifold.
We follow the procedure presented in \cite{Morrison:2004fr}.
In their notation, our orbifold, $\R^6/\Z_9^{(3,5,-6,-2)}$,
is given by $\mathbb{C}^3/\Z_9$, where the action of $\Z_9$ is defined by the weights $(1, 3, -8)_{\mathrm{hol}}$.
There is one relevant tachyon in the $(c_1,c_2,c_3)$ ring that survives the chiral GSO projection, $T_1$. We need not worry about the tachyons in the other chiral rings since $T_1$ in the $(c_1,c_2,c_3)$ ring is the most relevant.

We remark here that the singularity for $\mathbb{C}^3/\Z_9$ with weights $(1,3,-8)_{\mathrm{hol}}$ is a non-isolated singularity;
there is a codimension 6 singularity at the tip of the cone as well as a codimension 4 singularity which stretches out to the boundary $S^5 / \mathbb{Z}_9$. That being said, the singularity ``at infinity'' does not contribute any additional tachyons.

The orbifold $\mathbb{C}^3/\Z_9$ with weights $(1,3,-8)_{\mathrm{hol}}$
is a toric variety with fan spanned by the vertices:
\begin{equation}
    \alpha_1 = ((9,-3,8)),\qquad \alpha_2 = ((0,1,0)),\qquad \alpha_3 = ((0,0,1)).
\end{equation}
The tachyon $T_1$ corresponds to the lattice point $T_1 = ((1,0,1))$. Condensation of $T_1$ gives the residual subcones $C[T_1,\alpha_1,\alpha_2]$, $C[T_1,\alpha_1,\alpha_3]$, and $C[T_1,\alpha_2,\alpha_3]$, which correspond to smooth space, $\mathbb{C}^3/\Z_3$ with weights $(1,0,1)$, and smooth space, respectively. Observe that the
orbifold $\mathbb{C}^3/\Z_3$ with weights $(1,0,1)$ is supersymmetric, in a different complex structure (complex conjugate the $Z_3$ coordinate), and can also be presented as
$\mathbb{C} \times \mathbb{C}^2  / \mathbb{Z}_3$, in the obvious notation. This singularity realizes a 6D
$\mathfrak{su}(3)$ super Yang-Mills theory. The overall polarization, i.e., global form of the gauge group is independent of the other boundary data in the model.

\subsection{Codim. $6$ (multiple tachyons): $\R^6/\Z_{23}^{(4,7,-8,-3)}$}

In this section, we consider the orbifold $\R^6/\Z_{23}^{(4,7,-8,-3)}$. We begin by determining its 4D defect group. Letting $g$ denote the generator of $\Z_{23}$ and $\zeta$ a $23^{rd}$ root of unity, the action on the \textbf{4} of $SU(4)$ is induced via:
\begin{equation}
    r(g^n) = \diag(\zeta^{4n},\zeta^{7n},\zeta^{15n},\zeta^{20n})
\end{equation}
which determines the action on the \textbf{6} of $SO(6)$:
\begin{equation}
    R(g^n) = \diag(\zeta^{22n},\zeta^{19n},\zeta^{11n}, \zeta^{-22n},\zeta^{-19n},\zeta^{-11n})
\end{equation}

Extracting the quiver for a probe D0-brane, we extract the defect group via the adjacency matrix for the fermionic degrees of freedom:
\begin{equation}
    \mathbb{D}^{(1)} = \mathrm{Tor}(\mathrm{Coker} K) = \mathcal{A}^{(1)}_{\mathrm{elec}} \oplus \mathcal{A}^{(1)}_{\mathrm{mag}} \cong \Z_{23}\oplus \Z_{23}.
\end{equation}
The boundary geometry $S^5 / \mathbb{Z}_{23}$ has no singularities (the group acts freely on $S^5$)
and so $H_{1}(S^5 / \mathbb{Z}_{23}) \cong \mathbb{Z}_{23}$. Thus, the quiver based method and geometry based method predict the same
defect group.

We now turn to tachyon condensation. In this case, there is a sequence of tachyon condensations which we track in stages.
In the notation of \cite{Morrison:2004fr}, the orbifold we have been considering, $\R^6/\Z_{23}^{(4,7,-8,-3)}$, is given by $\mathbb{C}^3/\Z_{23}$, where $\Z_{23}$ acts with weights $(1,4,-11)_{\mathrm{hol}}$. The $(c_1,c_2,c_3)$ ring tachyons $T_1$, $T_2$, $T_8$, with R-charges $R_1 = \frac{17}{23}$, $R_2 = \frac{11}{23}$, and $R_8 = \frac{21}{23}$, respectively, survive the chiral GSO projection. Although there are GSO-preserved tachyons in the other rings, the most relevant tachyon is $T_2$.

The orbifold $\mathbb{C}^3/\Z_{23}$ with weights $(1,4,-11)_{\mathrm{hol}}$ is a toric variety with toric fan spanned by:
\begin{equation}
    \alpha_1 = ((23,-4,11)), \quad \alpha_2 = ((0,1,0)),\quad \alpha_3 = ((0,0,1)).
\end{equation}
The tachyons correspond to lattice vectors in the toric diagram: $T_1 = ((1, 0, 1))$, $T_2 = ((2, 0, 1))$, and $T_8 = ((8, -1, 4))$.
We blowup by the order of relevance, i.e. first $T_2$, then $T_1$, and finally $T_8$. Condensation of $T_2$ gives the subcones $C[T_2,\alpha_1,\alpha_2]$, $C[T_2, \alpha_2,\alpha_3]$, and $C[T_2, \alpha_1,\alpha_3]$, which correspond to $\mathbb{C}^3$, i.e.,
 smooth space; $\mathbb{C}^3/\Z_2$ with weights $(1, 0, -1)_{\mathrm{hol}}$; and $\mathbb{C}^3/\Z_8$ with weights $(1, 1, 2)_{\mathrm{hol}}$, respectively. Notice that there are non-isolated singularities. Furthermore, we note that in our usual notation
\begin{equation}
    C[T_2,\alpha_2,\alpha_3] \sim \R^6/\Z_2^{(1,0,-1,0)}\,, \qquad C[T_2,\alpha_1,\alpha_3] \sim \R^6/\Z_8^{(-1,-1,0,2)}
\end{equation}
We can determine the defect group of these cones by finding their quivers.
 An interesting part of this example comes from considering the defect group for $C[T_2,\alpha_1,\alpha_3]$. From the quiver, we find that the defect group is given by
\begin{equation}
\mathcal{A}^{(1)}_{\mathrm{elec}} \oplus \mathcal{A}^{(1)}_{\mathrm{mag}} \cong \Z_4 \oplus \Z_4
\end{equation}
Again, in an electric frame, we find a $2$-group symmetry characterized by the long exact sequence:
\begin{equation}
    0 \rightarrow \Z_4 \rightarrow \Z_8 \rightarrow SU(2) \rightarrow SO(3) \rightarrow 1,
\end{equation}
as captured in the geometry. This essentially follows from the following short exact sequence (detected in geometry via Mayer-Vietoris) being
non-split (see \cite{Cvetic:2022imb, DelZotto:2022joo}:
\begin{equation}
    1 \rightarrow \Z_4 \rightarrow \Z_8 \rightarrow \Z_2 \rightarrow 1 \,.
\end{equation}

The other tachyons are $T_1$ and $T_8$, but after $T_2$ condenses, it turns out that any remaining instabilities are absent. As explained in \cite{Morrison:2004fr}, the R-charges of $T_1$ and $T_8$ now shift so that $T_1$ is marginal and $T_8$ is irrelevant. Since $T_1$ is marginal we have actually landed on a geometry with no instability. We can, of course, still blowup by $T_1$. Further subdividing our fan by $T_1$ results in the subcones $C[T_1, \alpha_1,\alpha_3]$ and $C[T_1, T_2, \alpha_1]$, which correspond to the orbifolds $\mathbb{C}^3 / \Z_4$ with weights $(1, 0, 1)_{\mathrm{hol}}$ and $\mathbb{C}^3 / \Z_4$ with weights $(0, 1, 1)_{\mathrm{hol}}$, respectively. Both of these are actually supersymmetric, but in a different complex structure (complex conjugate the $Z_3$ coordinate).

\subsection{Codim. $6$ and $2$: $\R^6/\Z_9^{(-4,-2,1,5)}$}

We now turn to an example in which there are tachyons initially present at the boundary.
The geometry we consider supports a codimension 2 singularity in addition to the codimension 6 singularity at the tip of the cone.
As such, the we expect the defect group to be somewhat more intricate, as summarized by equation (\ref{eq:CoolFormula}).
We begin by studying the defect group before tachyon condensation.
The action on the \textbf{4} of $SU(4)$ is given by:
\begin{equation}
    r(g^n) = \diag(\zeta^{5n},\zeta^{7n},\zeta^{n},\zeta^{5n})
\end{equation}
where here we let $g$ denote the generator of $\Z_9$ and $\zeta$ a primitive $9^{th}$ root of unity.
This determines the action in the \textbf{6} of $SO(6)$:
\begin{equation}\label{(-4,-2,1)BosAc}
    R(g^n) = \diag (\zeta^{8n},\zeta^{6n},\zeta^{3n},\zeta^{-8n},\zeta^{-6n},\zeta^{-3n})
\end{equation}
The resultant quivers are summarized in figure \ref{(-4,-2,1)Quivs}.

\begin{figure}[]
\usetikzlibrary{arrows}
\centering
\begin{tikzpicture}
\node[draw=none,minimum size=7cm,regular polygon,regular polygon sides=9] (a) {};

\foreach \x in {1,2,...,9}
  \fill (a.corner \x) circle[radius=6pt, fill = none];

\begin{scope}[very thick,decoration={
    markings,
    mark=at position 0.8 with {\arrow[scale = 1.5, >=stealth]{>>}}}
    ]
\draw[postaction = {decorate}] (a.corner 1) to (a.corner 6);
\draw[postaction = {decorate}] (a.corner 2) to (a.corner 7);
\draw[postaction = {decorate}] (a.corner 3) to (a.corner 8);
\draw[postaction = {decorate}] (a.corner 4) to (a.corner 9);
\draw[postaction = {decorate}] (a.corner 5) to (a.corner 1);
\draw[postaction = {decorate}] (a.corner 6) to (a.corner 2);
\draw[postaction = {decorate}] (a.corner 7) to (a.corner 3);
\draw[postaction = {decorate}] (a.corner 8) to (a.corner 4);
\draw[postaction = {decorate}] (a.corner 9) to (a.corner 5);
\end{scope}
\begin{scope}[very thick,decoration={
    markings,
    mark=at position 0.65 with {\arrow[scale = 1.5, >=stealth]{>}}}
    ]
\draw[postaction = {decorate}] (a.corner 1) to (a.corner 2);
\draw[postaction = {decorate}] (a.corner 2) to (a.corner 3);
\draw[postaction = {decorate}] (a.corner 3) to (a.corner 4);
\draw[postaction = {decorate}] (a.corner 4) to (a.corner 5);
\draw[postaction = {decorate}] (a.corner 5) to (a.corner 6);
\draw[postaction = {decorate}] (a.corner 6) to (a.corner 7);
\draw[postaction = {decorate}] (a.corner 7) to (a.corner 8);
\draw[postaction = {decorate}] (a.corner 8) to (a.corner 9);
\draw[postaction = {decorate}] (a.corner 9) to (a.corner 1);
%
\draw[postaction = {decorate}] (a.corner 1) to (a.corner 8);
\draw[postaction = {decorate}] (a.corner 2) to (a.corner 9);
\draw[postaction = {decorate}] (a.corner 3) to (a.corner 1);
\draw[postaction = {decorate}] (a.corner 4) to (a.corner 2);
\draw[postaction = {decorate}] (a.corner 5) to (a.corner 3);
\draw[postaction = {decorate}] (a.corner 6) to (a.corner 4);
\draw[postaction = {decorate}] (a.corner 7) to (a.corner 5);
\draw[postaction = {decorate}] (a.corner 8) to (a.corner 6);
\draw[postaction = {decorate}] (a.corner 9) to (a.corner 7);
\end{scope}

\end{tikzpicture} \qquad \qquad
\begin{tikzpicture}
\node[draw=none,minimum size=7cm,regular polygon,regular polygon sides=9] (a) {};

\foreach \x in {1,2,...,9}
  \fill (a.corner \x) circle[radius=6pt, fill = none];

\begin{scope}[very thick,decoration={
    markings,
    mark=at position 0.7 with {\arrow[scale = 1.5, >=stealth]{>}}}
    ]
\draw[postaction = {decorate}][dashed][bend left] (a.corner 1) to (a.corner 4);
\draw[postaction = {decorate}][dashed][bend left] (a.corner 2) to (a.corner 5);
\draw[postaction = {decorate}][dashed][bend left] (a.corner 3) to (a.corner 6);
\draw[postaction = {decorate}][dashed][bend left] (a.corner 4) to (a.corner 7);
\draw[postaction = {decorate}][dashed][bend left] (a.corner 5) to (a.corner 8);
\draw[postaction = {decorate}][dashed][bend left] (a.corner 6) to (a.corner 9);
\draw[postaction = {decorate}][dashed][bend left] (a.corner 7) to (a.corner 1);
\draw[postaction = {decorate}][dashed][bend left] (a.corner 8) to (a.corner 2);
\draw[postaction = {decorate}][dashed][bend left] (a.corner 9) to (a.corner 3);
%
\draw[postaction = {decorate}][dashed] (a.corner 1) to (a.corner 9);
\draw[postaction = {decorate}][dashed] (a.corner 2) to (a.corner 1);
\draw[postaction = {decorate}][dashed] (a.corner 3) to (a.corner 2);
\draw[postaction = {decorate}][dashed] (a.corner 4) to (a.corner 3);
\draw[postaction = {decorate}][dashed] (a.corner 5) to (a.corner 4);
\draw[postaction = {decorate}][dashed] (a.corner 6) to (a.corner 5);
\draw[postaction = {decorate}][dashed] (a.corner 7) to (a.corner 6);
\draw[postaction = {decorate}][dashed] (a.corner 8) to (a.corner 7);
\draw[postaction = {decorate}][dashed] (a.corner 9) to (a.corner 8);
%
\draw[postaction = {decorate}][dashed] (a.corner 1) to (a.corner 7);
\draw[postaction = {decorate}][dashed] (a.corner 2) to (a.corner 8);
\draw[postaction = {decorate}][dashed] (a.corner 3) to (a.corner 9);
\draw[postaction = {decorate}][dashed] (a.corner 4) to (a.corner 1);
\draw[postaction = {decorate}][dashed] (a.corner 5) to (a.corner 2);
\draw[postaction = {decorate}][dashed] (a.corner 6) to (a.corner 3);
\draw[postaction = {decorate}][dashed] (a.corner 7) to (a.corner 4);
\draw[postaction = {decorate}][dashed] (a.corner 8) to (a.corner 5);
\draw[postaction = {decorate}][dashed] (a.corner 9) to (a.corner 6);
\end{scope}

\end{tikzpicture} \caption{\label{(-4,-2,1)Quivs} Left\,/\,Right: Fermionic\,/\,Bosonic quiver for $\R^6/\Z_9^{(-4,-2,1,5)}$.}
\end{figure}
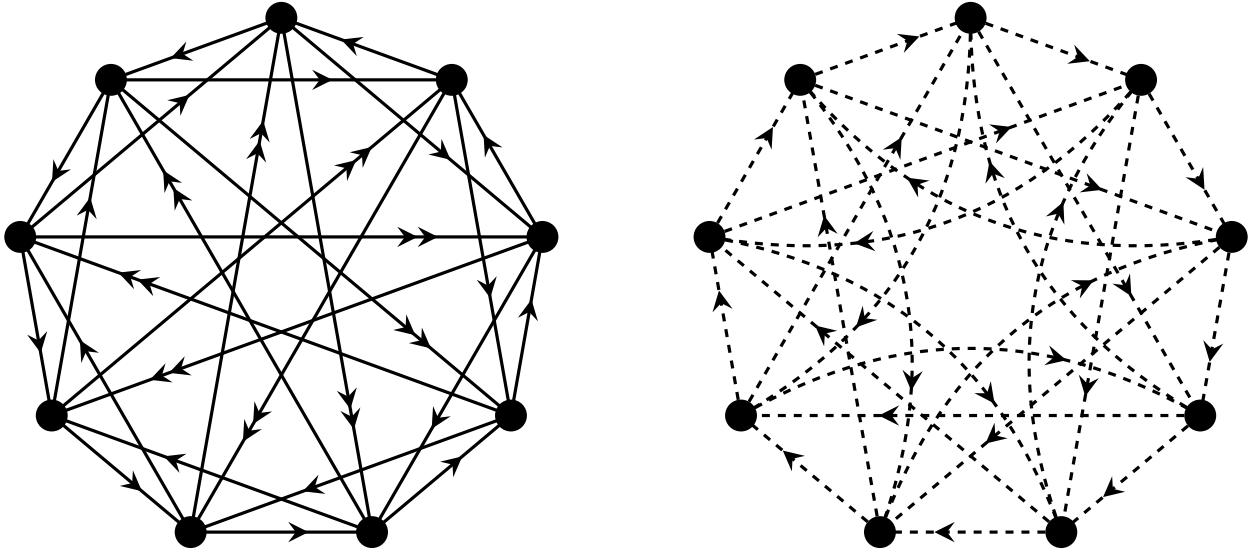

\begin{figure}
    \centering
    \scalebox{0.8}{\begin{tikzpicture}
	\begin{pgfonlayer}{nodelayer}
		\node [style=none] (4) at (-6, -1) {};
		\node [style=none] (5) at (-4, 2.25) {};
		\node [style=none] (6) at (-2, -1) {};
		\node [style=none] (7) at (-4, -2.5) {(i)};
		\node [style=none] (11) at (2, -1) {};
		\node [style=none] (12) at (4, 2.25) {};
		\node [style=none] (13) at (6, -1) {};
		\node [style=none] (14) at (4, -2.5) {(ii)};
		\node [style=none] (18) at (-4, 2.75) {$1$};
		\node [style=none] (19) at (-6.5, -1.5) {$\Z_3$};
		\node [style=none] (20) at (-1.5, -1.5) {$\Z_3$};
		\node [style=none] (21) at (-2.5, 0.75) {};
		\node [style=none] (22) at (-5.5, 0.75) {$1$};
		\node [style=none] (23) at (-4, -1.5) {$\Z_3$};
        \node [style=none] (24) at (-2.5, 0.75) {$1$};
		\node [style=none] (25) at (4, 0.25) {$T^3/\Z_{9}$};
		\node [style=none] (26) at (4, 2.75) {$S^1/\Z_{9}$};
		\node [style=none] (27) at (1.25, -1.5) {$S^1/\Z_{3}$};
		\node [style=none] (28) at (6.5, -1.5) {$S^1/\Z_{3}$};
		\node [style=none] (29) at (6.25, 0.75) {$S^3/\Z_{9}$};
		\node [style=none] (30) at (1.75, 0.75) {$S^3/\Z_{9}$};
		\node [style=none] (31) at (4, -1.5) {$S^3/\Z_{3}$};
	\end{pgfonlayer}
	\begin{pgfonlayer}{edgelayer}
		\draw [style=ThickLine] (4.center) to (6.center);
		\draw [style=ThickLine] (6.center) to (5.center);
		\draw [style=ThickLine] (5.center) to (4.center);
		\draw [style=ThickLine] (11.center) to (13.center);
		\draw [style=ThickLine] (13.center) to (12.center);
		\draw [style=ThickLine] (12.center) to (11.center);
	\end{pgfonlayer}
\end{tikzpicture}}
    \caption{We sketch in (i) the subgroups $H_k,H_{ij}$ and in (ii) singular loci for the quotient $S^5/\Gamma$ with $\Gamma=\Z_{9}$ and weights $(-4,-2,1,5)$.  }
    \label{fig:Labeleg5}
\end{figure}
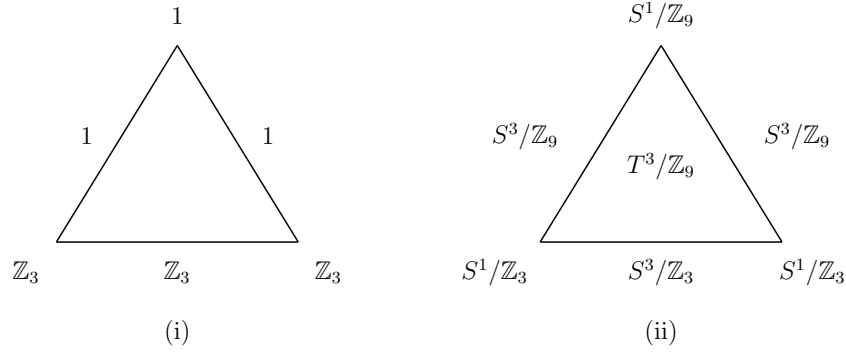

The defect group for the 4D theory is how this example distinguishes itself from those we have previously considered. Taking the Smith normal form of the fermionic adjacency matrix, we find that
\begin{equation}\label{eq:Z3fourtimes}
    \mathcal{A}^{(1)}_{\mathrm{elec}} \oplus \mathcal{A}^{(1)}_{\mathrm{mag}} \cong \textnormal{Tor}(\text{Coker}(K)) \cong \Z_3 \oplus \Z_3 \oplus \Z_3 \oplus \Z_3
\end{equation}
Distinguished from our previous examples, this is only partially predicted by the geometry. Indeed, the bosonic action from \eqref{(-4,-2,1)BosAc} induces an action on $S^5$ that has fixed points. Letting $\Gamma = \Z_9$, we find from Armstrong's theorem that
\begin{equation}
    H_1(S^5/\Gamma) \cong \text{Ab}[\pi_1(\Gamma/H)] \cong \Z_9/\Z_3 \cong  \Z_3\,,
\end{equation}
where $H$ denotes the normal subgroup which sweeps out a fixed point locus on the boundary.
A summary of the geometric contributions to the defect group is given in figure \ref{fig:Labeleg5}.
The remaining contributions to line (\ref{eq:Z3fourtimes}) follow from an application of line (\ref{eq:CoolFormula}).

We now study the tachyon condensation of our orbifold. In the notation of \cite{Morrison:2004fr}, our orbifold, $\R^6/\Z_9^{(-4,-2,1,5)}$ is given by $\mathbb{C}^3/\Z_9$ where $\Z_9$ acts in accordance to the weights $(1, 3,6)_{\mathrm{hol}}$.
Observe that in the holomorphic basis where the $\mathbb{Z}_9$ group acts via:
\begin{equation}
(Z_1,Z_2,Z_3) \mapsto (\omega Z_1 , \omega^3 Z_2 , \omega^6 Z_3),
\end{equation}
the $\mathbb{Z}_3 \subset \mathbb{Z}_9$ subgroup generated by $\omega^3$ leaves invariant the entire $X = 0$ locus, namely we have a codimension
$2$ singularity. The model has one tachyon in the $(c_1,c_2,c_3)$ ring that survives the chiral GSO projection, $T_3$.
We remark further that $T_3$ in the $(c_1, c_2, c_3)$ ring is the most relevant tachyon.

Condensation of $T_3$ is treated through toric geometry. The orbifold $\mathbb{C}^3/\Z_9$ with weights $(1,3,6)_{\mathrm{hol}}$
is a toric variety whose fan is described by the vertices
\begin{equation}
    \alpha_1 = ((9, -3,-6)),\qquad \alpha_2 = ((0,1,0)),\qquad \alpha_3 = ((0,0,1))
\end{equation}
The tachyon $T_3$ corresponds to the lattice vector $T_3 = ((3,-1,-2))$.
However, notice that $3T_3 = \alpha_1$, implying that blowing up by $T_3$ does not change the geometry of our orbifold. Since $T_3$ was the only relevant tachyon the $(c_1,c_2,c_3)$ ring, we conclude that $\mathbb{C}^3/\Z_9$ with weights $(1,3,6)_{\mathrm{hol}}$
resolves to a smooth space via non-chiral blowup modes.

\section{IIA Example: Non-Abelian Orbifolds} \label{sec:NONAB}

In this section we provide a few examples of how to extract the defect group
in the case of orbifold singularities:%
\begin{equation}
X=\mathbb{R}^{6}/\Gamma
\end{equation}
in which the group $\Gamma$ is a non-abelian finite subgroup
of $SU(4)$.\footnote{See \cite{Hanany:1999sp} for a more complete list of finite
subgroups of $SU(4)$.} Our aim here is not to be exhaustive, but rather to
just give a few illustrative examples of the general structure we expect to
find.

One way to generate examples is to first begin with a finite subgroup of
$SU(n)\subset SU(4)$ for $n=2,3$ and to then \textquotedblleft
twist\textquotedblright\ this embedding by an additional rephasing symmetry
which commutes with the original group action, i.e., a ``stacking and twisting'' procedure (see e.g., \cite{Berenstein:2000mb} for a general approach).
This provides a way to generate many examples of non-supersymmetric orbifolds where we can then read off the associated defect group. We illustrate
this procedure for the ADE series of finite subgroups of $SU(2) \subset SU(4)$.\footnote{Of course the 
A-type subgroups are abelian, but the D- and E-type subgroups are non-abelian.} See Appendix \ref{app:SU2stack} for another example.

It is worth noting that while these choices of finite subgroup are ``somewhat special'', the bigger the subgroup of $SU(4)$,
the more we expect defects to be screened by dynamical states. This is in accord with the fact that as the group grows, there is a bigger normal subgroup $H $ associated with fixed loci in the boundary geometry $S^5 / \Gamma$, so $\Gamma / H$ is consequently a smaller subgroup (often trivial). This same phenomenon was observed in reference \cite{DelZotto:2022fnw}.

\subsection{Stacking and Twisting Quivers with $SU(2)\subset SU(4)$}

To begin, consider $\Gamma_{SU(2)}\subset SU(2)$ a finite subgroup which acts
on the basis $(e_{1},e_{2};f_{1};f_{2})$ of the $\mathbf{4}$ as follows:%
\begin{equation}
\Gamma_{SU(2)}\text{ action:\ }e_{i}\mapsto g_{ij}e_{j}\text{ \ \ and
\ \ }f_{j}\mapsto f_{j}\text{.}%
\end{equation}
We can then perform a further \textquotedblleft twist\textquotedblright\ by a
$\mathbb{Z}_{N}$ via the action:%
\begin{equation}
\mathbb{Z}_{N}\text{ action:\ }e_{i}\mapsto\zeta e_{i}\text{ \ \ and
\ \ }f_{j}\mapsto\zeta^{-1}f_{j}\text{.}%
\end{equation}
The combined group action defines $\Gamma_{SU(4)}=\Gamma_{SU(2)}%
\times\mathbb{Z}_{N}$ a finite subgroup of $SU(4)$.

To extract the associated quiver, we exploit the fact that the two group
actions commute. Along these lines, we can first extract the quiver for the
case with just the $\Gamma_{SU(2)}$ quotient and then perform a further
quotient by the $\mathbb{Z}_{N}$ action.

With this in mind, denote by $Q$ the quiver for the probe theory obtained from
the finite subgroup of $SU(2)$. This comes with its own adjacency matrix for
bosons and fermions. We label the nodes in this quiver as $q$ and denote the
two adacency matrices as $A_{q,q^{\prime}}^{F}$ and $A_{q,q^{\prime}}^{B}$, in
the obvious notation. Since this is the adjacency matrix for a finite subgroup
of $SU(2)$, this adjacency matrix will be symmetric. In fact, the off-diagonal
entries match the adjacency matrix of the associated extended Dynkin diagram of the
corresponding ADE Lie group (via the McKay correspondence). The fermions and
bosons decompose into representations of $\mathfrak{su}(2)_{L}\times
\mathfrak{su}(2)_{R}$ as:%
\begin{align}
\mathfrak{su}(4) &  \supset\mathfrak{su}(2)_{L}\times\mathfrak{su}(2)_{R}\\
\mathbf{4} ~&  \mathbf{\rightarrow(2,1)\oplus(1,2)}\\ \label{eq:4decomp}
\mathbf{6} ~&  \mathbf{\rightarrow(2,2)\oplus(1,1)\oplus(1,1)}.
\end{align}
i.e., the fermionic singlets of $\mathfrak{su}(2)_{L}$ denote arrows which go
back to the same quiver node (i.e., the \textquotedblleft
gauginos\textquotedblright).\ We view the coordinate $q$ as a
\textquotedblleft horizontal coordinate\textquotedblright\ which moves one
within a given quiver.

Now, precisely because the $\mathbb{Z}_{N}$ group action commutes with
$\Gamma_{SU(2)}$, there is a natural group action on this quiver $Q$. In
particular, we can simply take all the nodes of the original quiver $Q$ and
copy the quiver nodes $N$ times. This copying procedure specifies for us the
basis of fractional branes and thus the collection of quiver nodes for
$\Gamma_{SU(4)}$. We refer to each such copy of quiver nodes as $\mathcal{Q}%
_{i}$ for $i=1,...,N$. There is both a \textquotedblleft horizontal
coordinate\textquotedblright\ labelling an irreducible representation of
$\Gamma_{SU(2)}$ as well as a \textquotedblleft vertical\textquotedblright%
\ coordinate labelling an irreducible representation of $%
\mathbb{Z}
_{N}$. We thus label a position in the collection of quiver nodes as $(q,i)$,
where $q$ ranges over irreducible representations of $\Gamma_{SU(2)}$ and $i$
ranges over the irreducible representations of $\mathbb{Z}_{N}$.

The connectivity of the quiver will, however, now be somewhat different. To
illustrate, observe that for the fermionic matter fields in the decomposition
$\mathbf{4\rightarrow(2,1)\oplus(1,2)}$ of line (\ref{eq:4decomp}), the elements of
the $\mathbf{(2,1)}$ will now be connected between adjacent $\mathcal{Q}_{i}$
and $\mathcal{Q}_{i+1}$ layers. Further, the elements of the $\mathbf{(1,2)}$
will now be connected between $\mathcal{Q}_{i}$ and $\mathcal{Q}_{i-1}$, with
multiplicity $2$. From this, we conclude that the fermionic adjacency matrix
is of the form:%
\begin{equation}
\mathbb{A}_{(qi);(q^{\prime}i^{\prime})}^{F}=A_{qq^{\prime}}^{F}%
\delta_{i,i^{\prime}+1}+2\delta_{qq^{\prime}}\delta_{i,i^{\prime}-1}\text{.}%
\end{equation}
The Dirac pairing is obtained by anti-symmetrizing on the indices:
\begin{align}
K_{(qi);(q^{\prime}i^{\prime})}  & =\mathbb{A}_{(qi);(q^{\prime
}i^{\prime})}^{F}-\mathbb{A}_{(q^{\prime}i^{\prime});(qi)}^{F}\\
& =\left(  A_{qq^{\prime}}^{F}-2\delta_{qq^{\prime}}\right)  (\delta
_{i,i^{\prime}+1}-\delta_{i,i^{\prime}-1})\\
& =C_{qq^{\prime}}^{\text{ADE}}(\delta_{i,i^{\prime}+1}-\delta_{i,i^{\prime
}-1}),
\end{align}
where we used the fact that $A_{qq'}^F$ is symmetric.
Here $C_{qq^{\prime}}^{\text{ADE}}$ denotes the adjacency matrix
of the corresponding extended Dynkin diagram with added entries
of $-2$ on the diagonal. Equivalently, it is the intersection pairing on the lattice of
two cycles of $\mathbb{C}^{2}/\Gamma_{SU(2)}$ with the affine node associated with the $B$-field adjoined (see e.g., \cite{Douglas:1996sw}).

From this, we extract the corresponding defect groups:%
\begin{align}
\mathbb{D}^{(1)}  & =
\left(
\mathbb{Z}
_{2}\oplus%
\mathbb{Z}
_{2}\right)  ^{N-1}
\text{\ \ \ \ \ \ \ \ \ \ \ \ \ \ \ \ \ for }D_{2M}\text{ series}
\\
\mathbb{D}^{(1)}  & =
\left(
\mathbb{Z}
_{L}\right)  ^{N-1}
\text{
\ \ \ \ \ \ \ \ \ \ \ \ \ \ \ \ \ \ \ \ \ \ \ otherwise},
\end{align}
where here $L = \vert Z(G_{\mathrm{ADE}}) \vert$, the order of the center for the corresponding ADE Lie group.

Let us discuss these results from a complementary geometric perspective. For this we first require the induced action on the ${\mathbf{6}}$ of $SO(6)$. As an action on $\mathbb{C}^3$ we find the action to decompose as $\mathbb{C}^2/\Gamma_{SU(2)}\times \mathbb{C}/\Z_N$. This orbifold has a single codimension 2 and codimension 4 locus. The codimension 4 locus has fixed points at infinity and does not contribute to the defect group by Armstrong's theorem. The codimension 2 singularity is supported on $S^3/\Gamma_{SU(2)}$ and following our counting, as for example in the discussion following \eqref{eq:CoolFormula}, we find exactly $N-1$ contribution of $\Z_L$ (or $\Z_2\times \Z_2$) to the defect group. Again, $N$ is odd by the requirement that the bosonic action be a faithful action of $\Z_N$.

Finally, let us turn to the structure of tachyon condensation in this model. Observe that the group action by $\Gamma_{SU(2)}$ on the geometry is, by itself, supersymmetric. In particular, it results in the quotient $\mathbb{R}^6 / \Gamma_{SU(2)} = \mathbb{R}^2 \times \mathbb{C}^2 / \Gamma_{SU(2)} $, in the obvious notation. Additionally, the group action of the $\mathbb{Z}_N$ factor acts only on the $\mathbb{R}^2$ factor of the target space. As such, the full quotient assumes the form:
\begin{equation}
\mathbb{R}^6 / \Gamma_{SU(4)} = (\mathbb{R}^2 / \mathbb{Z}_{N}) \times (\mathbb{C}^2 / \Gamma_{SU(2)}),
\end{equation}
in the obvious notation. As such, all of the tachyons are associated with the codimension loci originating from the $\mathbb{R}^2 / \mathbb{Z}_N$ factor. We can then simply borrow the analysis presented in reference \cite{Adams:2001sv}.

\section{IIB Case} \label{sec:IIB}

We now turn to the case of type IIB backgrounds $\mathbb{R}^{3,1} \times \mathbb{R}^{6} / \Gamma^{\mathbf{s}}$ with $N$ spacetime filling D3-branes. This system has been considered from various perspectives in \cite{Kachru:1998ys, Lawrence:1998ja, Hanany:1998sd, Adams:2001jb, Dymarsky:2005uh, Dymarsky:2005nc, Horowitz:2007pr, Pomoni:2009joh}. One can take a suitable limit to decouple the tachyonic closed string modes, leaving us with a 4D gauge theory defined by just the open string sector. In the large $N$ limit, the contribution to the gauge coupling beta functions is inherited from that of $\mathcal{N} = 4$ Super Yang-Mills theory, so the gauge coupling does not run at one-loop order.
Nevertheless, there always exists a non-vanishing beta function for a double trace operator which is not suppressed, even in the large $N$ limit \cite{Dymarsky:2005uh, Dymarsky:2005nc}. In particular, reference \cite{Pomoni:2009joh} found that regardless of whether the group action leads to an isolated singularity or instead has non-isolated singularities, there is a radiatively induced breaking pattern in which some scalar operators condense. This suggests a natural picture in which the time-dependence of the tachyonic IIA solution is now realized via non-trivial scale dependence in the IIB brane probe setting. In particular, we can simply track the quiver before and after various blowups, much as we did in our time-dependent analysis of the IIA case.\footnote{In fact, the time dependence is still there, it is just more implicit. Observe that in a radiatively generated potential, expanding around a local maximum naturally includes a tachyonic instability for the fields of the QFT degrees of freedom. The rolling motion of the fields to the nearby local minimum is time dependent and is simply how the time dependence is reflected in the QFT.}

In terms of the basis of fractional D-branes there is little change from the case of a probe D0-brane. Indeed, the procedure to produce a quiver gauge theory is identical (and in fact we derived the D0-brane worldvolume theory via dimensional reduction of the D3-brane case). On the other hand, the presence of S-duality, where we interchange F1- / D1-strings and NS5- / D5-branes means that the analysis of fractional branes will also have some limitations. To bypass some of these subtleties, in what follows we focus on the portion of the defect group and symmetries detected by $H_{\ast}(S^5 / \Gamma)$, i.e., the ``homology portion''. This leads to no loss of generality provided the tachyons are initially sequestered from the boundary geometry $S^5 / \Gamma$.

In fact, to keep the analysis streamlined we shall make the somewhat stronger assumption that all singularities present in the geometry are codimension 6, both before and after a tachyon condensation has taken place. We comment as appropriate how some features extend to cases where we have a locally supersymmetric codimension 4 singularity, but defer the most general case to future work.

Now, starting from a candidate defect of the IIA theory on $\mathbb{R}_{t} \times T^3$, we can T-dualize all three spatial directions. In the resulting IIB theory, then, we find that a line operator wrapped on a spatial cycle will turn into a surface operator (two-dimensional support). So, all of the candidate electric / magnetic 1-form symmetries of the IIA case will now become 2-form symmetries.\footnote{It is important in this argument that all of the fractional branes are associated with D-branes / boundary states of the worldsheet theory.} In more detail, we get surface operator defects from D3-branes wrapped on $\mathrm{Cone}(\gamma_1)$ for $\gamma_{1} \in H_{1}(S^5 / \Gamma)$, and D5-branes wrapped on $\mathrm{Cone}(\gamma_{3})$ for $\gamma_{3} \in H_{3}(S^5 / \Gamma)$. Observe also that we can also wrap NS5-branes on $\mathrm{Cone}(\gamma_{3})$ to produce another set of surface operators. There are corresponding 2-form symmetry operators which link with these objects which we can make explicit when  $\Gamma$ acts without fixed points on $S^5$. For example, the heavy defect D3-branes on $\mathrm{Cone}(\gamma_1)$ link with the symmetry operator D3-branes wrapped on elements of $H_{3}(S^5 / \Gamma)$. Likewise, we can wrap a D1-string (resp. F1-string) on a linking element of $H_1(S^5 / \Gamma)$ to get the symmetry operator for the D5-brane (resp. NS5-brane) wrapped on $\mathrm{Cone}(\gamma_3)$.

Much as in \cite{Heckman:2022xgu} we also find a collection of 0-form topological symmetry operators (codimension one in the 4D spacetime) which exhibit non-trivial braiding / fusion rules. These operators are obtained from D3-branes wrapped on cycles of $H_1(S^5 / \Gamma) = H_{D3}$ as well as D5-branes and NS5-branes wrapped on cycles of $H_{3}(S^5 / \Gamma) \equiv H_{D5} \equiv H_{NS5}$. The same braiding relations observed in reference \cite{Gukov:1998kn, Heckman:2022xgu} indicate that the full 0-form symmetry generated by these operators is a semi-direct product:
\begin{equation}
\label{eq:0FormSym}
G = (H_{D3} \times H_{D5})\rtimes  H_{NS5}=(H_{D3} \times H_{NS5})\rtimes  H_{D5}\,,
\end{equation}
where $H_{D3}$ commutes with $H_{NS5},H_{D5}$ and $H_{NS5},H_{D5}$ do not commute with each other. For $\Gamma\cong \Z_N$ acting fixed point free on $S^5$ we have
\be
G=(\Z_N\times \Z_N)\rtimes \Z_N
\ee
which is the unitriangular matrix group UT$(3,N)$. In the presence of codimension 4 fixed points, and characterizing the symmetry subgroups as the Pontryagin dual of defect subgroups, this becomes
\be\label{eq:Codim4Sym}
G=(\Gamma^\vee \times (\Gamma/H)^\vee)\rtimes (\Gamma/H)^\vee\,.
\ee

In addition to these zero-form symmetry operators, we also observe the presence of another codimension one topological interface by wrapping a 7-brane with constant axio-dilaton on the boundary $S^5 / \Gamma$. In the case of SCFTs with tuned axio-dilaton on the D3-brane stack, this implements duality defects.\footnote{See references \cite{Choi:2021kmx, Kaidi:2021xfk, Choi:2022zal, Kaidi:2022uux, Kaidi:2022cpf, Bashmakov:2022jtl} for the case of $\mathcal{N} = 4$ SYM, and references \cite{Heckman:2022xgu, Bashmakov:2022uek, Damia:2023ses} for various extensions with less supersymmetry.} In the present case, we do not really flow to a conformal fixed point, but at least in the large $N$ planar limit, the coupling constant does not run at one-loop order. As such, we can still speak of an approximate duality interface as inherited from the $\mathcal{N} = 4$ SYM case.

A general comment here is that since we can realize symmetry operators via ``branes at infinity,'' there is a sharp sense in which
we never lose any generalized symmetries. That being said, because the local geometry near the tip of the cone will certainly resolve / change,
the degrees of freedom charged under this symmetry will reorganize / change. We interpret this as spontaneous symmetry breaking.
This is in accord with the picture of radiatively induced contributions developed in references \cite{Adams:2001jb, Dymarsky:2005uh, Dymarsky:2005nc, Pomoni:2009joh}.

\subsection{Scale Dependent Considerations}

\begin{figure}
\centering
\scalebox{0.6}{\begin{tikzpicture}
	\begin{pgfonlayer}{nodelayer}
		\node [style=none] (0) at (-9, 6) {};
		\node [style=none] (1) at (-9, -6) {};
		\node [style=none] (2) at (7, 0) {};
		\node [style=none] (3) at (-7.75, 5.97) {};
		\node [style=none] (4) at (-5.25, 5.83) {};
		\node [style=none] (5) at (-7.75, -5.97) {};
		\node [style=none] (6) at (-5.25, -5.83) {};
		\node [style=none] (7) at (-2.75, -5.57) {};
		\node [style=none] (8) at (-2.75, 5.57) {};
		\node [style=none] (9) at (-0.25, 5.14) {};
		\node [style=none] (10) at (2.25, 4.48) {};
		\node [style=none] (11) at (4.75, 3.35) {};
		\node [style=none] (12) at (4.75, -3.35) {};
		\node [style=none] (13) at (2.25, -4.48) {};
		\node [style=none] (14) at (-0.25, -5.14) {};
		\node [style=none] (19) at (-6.5, 0.5) {};
		\node [style=none] (20) at (-6.5, -0.5) {};
		\node [style=none] (21) at (-6.55, -0.75) {};
		\node [style=none] (22) at (-6.55, 0.75) {};
		\node [style=none] (23) at (3.5, 0.5) {};
		\node [style=none] (24) at (3.5, -0.5) {};
		\node [style=none] (25) at (3.45, -0.75) {};
		\node [style=none] (26) at (3.45, 0.75) {};
		\node [style=none] (27) at (-1.5, 0.5) {};
		\node [style=none] (28) at (-1.5, -0.5) {};
		\node [style=none] (29) at (-1.55, -0.75) {};
		\node [style=none] (30) at (-1.55, 0.75) {};
		\node [style=none] (31) at (-6.5, -6.5) {\large $X_{0,1}$};
		\node [style=none] (32) at (-1.5, -6) {\large $X_{n-2,n-1}$};
		\node [style=none] (33) at (3.5, -4.75) {\large $X_{n-1,n}$};
		\node [style=none] (34) at (8, 0) {\large$\mathbb{R}^6/\Gamma_n$};
		\node [style=none] (36) at (-7.75, 6.5) {\large$S^5/\Gamma_0$};
		\node [style=none] (37) at (-5.25, 6.375) {\large$S^5/\Gamma_1$};
		\node [style=none] (38) at (-2.75, 6.125) {\large $S^5/\Gamma_{n-2}$};
		\node [style=none] (39) at (-0.25, 5.75) {\large $S^5/\Gamma_{n-1}$};
		\node [style=none] (40) at (2.25, 5.15) {\large $S^5/\Gamma_{n-1}$};
		\node [style=none] (41) at (4.75, 4.125) {\large $S^5/\Gamma_{n}$};
		\node [style=SmallCircleRed] (42) at (7, 0) {};
		\node [style=none] (43) at (-9.75, 6) {};
		\node [style=none] (44) at (-9.75, -6) {};
		\node [style=none] (45) at (6.5, -16) {};
		\node [style=none] (46) at (6.5, -11) {};
		\node [style=none] (47) at (7.5, -9) {};
		\node [style=none] (48) at (7.5, -14) {};
		\node [style=none] (49) at (3, -16) {};
		\node [style=none] (50) at (3, -11) {};
		\node [style=none] (51) at (4, -9) {};
		\node [style=none] (52) at (4, -14) {};
		\node [style=none] (53) at (-4.5, -12.5) {};
		\node [style=none] (54) at (-3, -12.5) {};
		\node [style=none] (55) at (1, -12.5) {\large$\mathcal{S}_{n-2,n-1}$};
		\node [style=none] (56) at (-8.75, -12.5) {\large$\mathcal{S}_{\textnormal{UV}}$};
		\node [style=none] (57) at (-2, -16) {};
		\node [style=none] (58) at (-2, -11) {};
		\node [style=none] (59) at (-1, -9) {};
		\node [style=none] (60) at (-1, -14) {};
		\node [style=none] (61) at (-7, -16) {};
		\node [style=none] (62) at (-7, -11) {};
		\node [style=none] (63) at (-6, -9) {};
		\node [style=none] (64) at (-6, -14) {};
		\node [style=none] (65) at (-9.75, -11) {};
		\node [style=none] (66) at (-8.75, -9) {};
		\node [style=none] (67) at (-9.75, -16) {};
		\node [style=none] (68) at (-8.75, -14) {};
		\node [style=none] (69) at (-1.5, -9.625) {};
		\node [style=none] (70) at (-6.5, -9.625) {};
		\node [style=none] (71) at (3.5, -9.625) {};
		\node [style=none] (72) at (7, -9.625) {};
		\node [style=none] (73) at (5.25, -12.5) {\large$\mathcal{S}_{n-1,n}$};
		\node [style=none] (74) at (-1.5, -16.75) {\large$\mathcal{J}_{n-2,n-1}$};
		\node [style=none] (75) at (-6.5, -16.75) {\large$\mathcal{J}_{0,1}$};
		\node [style=none] (76) at (3.5, -16.75) {\large$\mathcal{J}_{n-1,n}$};
		\node [style=none] (77) at (8.5, -12.5) {\large$\mathcal{B}^{E_{n\leq I}}$};
		\node [style=none] (78) at (-9.5, 0) {\large$\mathbb{R}^6/\Gamma_0$};
		\node [style=none] (79) at (-7.75, -6.35) {};
		\node [style=none] (80) at (-5.25, -6.125) {};
		\node [style=none] (81) at (-2.75, -5.875) {};
		\node [style=none] (82) at (-0.25, -5.5) {};
		\node [style=none] (83) at (2.25, -4.75) {};
		\node [style=none] (84) at (4.75, -3.75) {};
		\node [style=none] (85) at (-2, -17.5) {};
		\node [style=none] (86) at (-7, -9) {};
		\node [style=none] (87) at (-8, -11) {};
		\node [style=none] (88) at (-8, -16) {};
		\node [style=none] (89) at (-7, -14) {};
		\node [style=none] (90) at (7, -1.125) {};
		\node [style=none] (91) at (-4.25, 0) {};
		\node [style=none] (92) at (-3.75, 0) {};
	\end{pgfonlayer}
	\begin{pgfonlayer}{edgelayer}
		\draw [style=ThickLine, in=90, out=0, looseness=0.75] (0.center) to (2.center);
		\draw [style=ThickLine, in=0, out=-90, looseness=0.75] (2.center) to (1.center);
		\draw [style=PurpleLine, bend right=15, looseness=0.50] (10.center) to (13.center);
		\draw [style=PurpleLine, bend right=15, looseness=0.50] (11.center) to (12.center);
		\draw [style=PurpleLine, bend right=15, looseness=0.50] (9.center) to (14.center);
		\draw [style=PurpleLine, bend right=15, looseness=0.50] (8.center) to (7.center);
		\draw [style=PurpleLine, bend right=15, looseness=0.50] (4.center) to (6.center);
		\draw [style=PurpleLine, bend right=15, looseness=0.50] (3.center) to (5.center);
		\draw [style=DashedPurpleLine, bend left=15, looseness=0.50] (3.center) to (5.center);
		\draw [style=DashedPurpleLine, bend left=15, looseness=0.50] (4.center) to (6.center);
		\draw [style=DashedPurpleLine, bend left=15, looseness=0.50] (8.center) to (7.center);
		\draw [style=DashedPurpleLine, bend left=15, looseness=0.50] (9.center) to (14.center);
		\draw [style=DashedPurpleLine, bend left=15, looseness=0.50] (10.center) to (13.center);
		\draw [style=DashedPurpleLine, bend left=15, looseness=0.50] (11.center) to (12.center);
		\draw [style=PurpleLine, bend right=15] (19.center) to (20.center);
		\draw [style=PurpleLine, bend left=15] (22.center) to (21.center);
		\draw [style=PurpleLine, bend right=15] (23.center) to (24.center);
		\draw [style=PurpleLine, bend left=15] (26.center) to (25.center);
		\draw [style=PurpleLine, bend right=15] (27.center) to (28.center);
		\draw [style=PurpleLine, bend left=15] (30.center) to (29.center);
		\draw [style=PurpleLine, in=175, out=0] (3.center) to (4.center);
		\draw [style=PurpleLine, in=168, out=-7] (8.center) to (9.center);
		\draw [style=PurpleLine, in=150, out=-20] (10.center) to (11.center);
		\draw [style=PurpleLine, in=210, out=20] (13.center) to (12.center);
		\draw [style=PurpleLine, in=7, out=-168] (14.center) to (7.center);
		\draw [style=PurpleLine, in=0, out=-175] (6.center) to (5.center);
		\draw [style=DottedLine] (43.center) to (0.center);
		\draw [style=DottedLine] (44.center) to (1.center);
		\draw [style=DottedLine] (53.center) to (54.center);
		\draw [style=RedLine] (45.center) to (48.center);
		\draw [style=RedLine] (48.center) to (47.center);
		\draw [style=RedLine] (47.center) to (46.center);
		\draw [style=RedLine] (46.center) to (45.center);
		\draw [style=PurpleLine] (50.center) to (49.center);
		\draw [style=PurpleLine] (49.center) to (52.center);
		\draw [style=PurpleLine] (52.center) to (51.center);
		\draw [style=PurpleLine] (51.center) to (50.center);
		\draw [style=PurpleLine] (58.center) to (57.center);
		\draw [style=PurpleLine] (57.center) to (60.center);
		\draw [style=PurpleLine] (60.center) to (59.center);
		\draw [style=PurpleLine] (59.center) to (58.center);
		\draw [style=PurpleLine] (62.center) to (61.center);
		\draw [style=PurpleLine] (61.center) to (64.center);
		\draw [style=PurpleLine] (64.center) to (63.center);
		\draw [style=PurpleLine] (63.center) to (62.center);
		\draw [style=DottedLine] (66.center) to (63.center);
		\draw [style=DottedLine] (65.center) to (62.center);
		\draw [style=DottedLine] (68.center) to (64.center);
		\draw [style=DottedLine] (67.center) to (61.center);
		\draw (63.center) to (47.center);
		\draw (62.center) to (46.center);
		\draw (61.center) to (45.center);
		\draw [style=DashedLine] (64.center) to (48.center);
		\draw [style=DottedPurple, in=90, out=-90] (79.center) to (70.center);
		\draw [style=DottedPurple, in=-90, out=90] (70.center) to (80.center);
		\draw [style=DottedPurple, in=90, out=-90] (81.center) to (69.center);
		\draw [style=DottedPurple, in=-90, out=90] (69.center) to (82.center);
		\draw [style=DottedPurple, in=90, out=-90] (83.center) to (71.center);
		\draw [style=DottedPurple, in=-90, out=90] (71.center) to (84.center);
		\draw [style=ThickLine] (87.center) to (62.center);
		\draw [style=ThickLine] (86.center) to (63.center);
		\draw [style=ThickLine] (88.center) to (61.center);
		\draw [style=DashedLine] (89.center) to (64.center);
		\draw [style=DottedLine] (68.center) to (89.center);
		\draw [style=DottedRed] (90.center) to (72.center);
		\draw [style=DottedLine] (91.center) to (92.center);
	\end{pgfonlayer}
\end{tikzpicture}}
\caption{We sketch the SymTree relevant when descending from the energy scale $E_{\textnormal{UV}}$ to the scale $E_{n\leq I}$. The figure shows an idealized situation where a codimension 6 singularity repeatedly resolves to a single codimension 6 singularity, with the geometry shedding cobordisms $X_{n,n+1}$ (purple). The relative QFT $\mathcal{B}^{E_{n\leq I}}$ is engineered by D3-branes probing the residual singularity $\mathbb{R}^6/\Gamma_n$. Every cobordism $X_{n,n+1}$ results in a SymTree junction $\mathcal{J}_{n,n+1}$ which matches the symmetry theories in neighboring slabs. These symmetry theories originate from cylinders with cross section $S^5/\Gamma_{n}$. At the energy scale $E_{n\leq I}$ with effective geometry $\mathbb{R}^6/\Gamma_n$ all geometric structure beyond $S^5/\Gamma_n$ is ``at infinity". All interfaces are stacked onto the boundary conditions at infinity, resulting in a single bulk symmetry theory $\mathcal{S}_{n-1,n}$.
}
\label{fig:SymTree}
\end{figure}
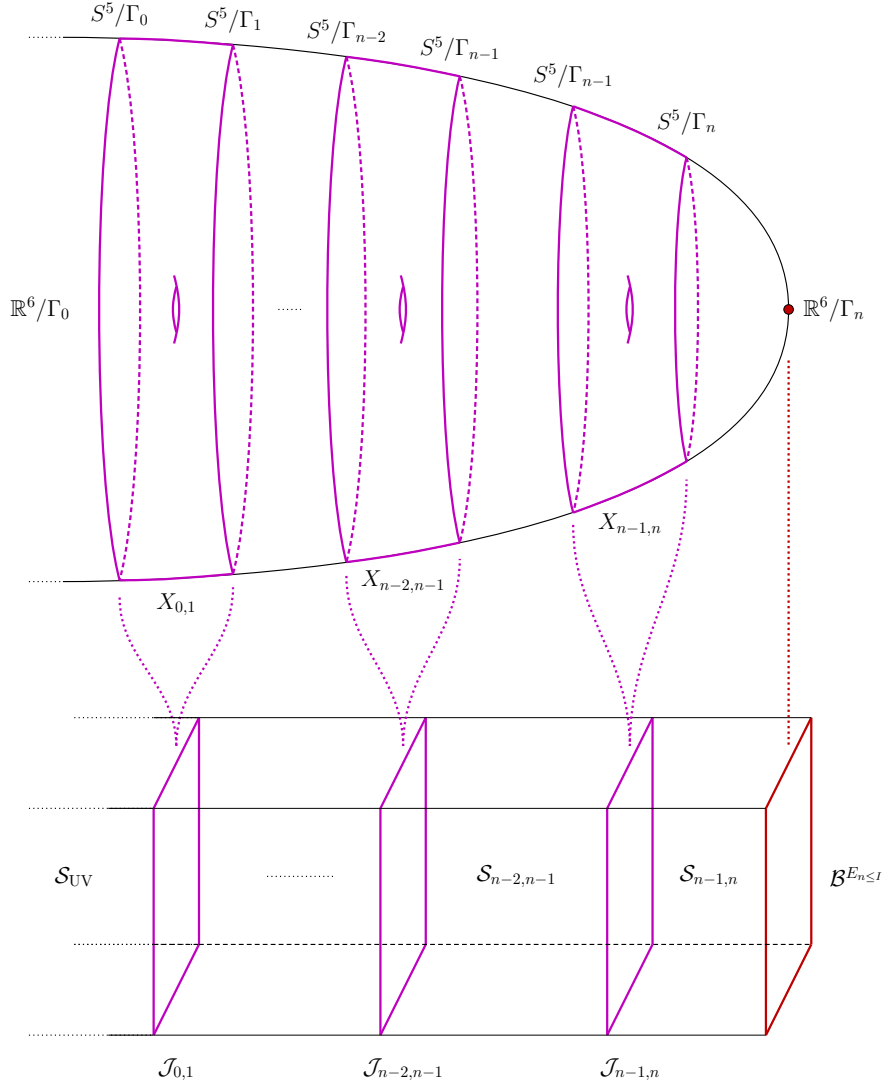

Let us now turn to the scale dependence of our system. As a proxy for this, we study the blowup moduli associated with tachyon condensation. This again gives us a sequence of quiver gauge theories (much as in the IIA case), but where now the evolution is in scale rather than time.

To track this, we can again introduce a bulk 5D system as given by a collection of SymTFTs.
These are partitioned up according to energy scales
\be
E_{\mathrm{UV}} = E_0 > E_1 >  ... > E_{I} = E_{\mathrm{IR}}\,,
\ee
in the obvious notation. In each partitioned region we have a SymTFT $\mathcal{S}_{n,n+1}$ (see figure \ref{fig:SymTree}).
There is a non-topological interface which in this case receives a contribution from the branes which have left the quiver (i.e., motion away from the singularity), and a contribution from the cobordism $X_{n,n+1}$ relating the boundary of the $n$-th and $(n+1)$-th local model. Even in the case where all branes remain in the system, the interface is non-topological, supporting abelian degrees of freedom resulting from supergravity reduced on $X_{n,n+1}$. In all cases, the role of the junction theory is simply to match the bulk modes on the two sides of the interface.

Of course, this setup is rather suggestive of holography, where we would identify the radial direction of $\mathrm{Cone}(\partial X)$ with a renormalization group scale. The non-supersymmetric case is somewhat subtle due to the possible presence of instabilities in the solution (see e.g.,
\cite{Horowitz:2007pr, Ooguri:2016pdq}) so we defer the existence of a possible holographic interpretation to future work.

\paragraph{0-form and 2-form Symmetries:} The cobordism $X_{n,n+1}$ characterizes the transition between the old and new asymptotic boundaries relevant in our description of the system at energies $E_n$ and $E_{n+1}$. In particular the 0-form and 2-form symmetries characterized by the asymptotic boundaries change, and the difference is precisely associated with degrees of freedom associated with $X_{n,n+1}$ decoupling. We now turn to discuss this process under the simplified assumption that the cobordism $X_{n,n+1}$ is smooth, with two connected boundary components, mapping between local model boundaries $S^5/\Gamma_n$ and $S^5/\Gamma_{n+1}$.

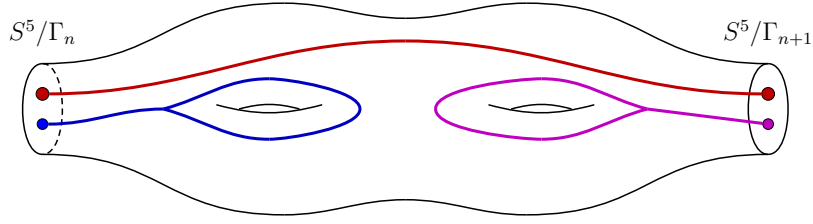
\begin{figure}
\centering
\scalebox{0.8}{
\begin{tikzpicture}
	\begin{pgfonlayer}{nodelayer}
		\node [style=none] (0) at (-6, 0.75) {};
		\node [style=none] (1) at (-6, -0.75) {};
		\node [style=none] (2) at (-2, 1.75) {};
		\node [style=none] (3) at (-2, -1.75) {};
		\node [style=none] (4) at (-6, 1.25) {$S^5/\Gamma_n$};
		\node [style=none] (5) at (6, 0.75) {};
		\node [style=none] (6) at (6, -0.75) {};
		\node [style=none] (7) at (6, 1.25) {$S^5/\Gamma_{n+1}$};
		\node [style=none] (8) at (2, 1.75) {};
		\node [style=none] (9) at (2, -1.75) {};
		\node [style=none] (10) at (0, 1.5) {};
		\node [style=none] (11) at (0, -1.5) {};
		\node [style=none] (12) at (-2.75, 0) {};
		\node [style=none] (13) at (-1.75, 0) {};
		\node [style=none] (14) at (-1.375, 0.07) {};
		\node [style=none] (15) at (-3.125, 0.07) {};
		\node [style=none] (16) at (1.75, 0) {};
		\node [style=none] (17) at (2.75, 0) {};
		\node [style=none] (18) at (3.125, 0.07) {};
		\node [style=none] (19) at (1.375, 0.07) {};
		\node [style=SmallCircleRed] (20) at (6, 0.25) {};
		\node [style=SmallCircleRed] (21) at (-6, 0.25) {};
		\node [style=SmallCircleBlue] (22) at (-6, -0.25) {};
		\node [style=SmallCirclePurple] (23) at (6, -0.25) {};
		\node [style=none] (24) at (0, 1.125) {};
		\node [style=none] (25) at (-0.75, 0) {};
		\node [style=none] (26) at (-2.25, 0.5) {};
		\node [style=none] (27) at (-2.25, -0.5) {};
		\node [style=none] (28) at (0.5, 0) {};
		\node [style=none] (29) at (2.25, 0.5) {};
		\node [style=none] (30) at (2.25, -0.5) {};
		\node [style=none] (31) at (-4, 0) {};
		\node [style=none] (32) at (4, 0) {};
	\end{pgfonlayer}
	\begin{pgfonlayer}{edgelayer}
		\draw [style=ThickLine, bend left=270, looseness=0.75] (0.center) to (1.center);
		\draw [style=DashedLine, bend left=90, looseness=0.75] (0.center) to (1.center);
		\draw [style=ThickLine, in=180, out=0, looseness=1.25] (0.center) to (2.center);
		\draw [style=ThickLine, in=0, out=-180, looseness=1.25] (3.center) to (1.center);
		\draw [style=ThickLine, bend left=270, looseness=0.75] (5.center) to (6.center);
		\draw [style=ThickLine, in=-180, out=0, looseness=1.25] (9.center) to (6.center);
		\draw [style=ThickLine, in=0, out=180, looseness=1.25] (5.center) to (8.center);
		\draw [style=ThickLine, in=-180, out=0] (2.center) to (10.center);
		\draw [style=ThickLine, in=180, out=0] (10.center) to (8.center);
		\draw [style=ThickLine, in=0, out=180] (9.center) to (11.center);
		\draw [style=ThickLine, in=0, out=180] (11.center) to (3.center);
		\draw [style=ThickLine, bend left=15] (12.center) to (13.center);
		\draw [style=ThickLine, bend right=15] (15.center) to (14.center);
		\draw [style=ThickLine, bend left=15] (16.center) to (17.center);
		\draw [style=ThickLine, bend right=15] (19.center) to (18.center);
		\draw [style=ThickLine, bend left=90, looseness=0.75] (5.center) to (6.center);
		\draw [style=RedLine, in=180, out=0] (21) to (24.center);
		\draw [style=RedLine, in=180, out=0] (24.center) to (20);
		\draw [style=BlueLine, in=180, out=0] (22) to (31.center);
		\draw [style=BlueLine, in=-180, out=15] (31.center) to (26.center);
		\draw [style=BlueLine, in=90, out=0, looseness=0.50] (26.center) to (25.center);
		\draw [style=BlueLine, in=0, out=-90, looseness=0.50] (25.center) to (27.center);
		\draw [style=BlueLine, in=345, out=-180] (27.center) to (31.center);
		\draw [style=PurpleLine, in=180, out=90, looseness=0.50] (28.center) to (29.center);
		\draw [style=PurpleLine, in=165, out=0] (29.center) to (32.center);
		\draw [style=PurpleLine] (32.center) to (23);
		\draw [style=PurpleLine, in=0, out=-165] (32.center) to (30.center);
		\draw [style=PurpleLine, in=-90, out=-180, looseness=0.50] (30.center) to (28.center);
	\end{pgfonlayer}
\end{tikzpicture}}
\caption{We sketch a cobordism $X_{n,n+1}$ realizing the transition from an old local model boundary $S^5/\Gamma_n$ relevant at energies $E_n$ to a new local model boundary $S^5/\Gamma_{n+1}$ relevant at energies $E_{n+1}$. In the process of shedding $X_{n,n+1}$ cycles, which can support defects, can be inherited (red) or removed (blue) or introduced (purple).}
\label{fig:defectplay}
\end{figure}

When the geometry $\mathbb{R}^6/\Gamma_n$ deforms via $X_{n,n+1}$ to $\mathbb{R}^6/\Gamma_{n+1}$ cycles wrapped to construct defects associated with the 0-form and 2-form symmetries can be added, removed or inherited (see figure \ref{fig:defectplay}). These defects are built from wrapped branes and so we will simply uniformly discuss their wrapping loci, as characterized by homology groups.

More precisely, we can characterize these three processes using homology groups of $X_{n,n+1}$ due to $\mathbb{R}^6/\Gamma$ being a cone whereby non-compact cycles in  $\mathbb{R}^6/\Gamma_n$ and  $\mathbb{R}^6/\Gamma_{n+1}$ are in one-to-one correspondence with with boundary cycles, which are also contained in $\partial X_{n,n+1}=S^5/\Gamma_n\sqcup S^5/\Gamma_{n+1}$. As such let us introduce the following notation for boundary components of the cobordism
\be \ba
\partial_{(\textnormal{old})}X_{n,n+1}&=S^5/\Gamma_{n}\,, \\
\partial_{(\textnormal{new})}X_{n,n+1}&= \overline{S^5/\Gamma_{n+1}}\,, \\
\partial X_{n,n+1}&=\partial_{(\textnormal{old})}X_{n,n+1}\sqcup \partial_{(\textnormal{new})}X_{n,n+1}\,,
\ea \ee
where $\overline{\bullet}$ denotes orientation reversal. In the completely general case the
boundaries labelled ``old'' and ``new'' consists of a disjoint union of 5-sphere quotients
and the discussion below generalizes in the obvious way.
The relevant groups associated with the three processes are the integer homology groups
\be\label{eq:Gps} \ba
R_{n,n+1}^{(k)}&\equiv H_k(X_{n,n+1},\partial X_{n,n+1})\,, \\
R_{n,n+1}^{\textnormal{($k$,old)}}&\equiv H_k(X_{n,n+1},\partial_{\textnormal{(old)}} X_{n,n+1})\,,  \\
R_{n,n+1}^{\textnormal{($k$,new)}}&\equiv H_k(X_{n,n+1},\partial_{\textnormal{(new)}} X_{n,n+1})\,,
\ea \ee
where inclusion lifts to a mapping of the latter pair into the first line. The combined cokernel with respect to both of these mappings characterizes cycles which run from the old boundary to the new boundary. Further, we also have the group $H_k(X_{n,n+1})$ which maps via the long exact sequence in relative homology separately into the latter two groups. The cokernel of these mappings characterizes relative cycles which connect to the old or the new boundaries.

We can formalize the above considerations by noting that the pair of triplets
\be\ba
\partial_{(\textnormal{old})} X_{n,n+1}&\subset \partial X_{n,n+1} \subset X_{n,n+1}\,, \\
\partial_{(\textnormal{new})} X_{n,n+1}&\subset \partial X_{n,n+1} \subset X_{n,n+1}\,,
\ea\ee
result in long exact sequences in relative homology. Due to the relevant 5-sphere quotients having no torsional 2- or 4-cycles these two long exact sequences decompose into a collection of short exact sequences. We can combine pairs of these short exact sequences into the cross:
\be\label{eq:cross}
\begin{array}{ccccc}
& & R_{n,n+1}^{\textnormal{($k$,new)}}  & & \\[0.35em]
& &\downarrow & & \\[0.5em]
 R_{n,n+1}^{\textnormal{($k$,old)}} & \rightarrow &R_{n,n+1}^{(k)}& \rightarrow &  \Gamma_{n+1}   \\[0.5em]
 & &\downarrow  & & \\[0.35em]
 & &\Gamma_n & & \\
\end{array}
\ee
Here, due to our simplifying assumption that $\Gamma_n,\Gamma_{n+1}$ are acting fixed point free, i.e., $X_{n,n+1}$ is smooth, and $ H_{k-1}(\partial X_{n,n+1},\partial_{\textnormal{(old)}} X_{n,n+1})\cong  \Gamma_{n+1}  $ and $ H_{k-1}(\partial X_{n,n+1},\partial_{\textnormal{(new)}} X_{n,n+1})\cong  \Gamma_{n}$. This is because in either case one of the 5-spheres
is collapsed to a point. For brevity, we also omitted zeros at either ends of both short exact sequences. We have two crosses, one for $k=2$ and one for $k=4$.

The two sequences of the cross induce a long exact sequence:
\be \label{eq:diagonal}
0~\rightarrow~H_k(X_{n,n+1})~\rightarrow~R_{n,n+1}^{\textnormal{($k$,old)}}\oplus R_{n,n+1}^{\textnormal{($k$,new)}}~\rightarrow~ R_{n,n+1}^{(k)}~\rightarrow~\Gamma_{n,n+1}~\rightarrow~0\,.
\ee
Here we have introduced
\be
\Gamma_{n,n+1}\equiv  R_{n,n+1}^{\textnormal{($k$,new)}}/(R_{n,n+1}^{\textnormal{($k$,old)}}\oplus R_{n,n+1}^{\textnormal{($k$,new)}})\,,
\ee
which is defined as the cokernel of the preceding map. This realizes the combined cokernel described above. Parametrizing as $\Gamma_n\cong \Z_{N_n}$ and $\Gamma_{n+1}\cong \Z_{N_{n+1}}$ we have $\Gamma_{n,n+1}=\Z_{g_{n,n+1}}$ with $g_{n,n+1}=\textnormal{gcd}(N_n,N_{n+1})$. The ordinary long exact sequence in relative homology informs us that $H_k(X_{n,n+1})$ is a subgroup of both $R_{n,n+1}^{\textnormal{($k$,old)}}$ and $R_{n,n+1}^{\textnormal{($k$,new)}}$ and as such we have introduced $H_k(X_{n,n+1})$ as an additional term into the diagonal sequence associated with the cross in order to maintain exactness and undo a double counting. From here we also naturally have the quotients
\be
R_{n,n+1}^{\textnormal{($k$,old)}} / H_k(X_{n,n+1}) \cong \Gamma_n \,, \qquad R_{n,n+1}^{\textnormal{($k$,new)}}/ H_k(X_{n,n+1})\cong \Gamma_{n+1}\,,
\ee
which removes all the bulk cycles and realizes the cokernels of the final two mapping described below \eqref{eq:Gps} which count the cycles removed and added by $X_{n,n+1}$ respectively. We evaluated these using the long exact sequence in relative homology. We see that we can naturally fill the cross to
\be
\begin{array}{ccccc}
 H_k(X_{n,n+1}) & \rightarrow & R_{n,n+1}^{\textnormal{($k$,new)}}  & \rightarrow & \Gamma_{n+1} \\[0.35em]
\downarrow & &\downarrow &&  \rotatebox{-90}{\!\!\!$\cong$}\\[0.5em]
 R_{n,n+1}^{\textnormal{($k$,old)}} & \rightarrow &R_{n,n+1}^{(k)}& \rightarrow &  \Gamma_{n+1}   \\[0.5em]
 \downarrow & &\downarrow  & &\downarrow \\[0.35em]
\Gamma_n &\cong &\Gamma_n & \rightarrow & 0 \\[0.75em]
\end{array}
\ee
organizing all of our data, the sequence \eqref{eq:diagonal} is implicit. The generalization to the case with multiple boundary components is similar but more involved; various entries are replaced by disjoint unions. Further, note that as we are discussing from a defect perspective, we also expect the above analysis to hold provided all tachyons are initially sequestered at the tip of the cone, i.e., it also holds even when local supersymmetry preserving codimension 4 singularities are present. On the other hand, when unsequestered tachyons are initially present at the boundary, then equation (\ref{eq:CoolFormula}) indicates that we should expect further physical contributions to be present.

Let us discuss the fate of the 0-form and 2-form symmetry when transitioning from energy $E_n$ to $E_{n+1}$. We uphold the simplifying assumptions from above, and as such we aim to describe how the 2-form defect group $\Gamma_{n}$ transitions to $\Gamma_{n+1}$ upon traversing the cobordism $X_{n,n+1}$. We focus explicitly on the 2-form symmetry, the 0-form analysis runs similarly.

The overall mechanism will formally resemble a Higgsing. To start, dualize the above analysis to cohomology and expand the supergravity field $C_4$ in
\be
H^2(X_{n,n+1},\partial X_{n,n+1})\cong \Z^{b_2}\oplus \dots
\ee
obtaining a collection of $b_2$ abelian 2-form fields. In many cases $X_{n,n+1}$ will be deformation equivalent to a weighted projective space, which does not have torsional cocycles, and as such we will assume here that the omitted torsional contribution indicated with $\dots$ vanishes.

Let us study screening effects in the partially resolved geometry relevant in taking the limit to $\mathbb{R}^6/\Gamma_{n+1}$. There, the 2-form symmetry defect group is still equal to $\Gamma_{n}$ and naively receives contributions from the weight lattice of $U(1)^{b_2}$ modulo the charge lattice associated with D3-brane wrappings on compact curves and the localized contribution from the residual singularity equal to $\Gamma_{n+1}$. Now, it can happen that an integer multiple of a non-compact curve of the latter has a compact representative, as made manifest by the map $ H_{k}(X_{n,n+1})\rightarrow H_{k}(X_{n,n+1},\partial X_{n,n+1})$. As such we have that the original defect group is given by
\be \label{eq:higgsed}
U(1)^{b_2}\times \Gamma_{n+1}\, /\sim
\ee
where $\sim$ describes the identification encoded in the above map. The screening by D3-branes wrapped on compact curves then reduces the above to $\Gamma_n$. Overall this parallels tracking the center group in adjoint Higgsings such as $SU(N_1)\rightarrow SU(N_2)\times U(1)^{b} /\Z_L$ for some integers $N_1,N_2,L,b$ (see \cite{Baume:2023kkf} for further discussion on this point).
The finite group $\Gamma_n$ is a subgroup of \eqref{eq:higgsed},
generically embedding both into the abelian factor and $\Gamma_{n+1}$.

Now, as the resolution proceeds to blowup, we reach energies $E_{n+1}$ and the cobordism $X_{n,n+1}$ hits the asymptotic boundary, the (co)cycles relevant for the abelian factors (become non-normalizable) decompactify and the abelian factors decouple. Of the original defect group isomorphic to $\Gamma_{n}$ the image of $\Gamma_n$ in $U(1)^{b_2}\times \Gamma_{n+1}\, /\sim$ projected to $\Gamma_{n+1}/\sim$ remains. Simultaneously, new wrapping loci have opened up, filling the defect group back up to $\Gamma_{n+1}$.

\paragraph{Duality Interfaces:}

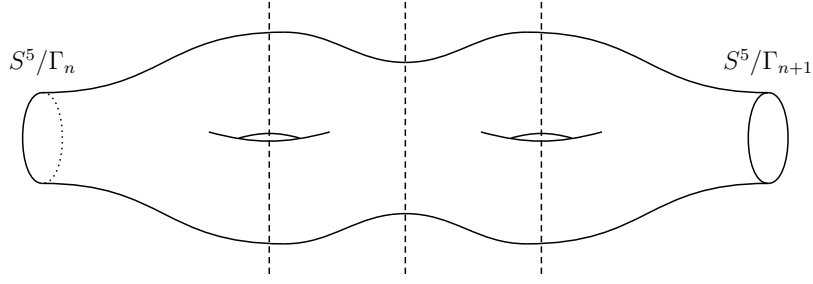
\begin{figure}
    \centering
    \scalebox{0.8}{
    \begin{tikzpicture}
    \begin{pgfonlayer}{nodelayer}
		\node [style=none] (0) at (-6, 0.75) {};
		\node [style=none] (1) at (-6, -0.75) {};
		\node [style=none] (2) at (-2, 1.75) {};
		\node [style=none] (5) at (-2, -1.75) {};
		\node [style=none] (7) at (-6, 1.25) {$S^5/\Gamma_n$};
		\node [style=none] (10) at (6, 0.75) {};
		\node [style=none] (11) at (6, -0.75) {};
		\node [style=none] (12) at (6, 1.25) {$S^5/\Gamma_{n+1}$};
		\node [style=none] (17) at (2, 1.75) {};
		\node [style=none] (18) at (2, -1.75) {};
		\node [style=none] (19) at (0, 1.25) {};
		\node [style=none] (20) at (0, -1.25) {};
		\node [style=none] (21) at (-2.75, 0) {};
		\node [style=none] (22) at (-1.75, 0) {};
		\node [style=none] (23) at (-1.25, 0.1) {};
		\node [style=none] (24) at (-3.25, 0.1) {};
		\node [style=none] (25) at (1.75, 0) {};
		\node [style=none] (26) at (2.75, 0) {};
		\node [style=none] (27) at (3.25, 0.1) {};
		\node [style=none] (28) at (1.25, 0.1) {};
		\node [style=none] (76) at (-2.25, 2.25) {};
		\node [style=none] (77) at (-2.25, -2.25) {};
		\node [style=none] (78) at (2.25, 2.25) {};
		\node [style=none] (79) at (2.25, -2.25) {};
		\node [style=none] (80) at (0, 2.25) {};
		\node [style=none] (81) at (0, -2.25) {};
	\end{pgfonlayer}
	\begin{pgfonlayer}{edgelayer}
		\draw [style=ThickLine, bend left=270, looseness=0.75] (0.center) to (1.center);
		\draw [style=DottedLine, bend left=90, looseness=0.75] (0.center) to (1.center);
		\draw [style=ThickLine, in=180, out=0, looseness=1.25] (0.center) to (2.center);
		\draw [style=ThickLine, in=0, out=-180, looseness=1.25] (5.center) to (1.center);
		\draw [style=ThickLine, bend left=270, looseness=0.75] (10.center) to (11.center);
		\draw [style=ThickLine, in=-180, out=0, looseness=1.25] (18.center) to (11.center);
		\draw [style=ThickLine, in=0, out=180, looseness=1.25] (10.center) to (17.center);
		\draw [style=ThickLine, in=-180, out=0] (2.center) to (19.center);
		\draw [style=ThickLine, in=180, out=0] (19.center) to (17.center);
		\draw [style=ThickLine, in=0, out=180] (18.center) to (20.center);
		\draw [style=ThickLine, in=0, out=180] (20.center) to (5.center);
		\draw [style=ThickLine, bend left=15] (21.center) to (22.center);
		\draw [style=ThickLine, bend right=15] (24.center) to (23.center);
		\draw [style=ThickLine, bend left=15] (25.center) to (26.center);
		\draw [style=ThickLine, bend right=15] (28.center) to (27.center);
		\draw [style=DashedLine] (76.center) to (77.center);
		\draw [style=DashedLine] (80.center) to (81.center);
		\draw [style=DashedLine] (78.center) to (79.center);
        \draw [style=ThickLine, bend left=90, looseness=0.75] (10.center) to (11.center);
	\end{pgfonlayer}
\end{tikzpicture}}
    \caption{We sketch the cobordism constructed by the partial resoltion of $\mathbb{R}^6/\Gamma_n$ between $S^5/\Gamma_n$ and  $S^5/\Gamma_{n+1}$. For any such cobordism there exists a handle presentation indicated by the dashed lines.}
    \label{fig:Sweep}
\end{figure}

It is also of interest to track the behavior of the duality interface given by a wrapped 7-brane on $S^5 / \Gamma$. For this discussion let us again consider the idealized setup in which a codimension 6 singularity resolves to exactly one codimension 6 singularity. We expect that this analysis extends to geometries with more general singularities.

Let $\R^6/\Gamma_n$ and $\R^6/\Gamma_{n+1}$ be two geometries with an isolated singularity such that $\R^6/\Gamma_{n+1}$ is a local model for the residual singularity in the partial resolution of $\R^6/\Gamma_n$ relevant for the transition from scale $E_n$ to $E_{n+1}$. The partially resolved geometry realizes a smooth cobordism between the asymptotic $S^5/\Gamma_n$ and $S^5/\Gamma_{n+1}$ explicitly obtained as a manifold with boundary $X_{n,n+1}$ by excising a ball centered on the residual singularity in the partially resolved geometry (see figure \ref{fig:Sweep}).
In the completely general case the cobordism $X_{n,n+1}$ would be a disjoint union of manifolds with boundaries each with more than two boundary components and which includes various codimension 2 and codimension 4 singularities.

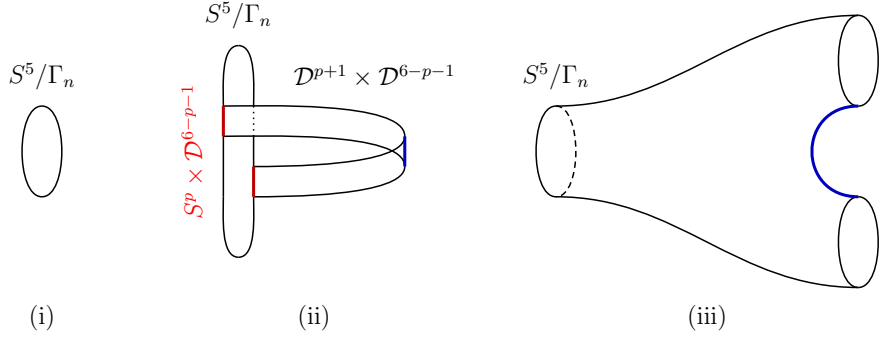
\begin{figure}
    \centering
    \scalebox{0.8}{
    \begin{tikzpicture}
	\begin{pgfonlayer}{nodelayer}
		\node [style=none] (0) at (4, 0.75) {};
		\node [style=none] (1) at (4, -0.75) {};
		\node [style=none] (2) at (9, 2.25) {};
		\node [style=none] (3) at (9, 0.75) {};
		\node [style=none] (4) at (9, -0.75) {};
		\node [style=none] (5) at (9, -2.25) {};
		\node [style=none] (13) at (6.5, -2.75) {(iii)};
		\node [style=none] (14) at (-1.25, 1.75) {};
		\node [style=none] (20) at (0, -2.75) {(ii)};
		\node [style=none] (21) at (-4.5, 0.75) {};
		\node [style=none] (22) at (-4.5, -0.75) {};
		\node [style=none] (27) at (-4.5, -2.75) {(i)};
		\node [style=none] (28) at (-1.5, 0.75) {};
		\node [style=none] (29) at (-1.5, 0.25) {};
		\node [style=none] (30) at (-1.25, -1.75) {};
		\node [style=none] (31) at (-1, -0.25) {};
		\node [style=none] (32) at (-1, 0.25) {};
		\node [style=none] (33) at (1.5, 0.25) {};
		\node [style=none] (34) at (1.5, -0.25) {};
		\node [style=none] (35) at (-1.5, -0.25) {};
		\node [style=none] (36) at (-1, 0.75) {};
		\node [style=none] (37) at (-1.5, -0.75) {};
		\node [style=none] (38) at (-1, -0.75) {};
		\node [style=none] (39) at (-4.5, 1.25) {$S^5/\Gamma_n$};
		\node [style=none] (40) at (-1.25, 2.25) {$S^5/\Gamma_n$};
		\node [style=none] (41) at (-2, 0) {\rotatebox[]{90}{\color{red}$S^p\times \mathcal{D}^{6-p-1}$}};
		\node [style=none] (42) at (1, 1.25) {$\mathcal{D}^{p+1}\times \mathcal{D}^{6-p-1}$};
		\node [style=none] (43) at (4, 1.25) {$S^5/\Gamma_n$};
		\node [style=none] (44) at (9, 2.75) {};
		\node [style=none] (45) at (9, -0.25) {};
        \node [style=none] (46) at (0, -3.5) {};
	\end{pgfonlayer}
	\begin{pgfonlayer}{edgelayer}
		\draw [style=ThickLine, bend left=270, looseness=0.75] (0.center) to (1.center);
		\draw [style=ThickLine, bend right=90, looseness=0.75] (2.center) to (3.center);
		\draw [style=ThickLine, bend left=270, looseness=0.75] (4.center) to (5.center);
		\draw [style=ThickLine, bend left=90, looseness=0.75] (2.center) to (3.center);
		\draw [style=ThickLine, bend left=90, looseness=0.75] (4.center) to (5.center);
		\draw [style=DashedLine, bend left=90, looseness=0.75] (0.center) to (1.center);
		\draw [style=ThickLine, in=-180, out=0] (0.center) to (2.center);
		\draw [style=ThickLine, in=0, out=180] (5.center) to (1.center);
		\draw [style=ThickLine, bend left=270, looseness=0.75] (21.center) to (22.center);
		\draw [style=ThickLine, in=-90, out=0, looseness=0.75] (30.center) to (38.center);
		\draw [style=ThickLine, in=180, out=-90, looseness=0.75] (37.center) to (30.center);
		\draw [style=ThickLine, in=-180, out=90, looseness=0.75] (28.center) to (14.center);
		\draw [style=ThickLine, in=90, out=0, looseness=0.75] (14.center) to (36.center);
		\draw [style=ThickLine] (28.center) to (37.center);
		\draw [style=ThickLine, in=90, out=0, looseness=0.50] (28.center) to (33.center);
		\draw [style=ThickLine, in=0, out=90, looseness=0.50] (34.center) to (29.center);
		\draw [style=ThickLine, in=-90, out=0, looseness=0.50] (38.center) to (34.center);
		\draw [style=ThickLine, in=-90, out=0, looseness=0.50] (31.center) to (33.center);
		\draw [style=RedLine] (28.center) to (29.center);
		\draw [style=RedLine] (31.center) to (38.center);
		\draw [style=BlueLine, bend right=90, looseness=1.75] (3.center) to (4.center);
		\draw [style=BlueLine] (33.center) to (34.center);
		\draw [style=ThickLine, bend left=90, looseness=0.75] (21.center) to (22.center);
		\draw [style=DottedLine] (36.center) to (32.center);
		\draw [style=ThickLine] (32.center) to (31.center);
	\end{pgfonlayer}
\end{tikzpicture}
    }
    \caption{Sketch of attaching the handle $\mathcal{H}^{(p)}$ to $S^5/\Gamma_n$. The sketch can be taken at face value for gluing a 0-handle to a circle. The initial configuration is shown in (i). In (ii) we glue $\mathcal{D}^{p+1}\times \mathcal{D}^{6-p-1}$ along $S^p\times \mathcal{D}^{6-p-1}$ to $S^5/\Gamma_n$. The resulting configuration is deformation equivalent to (iii). The gluing locus $S^p\times \mathcal{D}^{6-p-1}$ admits a retraction into the handle and can be collapsed to a single copy of $\mathcal{D}^{6-p-1}$ (blue).}
    \label{fig:HandleGluing}
\end{figure}

The cobordism $X_{n,n+1}$ is assumed to be smooth and therefore admits a handle presentation, i.e., $X_{n,n+1}$ is constructed by gluing $p\:\!$-handles
\be
\mathcal{H}^{(p)}= \mathcal{D}^{p+1}\times \mathcal{D}^{6-p-1}
\ee
to $S^5/\Gamma_n$. The gluing occurs along a copy of $S^{p}\times \mathcal{D}^{6-p-1}$. Here $\mathcal{D}^q$ denotes the $q$-disk. Note that within $S^5/\Gamma_n$ the disk $\mathcal{D}^{6-p-1}$ is contractible, while the sphere $S^{p}$ is not necessarily contractible. However, within the handle-attached space $S^5/\Gamma_n\cup \mathcal{H}^{(p)}$, the gluing locus becomes contractible to a copy of $\mathcal{D}^{6-p-1}$ within $\mathcal{H}^{(p)}$ (see figure \ref{fig:HandleGluing}).

\begin{figure}
    \centering
    \scalebox{0.7}{
    \begin{tikzpicture}
	\begin{pgfonlayer}{nodelayer}
		\node [style=none] (0) at (-3.75, 0.5) {};
		\node [style=none] (1) at (-3.75, -1) {};
		\node [style=none] (2) at (0, 1.5) {};
		\node [style=none] (3) at (0, -2) {};
		\node [style=none] (4) at (-3.75, 1) {$S^5/\Gamma_n$};
		\node [style=none] (5) at (3.75, 0.5) {};
		\node [style=none] (6) at (3.75, -1) {};
		\node [style=none] (7) at (3.75, 1) {$S^5/\Gamma_{n+1}$};
		\node [style=none] (10) at (3.75, 0.5) {};
		\node [style=none] (11) at (3.75, -1) {};
		\node [style=none] (12) at (-0.5, -0.25) {};
		\node [style=none] (13) at (0.5, -0.25) {};
		\node [style=none] (14) at (1, -0.15) {};
		\node [style=none] (15) at (-1, -0.15) {};
		\node [style=none] (21) at (-2.25, 0.75) {};
		\node [style=none] (22) at (-2.25, -1.25) {};
		\node [style=none] (23) at (-3.75, -12.25) {};
		\node [style=none] (24) at (-3.75, -13.75) {};
		\node [style=none] (25) at (0, -11.25) {};
		\node [style=none] (26) at (0, -14.75) {};
		\node [style=none] (27) at (-3.75, -11.75) {$S^5/\Gamma_n$};
		\node [style=none] (28) at (3.75, -12.25) {};
		\node [style=none] (29) at (3.75, -13.75) {};
		\node [style=none] (30) at (3.75, -11.75) {$S^5/\Gamma_{n+1}$};
		\node [style=none] (31) at (3.75, -12.25) {};
		\node [style=none] (32) at (3.75, -13.75) {};
		\node [style=none] (33) at (-0.5, -13.00) {};
		\node [style=none] (34) at (0.5, -13.00) {};
		\node [style=none] (35) at (1, -12.9) {};
		\node [style=none] (36) at (-1, -12.9) {};
		\node [style=none] (39) at (2.25, -12.00) {};
		\node [style=none] (40) at (2.25, -14.00) {};
		\node [style=none] (41) at (2, -13.5) {};
		\node [style=none] (42) at (2.45, -12.5) {};
		\node [style=none] (43) at (2, -12.75) {};
		\node [style=none] (44) at (2.5, -13.25) {};
		\node [style=none] (45) at (-5, 0) {(i)};
		\node [style=none] (46) at (-5, -13.00) {(iv)};
        \node [style=SmallCircleBlue] (47) at (2, -12.75) {};
		\node [style=SmallCircleBlue] (48) at (2.4, -13.25) {};
		\node [style=SmallCircleBlue] (49) at (2.42, -12.5) {};
		\node [style=SmallCircleBlue] (50) at (2.02, -13.5) {};
  \node [style=none] (51) at (-3.75, -16.5) {};
		\node [style=none] (52) at (-3.75, -18.00) {};
		\node [style=none] (53) at (0, -15.5) {};
		\node [style=none] (54) at (0, -19) {};
		\node [style=none] (55) at (-3.75, -16) {$S^5/\Gamma_n$};
		\node [style=none] (56) at (3.75, -16.5) {};
		\node [style=none] (57) at (3.75, -18) {};
		\node [style=none] (58) at (3.75, -16) {$S^5/\Gamma_{n+1}$};
		\node [style=none] (59) at (3.75, -16.5) {};
		\node [style=none] (60) at (3.75, -18) {};
		\node [style=none] (61) at (-0.5, -17.25) {};
		\node [style=none] (62) at (0.5, -17.25) {};
		\node [style=none] (63) at (1, -17.15) {};
		\node [style=none] (64) at (-1, -17.15) {};
		\node [style=none] (65) at (2.25, -16.25) {};
		\node [style=none] (66) at (2.25, -18.25) {};
		\node [style=none] (67) at (2, -17.75) {};
		\node [style=none] (68) at (2.5, -16.75) {};
		\node [style=none] (69) at (2, -17) {};
		\node [style=none] (70) at (2.5, -17.5) {};
		\node [style=none] (71) at (-5, -17.25) {(v)};
		\node [style=SmallCircleBlue] (72) at (2, -17) {};
		\node [style=SmallCircleBlue] (73) at (2.4, -17.5) {};
		\node [style=SmallCircleBlue] (74) at (2.42, -16.75) {};
		\node [style=SmallCircleBlue] (75) at (2, -17.75) {};
		\node [style=SmallCirclePurple] (76) at (1.25, -17.35) {};
		\node [style=SmallCirclePurple] (77) at (0, -17.13) {};
		\node [style=SmallCirclePurple] (78) at (0, -17.37) {};
		\node [style=SmallCirclePurple] (79) at (1.6, -17.1) {};
        \node [style=none] (80) at (-3.75, -20.75) {};
		\node [style=none] (81) at (-3.75, -22.25) {};
		\node [style=none] (82) at (0, -19.75) {};
		\node [style=none] (83) at (0, -23.25) {};
		\node [style=none] (84) at (-3.75, -20.25) {$S^5/\Gamma_n$};
		\node [style=none] (85) at (3.75, -20.75) {};
		\node [style=none] (86) at (3.75, -22.25) {};
		\node [style=none] (87) at (3.75, -20.25) {$S^5/\Gamma_{n+1}$};
		\node [style=none] (88) at (3.75, -20.75) {};
		\node [style=none] (89) at (3.75, -22.25) {};
		\node [style=none] (90) at (-0.5, -21.5) {};
		\node [style=none] (91) at (0.5, -21.5) {};
		\node [style=none] (92) at (1, -21.4) {};
		\node [style=none] (93) at (-1, -21.4) {};
		\node [style=none] (94) at (2.25, -20.5) {};
		\node [style=none] (95) at (2.25, -22.5) {};
		\node [style=none] (96) at (2, -22.00) {};
		\node [style=none] (97) at (2.5, -21.00) {};
		\node [style=none] (98) at (2, -21.25) {};
		\node [style=none] (99) at (2.5, -21.75) {};
		\node [style=none] (100) at (-5, -21.5) {(vi)};
		\node [style=SmallCirclePurple] (106) at (0, -21.37) {};
		\node [style=SmallCirclePurple] (107) at (0, -21.63) {};
		\node [style=SmallCircleBrown] (109) at (2.4, -21.00) {};
		\node [style=SmallCircleBrown] (110) at (2.05, -22.00) {};
		\node [style=none] (111) at (-3.75, -3.75) {};
		\node [style=none] (112) at (-3.75, -5.25) {};
		\node [style=none] (113) at (0, -2.75) {};
		\node [style=none] (114) at (0, -6.25) {};
		\node [style=none] (115) at (-3.75, -3.25) {$S^5/\Gamma_n$};
		\node [style=none] (116) at (3.75, -3.75) {};
		\node [style=none] (117) at (3.75, -5.25) {};
		\node [style=none] (118) at (3.75, -3.25) {$S^5/\Gamma_{n+1}$};
		\node [style=none] (119) at (3.75, -3.75) {};
		\node [style=none] (120) at (3.75, -5.25) {};
		\node [style=none] (121) at (-0.5, -4.5) {};
		\node [style=none] (122) at (0.5, -4.5) {};
		\node [style=none] (123) at (1, -4.4) {};
		\node [style=none] (124) at (-1, -4.4) {};
		\node [style=none] (131) at (-5, -4.5) {(ii)};
		\node [style=none] (136) at (-3.75, -8) {};
		\node [style=none] (137) at (-3.75, -9.5) {};
		\node [style=none] (138) at (0, -7) {};
		\node [style=none] (139) at (0, -10.5) {};
		\node [style=none] (140) at (-3.75, -7.5) {$S^5/\Gamma_n$};
		\node [style=none] (141) at (3.75, -8) {};
		\node [style=none] (142) at (3.75, -9.5) {};
		\node [style=none] (143) at (3.75, -7.5) {$S^5/\Gamma_{n+1}$};
		\node [style=none] (144) at (3.75, -8) {};
		\node [style=none] (145) at (3.75, -9.5) {};
		\node [style=none] (146) at (-0.5, -8.75) {};
		\node [style=none] (147) at (0.5, -8.75) {};
		\node [style=none] (148) at (1, -8.65) {};
		\node [style=none] (149) at (-1, -8.65) {};
		\node [style=none] (156) at (-5, -8.75) {(iii)};
		\node [style=none] (161) at (0, -0.25) {};
		\node [style=none] (162) at (0, -0.25) {};
		\node [style=none] (163) at (0, -0.25) {};
		\node [style=none] (164) at (0, -0.25) {};
		\node [style=none] (165) at (0, -0.25) {};
		\node [style=none] (166) at (0, -0.25) {};
		\node [style=none] (167) at (0, -0.25) {};
		\node [style=none] (168) at (0, -0.25) {};
		\node [style=none] (169) at (-0.5, -2.75) {};
		\node [style=none] (170) at (-0.5, -4.5) {};
		\node [style=none] (171) at (-0.5, -4.5) {};
		\node [style=none] (172) at (-0.5, -6.25) {};
		\node [style=none] (173) at (0, -7) {};
		\node [style=none] (174) at (0, -8.65) {};
		\node [style=none] (175) at (0, -8.8) {};
		\node [style=none] (176) at (0, -10.5) {};
		\node [style=none] (177) at (-0.25, -7.75) {};
		\node [style=none] (178) at (-0.25, -9.75) {};
		\node [style=SmallCircleBlue] (179) at (-0.2, -7.75) {};
		\node [style=SmallCircleBlue] (180) at (-0.2, -9.75) {};
	\end{pgfonlayer}
	\begin{pgfonlayer}{edgelayer}
		\draw [style=ThickLine, bend left=270, looseness=0.75] (0.center) to (1.center);
		\draw [style=DottedLine, bend left=90, looseness=0.75] (0.center) to (1.center);
		\draw [style=ThickLine, in=180, out=0, looseness=1.50] (0.center) to (2.center);
		\draw [style=ThickLine, in=0, out=-180, looseness=1.50] (3.center) to (1.center);
		\draw [style=ThickLine, bend left=270, looseness=0.75] (5.center) to (6.center);
		\draw [style=ThickLine, in=-180, out=0, looseness=1.50] (2.center) to (10.center);
		\draw [style=ThickLine, in=0, out=180, looseness=1.50] (11.center) to (3.center);
		\draw [style=ThickLine, bend left=15] (12.center) to (13.center);
		\draw [style=ThickLine, bend right=15] (15.center) to (14.center);
		\draw [style=RedLine, bend right=60, looseness=0.50] (21.center) to (22.center);
		\draw [style=DottedRed, bend left=45, looseness=0.50] (21.center) to (22.center);
		\draw [style=ThickLine, bend left=90, looseness=0.75] (5.center) to (6.center);
		\draw [style=ThickLine, bend left=270, looseness=0.75] (23.center) to (24.center);
		\draw [style=DottedLine, bend left=90, looseness=0.75] (23.center) to (24.center);
		\draw [style=ThickLine, in=180, out=0, looseness=1.50] (23.center) to (25.center);
		\draw [style=ThickLine, in=0, out=-180, looseness=1.50] (26.center) to (24.center);
		\draw [style=ThickLine, bend left=270, looseness=0.75] (28.center) to (29.center);
		\draw [style=ThickLine, in=-180, out=0, looseness=1.50] (25.center) to (31.center);
		\draw [style=ThickLine, in=0, out=-180, looseness=1.50] (32.center) to (26.center);
		\draw [style=ThickLine, bend left=15] (33.center) to (34.center);
		\draw [style=ThickLine, bend right=15] (36.center) to (35.center);
		\draw [style=ThickLine, bend left=90, looseness=0.75] (28.center) to (29.center);
		\draw [style=RedLine, bend right=60, looseness=0.50] (39.center) to (40.center);
		\draw [style=DottedRed, bend left=45, looseness=0.50] (39.center) to (40.center);
		\draw [style=BlueLine, in=60, out=180, looseness=0.50] (43.center) to (34.center);
		\draw [style=DottedBlue, in=180, out=-75, looseness=0.25] (34.center) to (44.center);
		\draw [style=DottedBlue, in=90, out=165, looseness=0.75] (42.center) to (33.center);
		\draw [style=BlueLine, in=180, out=-90, looseness=0.50] (33.center) to (41.center);
        \draw [style=ThickLine, bend left=270, looseness=0.75] (51.center) to (52.center);
		\draw [style=DottedLine, bend left=90, looseness=0.75] (51.center) to (52.center);
		\draw [style=ThickLine, in=180, out=0, looseness=1.50] (51.center) to (53.center);
		\draw [style=ThickLine, in=0, out=-180, looseness=1.50] (54.center) to (52.center);
		\draw [style=ThickLine, bend left=270, looseness=0.75] (56.center) to (57.center);
		\draw [style=ThickLine, in=-180, out=0, looseness=1.50] (53.center) to (59.center);
		\draw [style=ThickLine, in=0, out=-180, looseness=1.50] (60.center) to (54.center);
		\draw [style=ThickLine, bend right=15] (64.center) to (63.center);
		\draw [style=ThickLine, bend left=90, looseness=0.75] (56.center) to (57.center);
		\draw [style=RedLine, bend right=60, looseness=0.50] (65.center) to (66.center);
		\draw [style=DottedRed, bend left=45, looseness=0.50] (65.center) to (66.center);
		\draw [style=BlueLine, bend right=90, looseness=3.00] (72) to (75);
		\draw [style=DottedBlue, bend right=90, looseness=3.50] (74) to (73);
		\draw [style=BlueLine, bend right=15] (61.center) to (62.center);
		\draw [style=BlueLine, bend left=15] (61.center) to (62.center);
		\draw [style=PurpleLine, in=180, out=-90] (78) to (76);
		\draw [style=DottedPurple, in=90, out=-180, looseness=0.75] (79) to (77);
        \draw [style=ThickLine, bend left=270, looseness=0.75] (80.center) to (81.center);
		\draw [style=DottedLine, bend left=90, looseness=0.75] (80.center) to (81.center);
		\draw [style=ThickLine, in=180, out=0, looseness=1.50] (80.center) to (82.center);
		\draw [style=ThickLine, in=0, out=-180, looseness=1.50] (83.center) to (81.center);
		\draw [style=ThickLine, bend left=270, looseness=0.75] (85.center) to (86.center);
		\draw [style=ThickLine, in=-180, out=0, looseness=1.50] (82.center) to (88.center);
		\draw [style=ThickLine, in=0, out=-180, looseness=1.50] (89.center) to (83.center);
		\draw [style=ThickLine, bend right=15] (93.center) to (92.center);
		\draw [style=ThickLine, bend left=90, looseness=0.75] (85.center) to (86.center);
		\draw [style=RedLine, bend right=60, looseness=0.50] (94.center) to (95.center);
		\draw [style=DottedRed, bend left=45, looseness=0.50] (94.center) to (95.center);
		\draw [style=BlueLine, bend right=15] (90.center) to (91.center);
		\draw [style=BlueLine, bend left=15] (90.center) to (91.center);
		\draw [style=PurpleLine, in=180, out=-90] (107) to (110);
		\draw [style=DottedPurple, in=180, out=90] (106) to (109);
		\draw [style=ThickLine, bend left=270, looseness=0.75] (111.center) to (112.center);
		\draw [style=DottedLine, bend left=90, looseness=0.75] (111.center) to (112.center);
		\draw [style=ThickLine, in=180, out=0, looseness=1.50] (111.center) to (113.center);
		\draw [style=ThickLine, in=0, out=-180, looseness=1.50] (114.center) to (112.center);
		\draw [style=ThickLine, bend left=270, looseness=0.75] (116.center) to (117.center);
		\draw [style=ThickLine, in=-180, out=0, looseness=1.50] (113.center) to (119.center);
		\draw [style=ThickLine, in=0, out=-180, looseness=1.50] (120.center) to (114.center);
		\draw [style=ThickLine, bend left=15] (121.center) to (122.center);
		\draw [style=ThickLine, bend right=15] (124.center) to (123.center);
		\draw [style=ThickLine, bend left=90, looseness=0.75] (116.center) to (117.center);
		\draw [style=ThickLine, bend left=270, looseness=0.75] (136.center) to (137.center);
		\draw [style=DottedLine, bend left=90, looseness=0.75] (136.center) to (137.center);
		\draw [style=ThickLine, in=180, out=0, looseness=1.50] (136.center) to (138.center);
		\draw [style=ThickLine, in=0, out=-180, looseness=1.50] (139.center) to (137.center);
		\draw [style=ThickLine, bend left=270, looseness=0.75] (141.center) to (142.center);
		\draw [style=ThickLine, in=-180, out=0, looseness=1.50] (138.center) to (144.center);
		\draw [style=ThickLine, in=0, out=-180, looseness=1.50] (145.center) to (139.center);
		\draw [style=ThickLine, bend left=15] (146.center) to (147.center);
		\draw [style=ThickLine, bend right=15] (149.center) to (148.center);
		\draw [style=ThickLine, bend left=90, looseness=0.75] (141.center) to (142.center);
		\draw [style=RedLine, bend right=60, looseness=0.50] (169.center) to (170.center);
		\draw [style=DottedRed, bend left=45, looseness=0.50] (169.center) to (170.center);
		\draw [style=RedLine, bend right=60, looseness=0.50] (171.center) to (172.center);
		\draw [style=DottedRed, bend left=45, looseness=0.50] (171.center) to (172.center);
		\draw [style=RedLine, bend right=60, looseness=0.50] (173.center) to (174.center);
		\draw [style=DottedRed, bend left=45, looseness=0.50] (173.center) to (174.center);
		\draw [style=RedLine, bend right=60, looseness=0.50] (175.center) to (176.center);
		\draw [style=DottedRed, bend left=45, looseness=0.50] (175.center) to (176.center);
		\draw [style=BlueLine, in=-90, out=180] (178.center) to (149.center);
		\draw [style=BlueLine, in=-180, out=90] (149.center) to (177.center);
	\end{pgfonlayer}
\end{tikzpicture}
    }
    \caption{Deforming a 7-brane wrapping (red) across handles gives rise to a new 7-brane wrapping with brane-anti-brane fusion products $\mathcal{C}$ (blue) attached. See (i) $\rightarrow$ (ii). Potentially, there are additional junction operators (blue dots). See (ii). We can ``reconnect" these fusion products by fusing these (resulting in operators colored purple). See (iii). Then we contract the remaining brane-anti-brane fusion products resulting in topological operators localized to the 7-brane (brown dots). See (iv).  }
    \label{fig:FundamentalStep}
\end{figure}
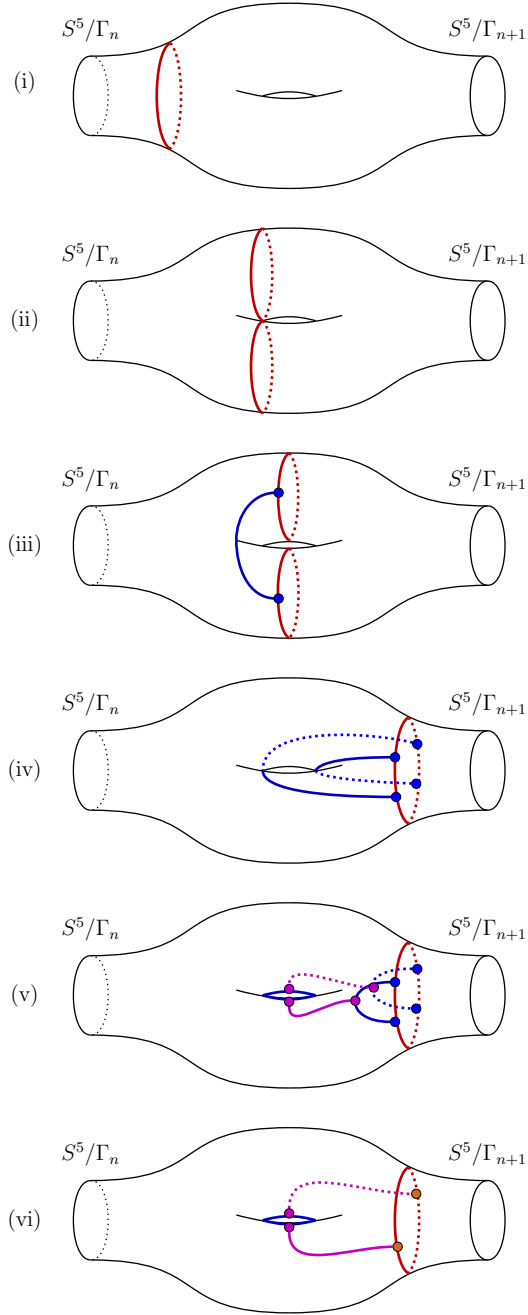

In the presence of a 7-brane wrapped on $S^5/\Gamma_n$ we see that, after attaching a handle, we can deform the 7-brane to wrap the new boundary at the cost of picking up a brane-anti-brane fusion products along a copy of $\mathcal{D}^{6-p-1}$. Proceeding in this manner we obtain brane-anti-brane wrappings along a collection of non-contractible subloci in $X_{n,n+1}$.
An intermediate result, after deforming across all handles, is the same type of 7-brane wrapping $S^5/\Gamma_{n+1}$ with 7-brane brane-anti-brane fusion products $\mathcal{C}$ stretching back into the cobordism, wrapping relative cycles in $X_{n,n+1}$ relative to the boundary component $S^5/\Gamma_{n+1}$. See subfigure (iv) in Figure \ref{fig:FundamentalStep}. From here we can, via further fusions, arrange for $\mathcal{C}$ to wrap internal cycles of the cobordism, i.e., cycles of the homology group $H_k(X_{n,n+1})$. See subfigure (v) and (vi) in figure \ref{fig:FundamentalStep}.

In the 4D theory a 7-brane wrapped on $S^5/\Gamma_n$ realizes a duality interface supporting a TFT at energies $E_n$. In \cite{Heckman:2022xgu} the TFT was determined by studying the line operators of the 3D interface theory, constructed from F1- / D1-strings attaching to the 7-brane. Then, using general results on 3D TFTs \cite{Hsin:2018vcg}, the TFT interacting with the 4D sector was deduced. In lowering the scale from $E_n$ to $E_{n+1}$ the 7-brane is deformed across the cobordism $X_{n,n+1}$ and the couplings of the original 3D TFT to the ambient 4D bulk are now deformed. It would be very interesting to describe these deformations in greater detail.

\subsection{Illustrative Example} \label{sec:IIBexamples}

Let us now turn to an example illustrating these general points. For this, we return to the example $\mathbb{R}^{6}/\Z_{17}^{(4,6,-7,-3)}$ of section \ref{sec:smthBordism} and make some of the scale dependent discussion explicit. Other examples with sequestered tachyons can be treated in a similar fashion.

Partially resolving  $\mathbb{R}^{6}/\Z_{17}^{(4,6,-7,-3)}$ we find the transition of local models
\be
\mathbb{R}^6/\Z_{17}~\rightarrow~\mathbb{R}^6/\Z_{3} \sqcup \mathbb{R}^6/\Z_{7}\,.
\ee
The stack of $N$ D3-branes probing $\mathbb{R}^6/\Z_{17}$ is partitioned between the two new singularities. The order of the quotient groups $\Z_3,\Z_7,\Z_{17}$ are all prime and consequently pairwise coprime. The cobordism $X_{0,1}$ has boundaries:\footnote{There is an orientation reversal on the last two factors of the disjoint union which we keep implicit here and in what follows.}
\be
\partial X_{0,1}={S}^5/\Z_{17} \sqcup \overline{S^5/\Z_{3}} \sqcup \overline{S^5/\Z_{7}}\,,
\ee
where the overline indicates orientation reversal, and we can decompose the partial resolution of $\mathbb{R}^6/\Z_{17}$ as
\be
\widetilde{\mathbb{R}^{6}/\Z_{17}}=\textnormal{Cone}(\overline{S^5/\Z_{7}})  \cup_{S^5/\Z_7} X_{0,1}\cup_{S^5/\Z_3} \textnormal{Cone}(\overline{S^5/\Z_{3}})
\ee
filling in two of the three boundaries, such that the asymptotic boundary $S^5/\Z_{17}$ remains.
Applying the Mayer-Vietoris sequence to the above decomposition, induced by the mappings
\be
\overline{S^5/\Z_3}\sqcup \overline{S^5/\Z_7}\rightarrow X_{0,1} \sqcup \textnormal{Cone}(\overline{S^5/\Z_{7}}) \sqcup \textnormal{Cone}(\overline{S^5/\Z_{3}})  \rightarrow \widetilde{\mathbb{R}^{6}/\Z_{17}}\,,
\ee
we learn, due to the lens space cones having no 2- and 4-cycles of their own,
and due to $\widetilde{\mathbb{R}^{6}/\Z_{17}}$ deformation retracting onto the weighted projective
space $\mathbb{WP}^2$ with weights $(1,3,7)$, that
\be
H_2(X_{0,1})\cong \Z\,, \qquad H_4(X_{0,1})\cong \Z\,,
\ee
where the groups are generated by the $|\Z_3\times \Z_7|=21$ multiple of the generating 2- and 4-cycle of $\mathbb{WP}^2$.
With this characterization we can also compute the various relative homology groups of the cross \eqref{eq:cross},
which are here simply:
\be
\begin{array}{ccccc}
\Z & \rightarrow &\Z & \rightarrow & \Z_3\oplus \Z_7 \\[0.35em]
\downarrow & &\downarrow &&  \rotatebox{-90}{\!\!\!$\cong$}\\[0.5em]
 \Z & \rightarrow &\Z & \rightarrow &  \Z_3\oplus \Z_7\\[0.5em]
 \downarrow & &\downarrow  & &\downarrow \\[0.35em]
\Z_{17}&\cong &\Z_{17}& \rightarrow & 0
\end{array}
\ee

Further, note that via Poincar\'e-Lefschetz duality we have $H_k(X_{0,1},\partial X_{0,1})\cong H^{6-n}(X_{0,1})\cong \Z$ and expanding the supergravity RR and NSNS gauge fields of degree $d_I$ in these cocycles we find the cobordism to compactify to a junction theory
supporting abelian $U(1)$ gauge theories with $(d_I-(6-n))$-form potentials. The cobordism $X_{0,1}$ gives rise to a junction theory where these abelian degrees of freedom localize in 4D. Brane wrappings which pass through $X_{0,1}$ result in defects of the 5D symmetry theory which pass through the junction and are dressed in the sense of \cite{Baume:2023kkf}. The group governing these dressings, in various degrees, is
\be
(U(1)\times \Z_3\times \Z_7)/ (\Z_3\times \Z_7)_{\textnormal{diag.}}
\ee
and the UV symmetry group $\Z_{17}$ interacts with the IR symmetry $\Z_3\times \Z_7$ via embedding purely into the $U(1)$ factor. Consequently, every defect which arises as a brane wrapping on a non-compact cycle, stretching from the singularities ${\mathbb{R}^{6}/\Z_{3}}$ and ${\mathbb{R}^{6}/\Z_{7}}$ within $\widetilde{\mathbb{R}^{6}/\Z_{17}}$ to the asymptotic boundary will be dressed, i.e., in the symTree passing the defect through the junction $\mathcal{J}_{0,1}$ will result in a non-trivial defect of the junction theory.

\section{Conclusions} \label{sec:CONC}

Symmetries provide important constraints on the dynamics of many quantum systems.
In this paper we have studied the generalized symmetries of a class of non-supersymmetric
backgrounds in type II string theory of the form $\mathbb{R}^{3,1} \times \mathbb{R}^{6} / \Gamma$. In the case
of type IIA backgrounds, we determined the generalized symmetries of ``pure geometry'' configurations. The quiver quantum mechanics of probe particles encodes the Dirac pairing of electric and magnetic states via the adjacency matrix for fermionic matter. This structure is directly visible in terms of the boundary orbifold structure of $S^5 / \Gamma$. In the case of type IIB backgrounds, we considered a stack of $N$ spacetime filling D3-branes probing the singularity. The same quiver data / geometric data encodes generalized symmetries for this theory. In both cases, many structures observed in related supersymmetric backgrounds naturally extend to this broader non-supersymmetric setting. We also encountered new phenomena, especially with regards to codimension 2 singularities and unsequestered tachyons in the boundary geometry. In the remainder of this section we discuss some potential avenues for future investigation.

While we have given a rather direct physical interpretation of the fermionic adjacency matrix in our quiver quantum mechanics / gauge theory, the role of the bosonic adjacency matrix is less clear. It would clearly be interesting to better understand the role (if any) of this structure.

In orbifolds with unsequestered tachyons we observed that the defect group
computed by quiver methods predicts extra structure beyond the homology groups $H_{\ast}(S^5 / \Gamma)$ for
$\Gamma$ a finite subgroup of $SU(4)$. While we motivated a plausible reason for these additional
contributions, it would be interesting to give a principled derivation.

It would also be of interest to study the structure of tachyon condensation in more general cases where $\Gamma$ is non-abelian.
A potential starting point would be to enumerate the finite subgroups of $SU(4)$ (see e.g., \cite{Hanany:1999sp} and references therein) and their associated quiver gauge theories.

A natural extension of the analysis presented here would be to directly track the strong coupling limit of our type IIA analysis, i.e., a possible M-theory uplift. In this setting, the dilaton obtained from circle compactification is still non-dynamical, so in principle we can lift all of these structures directly to a 5D quantum system. Of course, this will also require establishing a suitable uplift of tachyon condensation to the M-theory setting. Presumably the presence of candidate generalized symmetries can provide some insights into these instabilities.

In the case of supersymmetric backgrounds, $N$ D3-branes at the tip of a Calabi-Yau cone $\mathrm{Cone}(Y)$ yields a holographic dual of the form $\mathrm{AdS}_{5} \times Y$. In the case of $X = \mathrm{Cone}(Y)$ a non-supersymmetric orbifold there is an instability in such solutions \cite{Horowitz:2007pr}, so the connection with the worldvolume theory of the D3-branes is less clear. It would be interesting to track the fate of symmetry operators and defects to establish the existence / absence of a phase transition as one increases the 't Hooft coupling.

It would also be interesting to track the fate of generalized symmetries in the case of compact non-supersymmetric models with orbifold singularities, perhaps along the lines of \cite{Cvetic:2023pgm}. The presence of such topological structures and their dynamical counterparts once gravity is included could potentially shed light on the endpoint of tachyon condensation in such backgrounds.

\section*{Acknowledgements}

We thank M. Del Zotto, I.R. Klebanov, S.N. Meynet, E. Torres, and X. Yu for helpful correspondence and discussions.
The work of NB is supported by NSF GRFP DGE-2236662. The work of JJH is supported by DOE (HEP) Award
DE-SC0013528. The work of JJH and MH is supported in part by a University
Research Foundation grant at the University of Pennsylvania. The work of JJH
and MH is supported in part by BSF grant 2022100. The work of MH is also
supported by the Simons Foundation Collaboration grant \#724069 on
\textquotedblleft Special Holonomy in Geometry, Analysis and
Physics\textquotedblright. This research was supported in part by grant
NSF PHY-2309135 to the Kavli Institute for Theoretical Physics (KITP).

\appendix

\section{Further Details on Unsequestered Tachyons} \label{app:COOLNESS}

In this Appendix we provide some further discussion and examples centered on the defect group in the case of
unsequestered tachyons. In particular we provide further details on the
motivation for equation (\ref{eq:CoolFormula}) which we reproduce here for convenience of the reader:
\be
\mathbb{D}^{(1)} = \mathrm{Tor} ( \mathrm{Coker} K ) \cong (\Gamma/H)^2\oplus \lb  \bigoplus_{i} \, (\Gamma/H_{\mathscr{S}_{i}})^{|H_i|-1}\rb\,.
\ee
We focus on the case of $\Gamma = \mathbb{Z}_N$ abelian, acting on $\mathbb{C}^3$ with
integer weights $(v_1,v_2,v_3)_{\mathrm{hol}}\in \Z^3$ as:
\be
(Z_1,Z_2,Z_3)~\sim~(\zeta^{v_1}Z_1,\zeta^{v_2}Z_2,\zeta^{v_3}Z_3)
\ee
where $\zeta$ is an $N^{\textnormal{th}}$ primitive root of unity. We restrict ourselves to weights which are such that $\textnormal{gcd}(N,v_1,v_2,v_3)=1$, which implies that our action is faithful. If $v_1+v_2+v_3=0$ modulo $N$, then $\Gamma\subset SU(3)$ and otherwise $\Gamma\subset U(3)$.  Note that although we have been considering orbifolds of the form $\mathbb{R}^6/\Gamma^{\mathbf{s}}$, where $\mathbf{s}$ indicates the group action on the spinor representation of $SU(4)$, we can equivalently study orbifolds of the above form by changing to a complex basis and considering the induced action on the vector representation of $SO(6)$.

We define subgroups of $\Gamma$ given triplets $(i,j,k)$ drawn from $\{i,j,k\}=\{1,2,3\}$. Given this labelling, we let $H_{ij}\subset \Gamma$ denote the three subgroups of order $g_k=\textnormal{gcd}(N,v_k)$, i.e., we have
\be \label{eqn:Hij}
H_{ij}\cong \Z_{g_k}\subset \Z_N\,,
\ee
which is invariant under interchange of indices: $H_{ij}=H_{ji}$.
Further, let $H_{k}\subset \Gamma$ denote the three subgroups of
order $g_{ij}=\textnormal{gcd}(N,v_i,v_j)$, i.e., we have
\be\label{eqn: Hk}
H_{k}\cong \Z_{g_{ij}}\subset\Z_N\,.
\ee
We remark that the subgroups $H_{ij}$ and $H_k$ are not necessarily distinct subgroups of $\Z_N$. Furthermore, there are natural subgroup relations
\be
H_k\subseteq H_{ik}=H_{ki}\,,
\ee
and faithfulness of the action implies $\textnormal{gcd}(g_1,g_2,g_3)=1$. Finally, let $H = \langle H_{ij}, H_k\rangle$ denote the subgroup generated by all of these subgroups and
\be
H^{(1)}=\langle H_{k} \rangle \cong H_1\times H_2 \times H_3
\ee
where the isomorphism follows from the $g_{ij}$ being pairwise coprime by our faithfulness assumption.

These subgroups have various geometric interpretations within the asymptotic boundary of $X=\mathbb{C}^3/\Gamma$ which is induced from some $\mathbb{R}^6/\Gamma^{\mathbf{s}}$. 
First, we have via Armstrong's theorem
\be
\pi_1(\partial X)\cong \Gamma/H\,.
\ee
The subgroups $H_{ij}$ are generated by elements which fix the locus $Z_i=Z_j=0$. Similarly, the $H_{k}$ are generated by elements which fix the locus $Z_k=0$. The latter results in codimension 2 singularities at infinity with the model $\mathbb{C}/H_k$, and there is a full $\mathscr{S}_{ij}=S^3_{ij}/(\Gamma/H_k)$ worth of such singularities in $\partial X$. The model  $\mathbb{C}/H_k$ is non-trivially fibered over $\mathscr{S}_{ij}$ according to the extension
\be
0~\rightarrow~H_k~\rightarrow~\Gamma~\rightarrow~\Gamma/H_k ~\rightarrow~0\,.
\ee
Similar comments hold for codimension 4 singularities associated with $H_{ij}$ supported on $\mathscr{S}_{ij}=S^1_k/(\Gamma/H_{ij})$.

Given a codimension 2 singularity $\mathscr{S}_{ij}$ it may be intersected by other codimension 2 singularities $\mathscr{S}_{jk},\mathscr{S}_{ik}$ or contain codimension 4 singularities $\mathscr{S}_i,\mathscr{S}_j$. Note that any codimension 2 singularities always intersect pairwise, so given any codimension 2 locus the subgroup $H_{\mathscr{S}_{k}}\subset \Gamma$ generated by elements that fix some element of the associated $S^3_{ij}$ contains $H^{(1)}$. Further, we have contributions from codimension 4 structures, overall resulting in
\be
H_{\mathscr{S}_{k}}=\langle H^{(1)}, H_{ik}, H_{jk}\rangle\,.
\ee
We can characterize $H_{\mathscr{S}_{k}}$ conversely by noting that the only elements with fixed points not contributing to it are
\be
H_{ij}/\langle H_i, H_j\rangle
\ee
which is associated with the codimension 4 singularity linking $\mathscr{S}_{k}$ in $\partial X$. The quotient $H_{ij}/\langle H_i, H_j\rangle$ is precisely associated with those elements which have fixed point only along $S^1_k$, and which are not associated with codimension 4 structures along $S^3_{ik}$ and $S^3_{jk}$ which intersect in $S^1_k$. Therefore,
\be
H_{\mathscr{S}_{k}}\cong \Z_{\ell_k} \,, \qquad \ell_k= \textnormal{lcm}(g_{i},g_{j})\,.
\ee

Toric geometry suggests a simple presentation of the above data. First, note that $S^5$ admits a projection to a triangle $\Delta$ cut out by $|Z_1|^2+|Z_2|^2+|Z_3|^2=1$ which is a hyperplane in the positive octant $\mathbb{R}^3_{\geq 0}$. The fibers projecting onto the interior of $\Delta$ are $T^3$, the fibers projecting onto the edge $|Z_k|=0$ are $T^2$, and the fibers projecting onto the corner $|Z_i|,|Z_j|=0$ are $S^1$. The corners lift to circles, and the edges lift to 3-spheres.

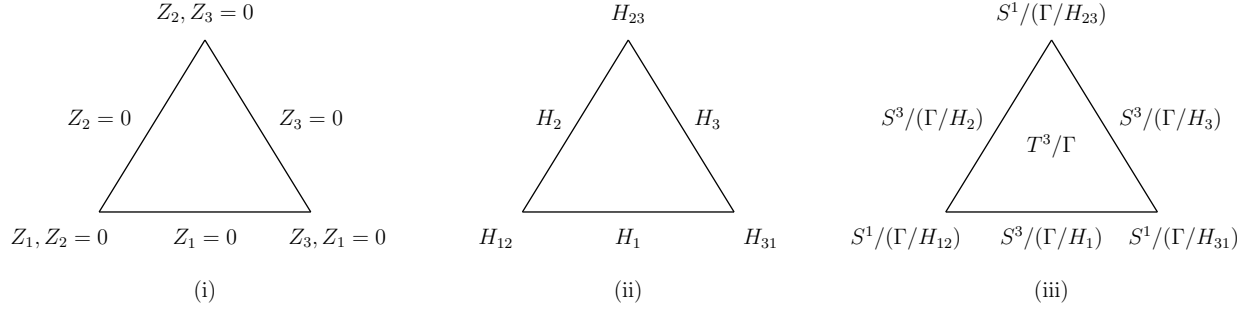
\begin{figure}
    \centering
    \scalebox{0.7}{\begin{tikzpicture}
	\begin{pgfonlayer}{nodelayer}
		\node [style=none] (0) at (-8, 0) {};
		\node [style=none] (1) at (-6, 3.25) {};
		\node [style=none] (2) at (-4, 0) {};
		\node [style=none] (3) at (-6, -1.5) {(i)};
		\node [style=none] (4) at (0, 0) {};
		\node [style=none] (5) at (2, 3.25) {};
		\node [style=none] (6) at (4, 0) {};
		\node [style=none] (7) at (2, -1.5) {(ii)};
		\node [style=none] (8) at (-6, 3.75) {$Z_2,Z_3=0$};
		\node [style=none] (9) at (-8.75, -0.5) {$Z_1,Z_2=0$};
		\node [style=none] (10) at (-3.5, -0.5) {$Z_3,Z_1=0$};
		\node [style=none] (11) at (8, 0) {};
		\node [style=none] (12) at (10, 3.25) {};
		\node [style=none] (13) at (12, 0) {};
		\node [style=none] (14) at (10, -1.5) {(iii)};
		\node [style=none] (18) at (-4, 1.75) {$Z_3=0$};
		\node [style=none] (19) at (-8, 1.75) {$Z_2=0$};
		\node [style=none] (20) at (-6, -0.5) {$Z_1=0$};
		\node [style=none] (21) at (2, 3.75) {$H_{23}$};
		\node [style=none] (22) at (-0.5, -0.5) {$H_{12}$};
		\node [style=none] (23) at (4.5, -0.5) {$H_{31}$};
		\node [style=none] (24) at (3.5, 1.75) {$H_3$};
		\node [style=none] (25) at (0.5, 1.75) {$H_2$};
		\node [style=none] (26) at (2, -0.5) {$H_1$};
		\node [style=none] (27) at (10, 1.25) {$T^3/\Gamma$};
		\node [style=none] (28) at (10, 3.75) {$S^1/(\Gamma/H_{23})$};
		\node [style=none] (29) at (7.25, -0.5) {$S^1/(\Gamma/H_{12})$};
		\node [style=none] (30) at (12.5, -0.5) {$S^1/(\Gamma/H_{31})$};
		\node [style=none] (31) at (12.25, 1.75) {$S^3/(\Gamma/H_{3})$};
		\node [style=none] (32) at (7.75, 1.75) {$S^3/(\Gamma/H_{2})$};
		\node [style=none] (33) at (10, -0.5) {$S^3/(\Gamma/H_{1})$};
	\end{pgfonlayer}
	\begin{pgfonlayer}{edgelayer}
		\draw [style=ThickLine] (0.center) to (2.center);
		\draw [style=ThickLine] (2.center) to (1.center);
		\draw [style=ThickLine] (1.center) to (0.center);
		\draw [style=ThickLine] (4.center) to (6.center);
		\draw [style=ThickLine] (6.center) to (5.center);
		\draw [style=ThickLine] (5.center) to (4.center);
		\draw [style=ThickLine] (11.center) to (13.center);
		\draw [style=ThickLine] (13.center) to (12.center);
		\draw [style=ThickLine] (12.center) to (11.center);
	\end{pgfonlayer}
\end{tikzpicture}}
    \caption{Sketch of the projection $S^5/\Gamma\rightarrow \Delta$ characterizing the orbifold data of $S^5/\Gamma$. In (i) we show the triangle base $\Delta$ and give conventions for cutting out its edges and corners. In (ii) we associate the subgroups $H_k,H_{ij}$ with these edges and corners. In (iii) we label the edges and corners by their preimage with respect to the projection $S^5/\Gamma\rightarrow \Delta$. }
    \label{fig:Labelling}
\end{figure}

The projection to $\Delta$ factors through the quotient by $\Gamma$ and we obtain a similar fibration structure for $S^5/\Gamma\rightarrow \Delta$. The corners lift to $S^1/(\Gamma/H_{ij})$ and the edges lift to $S^3/(\Gamma/H_k)$. As such, we can represent the full orbifold structure of $S^5/\Gamma$ by labelling the triangle base $\Delta$ as in figure \ref{fig:Labelling}.
Empirically, then, we find equation \eqref{eq:CoolFormula}.

\subsection{Illustrative Example}

\begin{figure}
    \centering
    \scalebox{0.8}{\begin{tikzpicture}
	\begin{pgfonlayer}{nodelayer}
		\node [style=none] (4) at (-6, -1) {};
		\node [style=none] (5) at (-4, 2.25) {};
		\node [style=none] (6) at (-2, -1) {};
		\node [style=none] (7) at (-4, -2.5) {(i)};
		\node [style=none] (11) at (2, -1) {};
		\node [style=none] (12) at (4, 2.25) {};
		\node [style=none] (13) at (6, -1) {};
		\node [style=none] (14) at (4, -2.5) {(ii)};
		\node [style=none] (18) at (-4, 2.75) {$\Z_{33}$};
		\node [style=none] (19) at (-6.5, -1.5) {$\Z_{21}$};
		\node [style=none] (20) at (-1.5, -1.5) {$\Z_{35}$};
		\node [style=none] (21) at (-2.5, 0.75) {};
		\node [style=none] (22) at (-5.5, 0.75) {$\Z_3$};
		\node [style=none] (23) at (-4, -1.5) {$\Z_7$};
        \node [style=none] (24) at (-2.5, 0.75) {$1$};
		\node [style=none] (25) at (4, 0.25) {$T^3/\Z_{1155}$};
		\node [style=none] (26) at (4, 2.75) {$S^1/\Z_{35}$};
		\node [style=none] (27) at (1.25, -1.5) {$S^1/\Z_{55}$};
		\node [style=none] (28) at (6.5, -1.5) {$S^1/\Z_{33}$};
		\node [style=none] (29) at (6.25, 0.75) {$S^3/\Z_{1155}$};
		\node [style=none] (30) at (1.75, 0.75) {$S^3/\Z_{385}$};
		\node [style=none] (31) at (4, -1.5) {$S^3/\Z_{165}$};
	\end{pgfonlayer}
	\begin{pgfonlayer}{edgelayer}
		\draw [style=ThickLine] (4.center) to (6.center);
		\draw [style=ThickLine] (6.center) to (5.center);
		\draw [style=ThickLine] (5.center) to (4.center);
		\draw [style=ThickLine] (11.center) to (13.center);
		\draw [style=ThickLine] (13.center) to (12.center);
		\draw [style=ThickLine] (12.center) to (11.center);
	\end{pgfonlayer}
\end{tikzpicture}}
    \caption{We sketch in (i) the subgroups $H_k,H_{ij}$ and in (ii) singular loci for the quotient $S^5/\Gamma$ with $\Gamma=\Z_{1155}$ and weights $(33,35,42)$.  }
    \label{fig:Labelling2}
\end{figure}
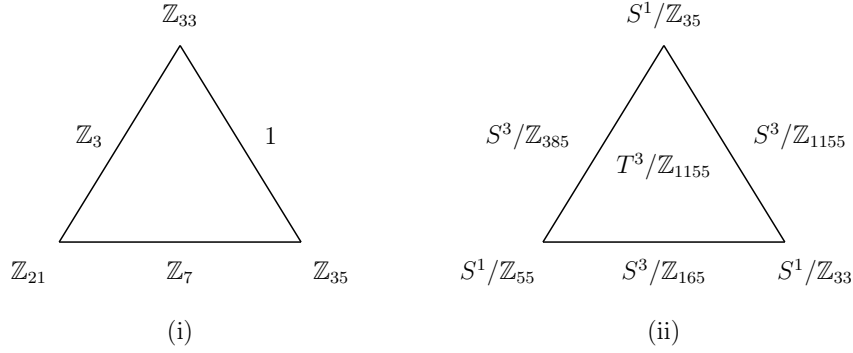

At this point, we give an example. Consider $\Z_N$ with $N=3\times 5\times 7\times 11=1155$ with weights $(v_1,v_2,v_3)_{\textnormal{hol}}=(3\times 11, 5\times 7, 2\times 3\times 7 )=(33,35,42)$. The various subgroups are
\be \ba
H_1=\Z_7 \,, &&\quad  H_2=\Z_3\,, && \quad H_3=1\,, \\
H_{23}=\Z_{33}\,, && \quad H_{31}=\Z_{35}\,, &&\quad H_{12}=\Z_{21}\,, \\
\ea \ee
and $H=\Z_N$. Further, the groups generated by elements with fixed points contained in various three-spheres associated with codimension 2 singularities are
\be
H_{\mathscr{S}_{1}}=\Z_{105}\,, \qquad H_{\mathscr{S}_{2}}=\Z_{231} \,, \qquad H_{\mathscr{S}_{3}}=\Z_{1155}\,.
\ee
Therefore, $\Gamma/H=1$ and
\be
\mathbb{D}^{(1)} \cong  \Z_{11}^6\oplus \Z_5^2\,,
\ee
which agrees with what we find by computing $K$ using quiver based methods.
A summary of the geometric data is given in figure \ref{fig:Labelling2}.

\section{Worldsheet Considerations} \label{app:WORLDSHEET}

In this Appendix, we will review the worldsheet analysis of tachyon condensation for non-supersymmetric orbifolds of the form $\mathbb{R}^6 / \Gamma$, closely following the treatment given in references \cite{Lee:2003ar, Morrison:2004fr}, to which we refer the interested reader for further details / explanation. Our main aim here will be to briefly summarize how to read off the relevant tachyonic operators in the twisted sectors of the theory.

To set notation, we work in the ``holomorphic conventions'' of \cite{Lee:2003ar, Morrison:2004fr} where one introduces local coordinates $(Z_1,Z_2,Z_3)$ on $\mathbb{C}^3$. The group action of a $\mathbb{Z}_N$ orbifold is assumed to act via a primitive $N^{th}$ root of unity $\omega$
on the coordinates as follows:
\begin{equation}\label{eq:groupaction}
(Z_1, Z_2, Z_3) \mapsto (\omega^{k_1} Z_1, \omega^{k_2} Z_2, \omega^{k_3} Z_3),
\end{equation}
and we denote this as $\mathbb{C}^3 / \mathbb{Z}_N$ with weights $(k_1, k_2, k_3)_{\mathrm{hol}}$. Throughout, we focus on examples where $\textnormal{gcd}(k_i,N)=1$, for some $k_i$, allowing us to redefine the generator of $\Z_N$ such that, after possibly additionally relabelling coordinates, the first weight becomes 1. These cases are therefore captured by restricting to the cases with $k_1=1$ on which we focus going forward.

Observe that at the level of the worldsheet CFT we can independently treat the orbifolds for the $Z_1$, $Z_2$ and $Z_3$ states (and their worldsheet superpartners). With this in mind, we get a chiral / anti-chiral ring of operators for each holomorphic coordinate separately. Following \cite{Morrison:2004fr}, we work in conventions where the most relevant tachyonic deformation is in the $(c_1, c_2, c_3)$ ring.
In this ring, the GSO action will project out twisted sector operators in the $j$-th twisted sector ($j=1,\dots,N-1$) if the GSO exponent $E_j$ for worldsheet scalars $X_{j}(\sigma + 2 \pi) = (-1)^{E_j} X_{j}(\sigma)$ is even. Here, $E_j$ is specified in this ring via:
\begin{equation}
    E_j = \sum_{i=1}^3 \textrm{Int}\left( \frac{jk_i}{N} \right) =  \textrm{Int}\left( \frac{jk_2}{N} \right) + \textrm{Int}\left( \frac{jk_3}{N} \right), \label{eq:opGSO}
\end{equation}
where $\textrm{Int}(x)$ refers to the integer part of $x$. Alternatively, one could have chose to study the ring of operators with anti-chiral action on one (or more) of the target space coordinates (e.g. the $(c_1,c_2,\overline{c_3})$ ring). In this setting, the GSO action would project out twisted sector operators if $E_j$ is odd. Similar conditions follow for the remaining rings of operators.

The twisted sector operators that survive the GSO action are then characterized according to their R-charge:
\begin{equation}
    R_j = \sum_{i=1}^3 \textrm{Frac}\left( \frac{jk_i}{N} \right) = \frac{j}{N} + \textrm{Frac}\left( \frac{jk_2}{N} \right) + \textrm{Frac}\left( \frac{jk_3}{N} \right). \label{eq:oprcharge}
\end{equation}
Here $\textrm{Frac}(x)$ refers to the fractional part of $x$.\footnote{Note that if $x$ is positive, then $\textrm{Frac}(-x) = 1- \textrm{Frac}(x)$ and $\textrm{Int}(-x) = -1- \textrm{Int}(x)$.} Each operator is characterized on the worldsheet as follows:
\begin{itemize}
    \item $R_j<1$: the operator is tachyonic.
    \item $R_j=1$: the operator is marginal.
    \item $R_j>1$: the operator is irrelevant.
\end{itemize}
Resolutions of the target space geometry are controlled by the tachyonic and marginal operators.\footnote{Tachyonic operators are a feature of non-supersymmetric orbifold singularities. Supersymmetric configurations, such as a Calabi-Yau threefold will only have marginal operators which can be used to resolve the singularity.} The most relevant tachyonic operator is that with the lowest R-charge and will drive the first deformation. This will be followed by subsequent deformations due to any remaining tachyonic operators, where the order of these deformations is determined again by the corresponding renormalized R-charge \cite{Morrison:2004fr}. Importantly, a non-supersymmetric orbifold that is compatible with a type II GSO projection will always have at least one twisted sector tachyon. As such, the endpoint of the RG-flow associated to tachyon condensation is always either a smooth space, or a singular space with enhanced supersymmetry where there are non-chiral metric blow-up modes to resolve the space. Throughout the note, we label tachyonic operators in the $j$-th twisted sector as:
\begin{equation}
    T_j = \left( \frac{j}{N},\textrm{Frac}\left(\frac{jk_2}{N}\right),\textrm{Frac}\left(\frac{jk_3}{N}\right) \right)_{\mathrm{hol}},
\end{equation}
where the weights follow the relations in (\ref{eq:groupaction}).

We are now ready to review the resolution of non-supersymmetric orbifolds as driven by tachyon condensation.
We have been studying non-supersymmetric orbifold singularities of the form $\mathbb{R}^6/\Gamma_{SU(4)}$ for
$\Gamma_{SU(4)}$ a finite abelian subgroup of $SU(4)$. As such, we can study the resolution of the singularity
as described by its toric geometry.

The toric fan $\Sigma$ associated to $X$ is defined by the three edges
\begin{equation}
    \alpha_1 = ((N,-k_2,-k_3)), \;\; \alpha_2 = ((0,1,0)), \;\; \alpha_3 = ((0,0,1)).
\end{equation}

These vectors $\alpha_i$ are the vertices of the simplex $\Delta$ that defines the fan of cones subtended with the origin.

\begin{figure}
    \centering
    \scalebox{1}{
    \begin{tikzpicture}
	\begin{pgfonlayer}{nodelayer}
		\node [style=none] (0) at (-6, -8) {};
		\node [style=none] (1) at (2, -8) {};
		\node [style=none] (2) at (0, -4) {};
		\node [style=none] (3) at (-0.25, -6.5) {};
		\node [style=none] (4) at (-1.5, -6.5) {};
		\node [style=none] (5) at (-6.25, -8.5) {$\alpha_1$};
		\node [style=none] (5) at (2.25, -8.5) {$\alpha_2$};
		\node [style=none] (6) at (0.5, -3.75) {$\alpha_3$};
		\node [style=none] (7) at (-1.75, -6.25) {$T$};
	\end{pgfonlayer}
	\begin{pgfonlayer}{edgelayer}
		\draw [style=ThickLine] (0.center) to (1.center);
		\draw [style=ThickLine] (0.center) to (2.center);
		\draw [style=ThickLine] (2.center) to (1.center);
		\draw [style=DashedLine] (2.center) to (3.center);
		\draw [style=DashedLine] (0.center) to (3.center);
		\draw [style=DashedLine] (3.center) to (1.center);
		\draw [style=BlueLine] (0.center) to (4.center);
		\draw [style=BlueLine] (4.center) to (2.center);
		\draw [style=BlueLine] (4.center) to (3.center);
		\draw [style=BlueLine] (4.center) to (1.center);
	\end{pgfonlayer}
\end{tikzpicture}}
    \caption{Depiction of a general simplex $\Delta$ (black lines). The most relevant tachyon, which is located at $T$ subdivides $\Delta$ into subcones (blue lines), each corresponding to a space of the form $\mathbb{R}^6/\Gamma$.}
    \label{fig:toricpic}
\end{figure}
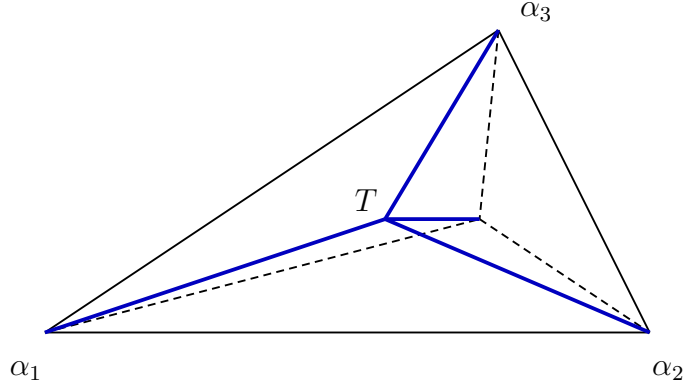

There is a correspondence between operators in the orbifold theory and points in the simplex. In particular, an operator $\mathcal{O}_j$ with $R$-charge $R_j$ corresponds to a lattice point $T_j = \left(\left(j,-\textrm{Int}\left(\frac{jk_2}{N}\right), -\textrm{Int}\left( \frac{jk_3}{N}\right)\right)\right)$. Tachyonic and marginal operators will appear in the toric variety and subdivide the simplex into subcones. Tachyonic operators will appear in the interior of $\Delta$, while marginal operators will appear on the boundary. All of the computations in this paper take the limit where the most tachyonic operator (i.e. the operator with the lowest R-charge) will condense first, and subdivide $\Delta$ with this operator (see figure \ref{fig:toricpic}).\footnote{This procedure does not generalize to cases where there are multiple tachyons with equal R-charges in the same ring of operators.} After condensation, there will be three subcones that are each less singular than the starting cone. Each of these subcones will correspond to an orbifold singularity of the form $\mathbb{R}^6/\Gamma$ that can be read off by studying the toric fan of the subcone. If any of these subcones still correspond to a non-supersymmetric orbifold, then there will be a tachyon with renormalized R-charge to facilitate subsequent blowups.\footnote{Importantly, it can be shown that if the original non-supersymmetric orbifold $\mathbb{R}^6/
\Gamma$ admits a type II GSO projection, then the orbifold singularities after a blow-up will as well.} The computation of renormalized R-charge is given in \cite{Morrison:2004fr}.

To be precise, we label the location of the most relevant tachyon by $T$. Then, the original cone, which is given by $C[\alpha_1,\alpha_2,\alpha_3]$, is divided into three subcones:
$C[T,\alpha_1,\alpha_2]$, $C[T,\alpha_2,\alpha_3]$, and $C[T,\alpha_3,\alpha_1]$.

The data of a toric variety can also be encoded as the ground states of a gauged linear sigma model (GLSM) \cite{Hori:2003ic}. The GLSM allows one to study the tachyon condensation throughout the flow instead of just at the endpoints as we have done in this paper.

\section{More Non-Abelian Examples: Double Twisting} \label{app:SU2stack}
In this appendix we give further examples of non-Abelian orbifolds, generalizing the ideas of section \ref{sec:NONAB}, by taking an additional Abelian twist and considering the geometry
\be
\mathbb{R}^6/(\Z_M \times \Z_N\times \Gamma_{SU(2)})\,,
\ee
with $\Gamma^{\mathbf{s}}=\Z_M \times \Z_N\times \Gamma_{SU(2)}$ acting faithfully. Here $\Gamma_{SU(2)}$ is a finite subgroup of $SU(2)$ and $N,M$ are odd. We denote by 
$(e_{1},e_{2};f_{1};f_{2})$ an ordered basis of the $\mathbf{4}$ of $SU(4)$. The orbifolding we consider is
\begin{equation}\ba
\mathbb{Z}_M \text{ action:\ } &\qquad (e_1,e_2,f_1,f_2) \mapsto (\zeta e_1,\zeta e_2,\zeta^{-2}f_1,f_2)\\
\mathbb{Z}_{N}\text{ action:\ } &\qquad  (e_1,e_2,f_1,f_3) \mapsto (\xi e_1,\xi e_2,f_1,\xi^{-2}f_2)\\
\Gamma_{SU(2)}\text{ action:\ } &\qquad  (e_1,e_2,f_1,f_3) \mapsto (g_{ij}e_j,f_1,f_2)
\ea \end{equation}
where $\zeta,\xi$ are a primitive $M^{th},N^{th}$ root of unity respectively and $g_{ij}$ the matrix representation for an ADE subgroup of $SU(2)$.

We now derive the probe theory for $\mathbb{R}^6/\Gamma^{\mathbf{s}}$. To begin, we denote by $Q$ the quiver for the probe theory for $\R^6/\Gamma_{SU(2)}$ which comes with its own adjacency matrix for bosons and fermions, the latter we denote by $A_{qq'}^F$. The effect of taking an additional quotient by $\Z_N, \Z_M$ now amounts to decomposing the bosons and fermions of the probe theory associated with $\R^6/\Gamma_{SU(2)}$. The fermions decompose as
\be 
{\mathbf 4}\rightarrow {\mathbf 2}\oplus {\mathbf 1}\oplus {\mathbf 1}\,.
\ee 
Further, we need to introduce a total of $NM$ copies of this quiver, denoted as $\mathcal{Q}_{ij}$ with $i=1,\dots,N$ and $j=1,\dots,M$ labelling the irreducible representations of $\Z_N\times \Z_M$ (see \cite{Berenstein:2000mb} for a general approach to such examples). Next, we reconnect the nodes following \cite{Douglas:1996sw} according to the above group action, analogous to our discussion in section \ref{sec:NONAB}. Overall,
the adjacency matrix for the fermionic quiver computes to
\be
\mathbb{A}_{(q,ij),(q',i'j')}^F=A_{q,q'}^F\lb \delta_{i,i'}\delta_{j,j'+1}+\delta_{i,i'+1}\delta_{j,j'}\rb+\delta_{q,q'}\lb \delta_{i,i'-2}\delta_{j,j'}+ \delta_{i,i'}\delta_{j,j'-2}\rb\,.
\ee

With this result in hand, let us consider
$\mathbb{R}^6/\Gamma^{\mathbf{s}}$ as a geometric background in IIA as in section \ref{sec:IIA}. We compute from here the 4D defect group of lines:
\begin{equation}
\label{eq:triplequotient}
    \mathbb{D}^{(1)} = \left( \mathbb{Z}_{M/g} \oplus \mathbb{Z}_{N/g} \right)^2 \oplus (\mathbb{Z}_l\times Z_{SU(2)})^{g-1}.
\end{equation}
Here $g=\textnormal{gcd}(N,M)$ and $l=\textnormal{lcm}(N,M)$ and $Z_{SU(2)}$ is the center subgroup of the simply connected Lie group associated with $\Gamma_{SU(2)}$ via the McKay correspondence.

We turn to give a geometric derivation of this result. To simplify the discussion let us first consider the case $\Gamma_{SU(2)}=1$, for which the orbifolding is only by $\Z_N\times \Z_M$. For this case, in the obvious notation, we have
\begin{equation}\label{eq:2abelian}\ba
    \mathbb{A}_{(ij);(i'j')}^F &=2\lb \delta_{i,i'}\delta_{j,j+1}+\delta_{i,i'+1}\delta_{j,j'} \rb+\delta_{i,i'-2}\delta_{j,j'}+\delta_{i,i'}\delta_{j,j'-2} \\[0.5em]
     \mathbb{D}^{(1)} &= \left( \mathbb{Z}_{M/g} \oplus \mathbb{Z}_{N/g} \right)^2 \oplus (\mathbb{Z}_l)^{g-1},
    \ea
\end{equation}
which we now discuss in greater detail. Afterwards we will turn the non-Abelian quotient back on.

First, we observe that while the Abelian actions by $\Z_N,\Z_M$ are fixed point free, there exist (anti)diagonal subgroups which have codimension 2 and 4 fixed points.

We find one codimension 2 singularity localized at $Z_1=0$. For the above phase rotations to cancel we need to consider the diagonal $\Z_g$ subgroups generated by $\zeta^{M/g}$ and $\xi^{N/g}$. This determines $H_{1}\cong \Z_g^{\textnormal{diag}}$ and this contributes
\be \label{eq:Contribution1}
\lbb \lb \Z_M\oplus\Z_N \rb/ \Z_g^{\textnormal{diag}}\rbb^{g-1}\cong\Z_l^{g-1}
\ee
to the defect group following our general analysis in section \ref{sec:IIA}.

Further, we find one codimension 4 singularity localized at $Z_2,Z_3=0$. Taking analogous steps as in the above analysis we find the anti-diagonal subgroup $\Z_g^{\overline{\mathrm{diag}}}=H_{23}$ to have fixed points. The subgroup of $\Z_N\times \Z_M$ generated by all group elements with fixed points is therefore $H=\Z_g^{\textnormal{diag}}\times\Z_g^{\overline{\mathrm{diag}}}$ and via Armstrong's theorem we find an additional contribution to the defect group equal to
\be \label{eq:Contribution2}
\lbb \lb \Z_M\oplus\Z_N \rb/ \lb \Z_g^{\textnormal{diag}}\oplus \Z_g^{\overline{\mathrm{diag}}}\rb \rbb ^{2}\cong \lbb \Z_{M/g}\oplus \Z_{N/g}\rbb^2\,.
\ee
Overall, taking the direct sum of \eqref{eq:Contribution1} and \eqref{eq:Contribution2}, we find exactly \eqref{eq:2abelian}.

Let us now turn the non-Abelian quotient back on. Note that all of $\Gamma_{SU(2)}$ fixes a circle of the bosonic $S^5\subset \mathbb{R}^6$, which is also fixed by $\Z_g^{\overline{\mathrm{diag}}}$ and that the center subgroup $Z_{SU(2)}\subset \Gamma_{SU(2)}$ and $\Z_g^{\overline{\mathrm{diag}}}$ can have overlap in the above parametrization in a common subgroup $Z$, and in this case the group acting faithfully on the geometry $\mathbb{R}^6$ is $\lbb (\Z_g^{\overline{\mathrm{diag}}}\times Z_{SU(2)}) / Z\rbb \times \Z_l$. Consider the case in which the initial group action is faithful, i.e., $Z=0$. In this case our geometric analysis computes the defect group \eqref{eq:triplequotient} as the codimension 2 locus is quotiented further by $\Gamma_{SU(2)}$, in particular, via Armstrong's theorem, there is no contribution to the homology portion of the result.

Next, we turn to discuss the non-compactly supported tachyons. For this, note first that $\Gamma_{SU(2)}$ acts supersymmetrically, therefore, there are now instabilities associated with singularities arising from this quotient and it will be sufficient to discuss the ones associated with the quotient of $\Z_N\times \Z_M$.

 The two subgroups of $\Z_N\times \Z_M$ which give rise to the codimension 2 and 4 singularities are $\Z_g^{\textnormal{diag}}$ and $\Z_g^{\overline{\mathrm{diag}}}$ respectively. As the $\Z_M$ action is a subgroup of $SU(3)$, while $\Z_N$ is not, we have that tachyons are supported on the singularities associated to both. 
 
 The fixed point sets of $\Z_g^{\textnormal{diag}}$ and $\Z_g^{\overline{\mathrm{diag}}}$ are disjoint and we can study their decay independently of another. This is made explicit by noting that the bosonic geometry $\mathbb{R}^6/(\Z_N\times \Z_M)$, due to our choices of weights, can be rewritten into the form
\be
\mathbb{R}^6/(\Z_N\times \Z_M)=(\mathbb{R}^4\times (\mathbb{R}^2/\Z_g^{\textnormal{diag}}))/\Z_l
=((\mathbb{R}^4/\Z_g^{\overline{\mathrm{diag}}})\times \mathbb{R}^2)/\Z_l'\,,
\ee
from which we immediately derive the local geometry for each of the singular loci. More precisely, the total space can be viewed as a $\mathbb{R}^2/\Z_g^{\textnormal{diag}}$ or $\mathbb{R}^4/\Z_g^{\overline{\mathrm{diag}}}$ bundle over a $\mathbb{R}^4/\Z_l$ or $\mathbb{R}^2/\Z_l'$ base respectively. Taking the first perspective we can analyze the codimension 2 tachyon, which is then localized in the fiber and decays at infinity via dilaton pulses following our discussion in section \ref{sec:IIA}. Taking the second perspective we can analyze the codimension 4 tachyon which triggers a toric blowup of the $\mathbb{R}^4/\Z_g^{\overline{\mathrm{diag}}}$ at infinity, again resolving the fiber. Away from infinity, at the tip of the cone, these instabilities interact.

Let us highlight the main features of this example. First, the codimension 4 singularities are not supersymmetric and so the asymptotic geometry changes as a function of time (when used as IIA background) or scale (when used as IIB background). Second, we clearly see that given a fixed point free action (in this example start either with $\Z_N$ or $\Z_M$) which gives rise to a `small' defect group we expect to drastically enlarge the defect group by any subsequent quotients, unless we make precise arithmetic choices, e.g., take $N,M$ to be coprime.

\newpage

\bibliographystyle{utphys}
\bibliography{NonSUSYOrb}

\end{document}